\documentclass[aps,prb,twocolumn,10pt,showkeys,superscriptaddress,floatfix]{revtex4-2}
\usepackage{times}
\usepackage{amsmath}
\usepackage{amssymb}
\usepackage{amsfonts}
\usepackage{multirow}
\usepackage{mathrsfs}
\usepackage{graphicx}
\usepackage{subfigure}
\usepackage{booktabs}
\usepackage{rotating}
\usepackage{longtable} 
\usepackage{bm}

\usepackage{epstopdf}
\usepackage{comment}
\usepackage{graphicx}
\usepackage{subfigure}
\usepackage{amsmath}
\usepackage{orcidlink}

\makeatletter

\newcommand{\Rmnum}[1]{\expandafter\@slowromancap\romannumeral #1@} 
\makeatother
\makeatletter
\newcommand*{\rom}[1]{\expandafter\@slowromancap\romannumeral #1@}
\makeatother

\makeatletter

\hypersetup{
  colorlinks   = true, 	
  urlcolor     = blue, 	
  linkcolor    = blue, 	
  citecolor   = blue 
} 
\usepackage{hyperref}
\begin{document}
\title{ Letelier-AdS Black Hole Surrounded by a Perfect Fluid Dark Matter in the presence of Quintessence}

\author{Faizuddin Ahmed\orcidlink{0000-0003-2196-9622}}
\email{faizuddinahmed15@gmail.com}
\affiliation{Department of Physics,  The Assam Royal Global University, Guwahati, 781035, Assam, India}
\author{Edilberto O. Silva\orcidlink{0000-0002-0297-5747}}
\email{edilberto.silva@ufma.br (Corresp. author)}
\affiliation{Programa de P\'os-Gradua\c c\~ao em F\'{\i}sica \& Coordena\c c\~ao do Curso de F\'{\i}sica -- Bacharelado, Universidade Federal do Maranh\~{a}o, 65085-580 S\~{a}o Lu\'{\i}s, Maranh\~{a}o, Brazil}

\begin{abstract}
This study investigates a Schwarzschild-anti-de Sitter black hole coupled to a cloud of strings featuring only the electric-like component of the string bivector, embedded in a perfect fluid dark matter and a quintessence field. We examine the dynamics of photons and massive particles, focusing on trajectories, photon spheres, BH shadows, their topological characteristics, and innermost stable circular orbits (ISCOs), and emphasizing the influence of string cloud, perfect-fluid dark matter, and quintessence-like field parameters. Additionally, we explore the black hole’s thermodynamics, deriving the Hawking temperature, Gibbs free energy, and specific heat, and discuss the modified first law of thermodynamics and thermodynamic topology under external matter fields. We demonstrate that the presence of a string cloud, perfect-fluid dark matter, and a quintessence-like field together modifies the geodesic structure and thermodynamic properties, thereby shifting the results relative to the standard Schwarzschild BH solution. 
\end{abstract}

\keywords{Black hole; King dark matter; isotropic configuration; accretion disk; topological charge
}
\maketitle

\small 

\section{Introduction} \label{Sec:1}

The study of black holes has gained unprecedented significance following recent breakthroughs in observational astronomy. The Event Horizon Telescope (EHT) collaboration has provided the first-ever direct images of the shadow of a supermassive black hole, initially capturing M87* in 2019 \cite{EHT2019} and more recently Sagittarius A* (Sgr A*), the supermassive black hole at the center of our galaxy \cite{EHT2022}. These observations not only offer compelling evidence supporting the existence of black holes predicted by general relativity but also open new avenues for testing fundamental physics in extreme gravitational regimes. The imaging of M87* revealed a bright emission ring surrounding a dark shadow consistent with theoretical models of a Kerr black hole, thus enabling stringent tests of gravity near the event horizon. Similarly, the image of Sgr A* provides crucial insights into the dynamics and environment of our galactic nucleus. These experimental breakthroughs highlight the importance of studying black hole properties-geometry, thermodynamics, and interactions with surrounding matter fields-to better understand their role in astrophysics and fundamental physics. Such studies are essential for interpreting observational data and for advancing our knowledge of gravity, quantum effects near horizons, and the nature of spacetime itself. 

Cosmic strings are one-dimensional topological defects that can form during symmetry-breaking phase transitions in the early universe \cite{Kibble1976,Vilenkin2000}. These objects have been extensively studied due to their potential impact on large-scale structure formation. Specifically, cosmic strings may have acted as seeds for primordial density fluctuations, thereby influencing the formation of galaxies and galaxy clusters \cite{Mitchell1987}. An influential model to describe such objects within the framework of General Relativity (GR) is the Letelier cloud of strings model \cite{Letelier1979}, which treats a cloud of strings as a pressureless perfect fluid. This approach has been widely adopted to construct black hole solutions not only in classical GR but also in various extended theories of gravity such as Einstein-Gauss-Bonnet and Lovelock gravity \cite{Letelier1981,Letelier1983,Singh2020,Nascimento2024a,Nascimento2024b,Chabab2020,Dadhich2022}. These solutions extend and generalize classical results, such as the Schwarzschild metric, by incorporating the effects of cosmic strings, thereby enriching our understanding of gravitational phenomena in their presence. The string cloud framework offers a robust alternative to conventional point-particle models by aligning with the predictions of string theory, which posits that the universe's fundamental constituents may be one-dimensional strings rather than zero-dimensional particles. Recent studies have further highlighted the gravitational and astrophysical importance of both cosmic strings and fundamental strings, underscoring their relevance in modern cosmology and astrophysics \cite{Synge1960}. These investigations illuminate the role these strings might play in phenomena ranging from gravitational lensing to cosmic microwave background anisotropies, making them key elements in the quest to unify high-energy physics with cosmological observations.

One of the most significant challenges in modern physics is understanding the nature of dark matter. Although dark matter accounts for approximately 27\% of the universe's total energy density, its direct detection remains elusive, making it one of the greatest unresolved mysteries in cosmology and particle physics. Understanding dark matter is crucial not only for explaining the formation and evolution of large-scale cosmic structures and galaxies but also for studying its interaction with dark energy, the driver of the universe’s accelerated expansion. The perfect fluid dark matter model provides a compelling framework for investigating dark matter's role in astrophysical and cosmological contexts, particularly in relation to black holes. Unlike standard particle-based models, perfect fluid dark matter treats dark matter as a continuous, non-viscous fluid governed by specific equations of state \cite{Kiselev2003a}. This fluid-like depiction enables the modeling of dark matter distributions around BHs and allows examination of how these distributions modify the observable characteristics of BH spacetimes \cite{Qiao2023}. 

Several studies have shown that incorporating perfect fluid dark matter alters the metrics of well-known black hole solutions. Modifications have been predicted and analyzed for the Schwarzschild \cite{Rayimbaev2021,Hamil2025}, Bardeen \cite{Zhang2021}, Kerr \cite{Hou2018,Rizwan2019}, Reissner-Nordström (RN) \cite{Hu2019}, and Euler-Heisenberg \cite{Ma2024,Shahzad2025} solutions. These alterations have significant implications for gravitational wave signatures \cite{Shaymatov2021,Li2022}, potentially affecting geodesic motion \cite{Das2021}, spacetime stability, and thermodynamic phase transitions \cite{Mustafa2024,Zhang2024,Ashraf2025}. Furthermore, perfect fluid dark matter influences several astrophysical phenomena associated with black hole's, including shadows \cite{Atamurotov2022,Atamurotov2024}, gravitational lensing \cite{Atamurotov2021}, accretion disk properties \cite{HeydariFard2023}, and deflection angles of light \cite{Yildiz2024}. The thermodynamic behavior of BHs immersed in perfect fluid dark matter environments has also been extensively studied \cite{Rakhimova2023,Rizwan2023}, along with their quasinormal modes \cite{Das2023,Chen2024} and Gibbs free energy functions \cite{AmaTulMughani2022,Javed2024,Fatima2025}. Moreover, the structure of event horizons \cite{Feng2024} and wormhole solutions \cite{Sekhmani2025} have been shown to be sensitive to the presence of perfect fluid dark matter. Recently, Sekhmani {\it et al.}~\cite{Sekhmani2025} investigated the thermodynamics and phase transitions of asymptotically ModMax anti-de Sitter (AdS) black holes in the presence of perfect fluid dark matter. Shaymatov {\it et al.} \cite{Shaymatov2021} investigated the effects of perfect fluid dark matter on particle motion around a static black hole immersed
in an external magnetic field. Moreover, phase structure and critical behavior of charged-AdS black holes \cite{PDU}, photon orbits and phase transition in Letelier AdS black holes\cite{AS}, and thermodynamics and geodesic structure \cite{BH} in Letelier AdS black holes, both immersed in perfect fluid dark matter, have also been investigated.

The presence of neighboring matter fields prevents black holes from existing in a perfect vacuum. Furthermore, recent astronomical observations provide strong evidence for the accelerated expansion of the universe \cite{Garnavich1998,Riess1998,Dalal2000}, suggesting a scenario dominated by negative pressure. Two leading explanations for this negative pressure have been proposed. The first is the cosmological constant, and the second is the concept of quintessence. The cosmological constant, \(\Lambda\), accelerates cosmic expansion through vacuum energy, often identified as dark energy in Einstein's equations. However, it faces the so-called cosmological constant problem, which asks why the observed cosmological constant is so small, on the order of the current Universe's critical density. On the other hand, quintessence is a candidate form of dark energy characterized by the equation of state \(p_{\rm QF} = w \rho_{\rm QF}\), where \(p_{\rm QF}\) is the pressure, \(\rho_{\rm QF}\) the energy density, and the state parameter \(w \in (-1, -1/3)\) \cite{Caldwell2009,Hellerman2001}. In \cite{Kiselev2003}, Kiselev derived a Schwarzschild-like solution in General Relativity describing a black hole surrounded by quintessence dark energy using the corresponding stress-energy tensor.

In this study, we investigate a Schwarzschild-AdS black hole solution coupled to a cloud of strings carrying only the \emph{electric-like} component of the string bivector, and surrounded by a perfect-fluid dark matter distribution in the presence of a quintessence-like field. We begin by analyzing the geodesic structure, both massless and massive test particles, focusing on the photon sphere, the BH shadows, circular null orbits, and the ISCO. We demonstrate how geometric and physical parameters governing the space-time curvature significantly affect the dynamics of photons and particles, including shifts in the photon sphere radius, the shadow size, and the ISCO location. Subsequently, we turn our attention to the thermodynamic properties of black holes, with a particular focus on the sparsity of Hawking radiation. A thorough understanding of black hole thermodynamics and its deep connection to event horizon characteristics relies fundamentally on the concept of black hole entropy. This investigation is especially timely and significant because black hole solutions within the framework of general relativity, when coupled simultaneously with a cloud of strings, a quintessence field, and perfect fluid dark matter, remain largely unexplored. Studying the thermodynamic behavior of such composite systems offers valuable insights into their physical nature and stability. Beyond the geometric and dynamic features, thermodynamic properties provide an essential complementary perspective, enabling us to probe the underlying microphysics and reveal new facets of black hole behavior.

The paper is organized as follows: In Section \ref{sec:2}, we present the Einstein field equations coupled with a cloud of strings in the presence of a quintessence field and perfect fluid dark matter. In Section \ref{sec:3}, we analyze the geodesic motion of both massless and massive test particles, with particular emphasis on the photon sphere, as well as the innermost stable circular orbit radius. In Section \ref{sec:4}, we investigate the topological characteristics of the photon sphere. Section \ref{sec:5} is dedicated to the thermodynamic study of the black hole, in which we explore the combined effects of the string cloud, perfect-fluid dark matter, and the quintessence field on various thermodynamic quantities and discuss the results in detail. In Section \ref{sec:6}, the thermodynamic topology of the black hole is investigated. Finally, in Section \ref{sec:7}, we summarize our findings and present the conclusions.

\section{PFDM BH solution with a Cloud of Strings in the presence of QF}\label{sec:2}

The Einstein-Hilbert action coupled to a string cloud and surrounded by matter sources: perfect fluid, dark matter, and quintessence-like field is given by (setting $8\pi G=1$) \cite{MHL,Kiselev2003}
\begin{equation}
S =\int d^4x\sqrt{-g}\left[\frac{R}{2}+\frac{3}{\ell^2_p}+\mathcal{L}^{\text{PFDM}}+\mathcal{L}^{\text{QF}}\right] + S_M,\label{aa1}
\end{equation}
where $R$ is the Ricci scalar curvature, $\mathcal{L}^{\text{PFDM}}$ is the perfect fluid dark matter Lagrangian density, $\mathcal{L}^{\text{QF}}$ is the Lagrangian density of the quintessence-like field, and $S_M$ represents the matter action of the string cloud. 

For a cloud of strings, the matter content is described by the Nambu-Goto action \cite{Letelier1979,YBZ}:
\begin{equation}
S_M = \mathcal{M}\,\int_\Sigma \sqrt{-\gamma}\,d\lambda^0 d\lambda^1 = \mathcal{M}\, \int_\Sigma \left[-\tfrac{1}{2}\Sigma_{\mu\nu}\,\Sigma^{\mu\nu}\right]^{1/2} d\lambda^0 d\lambda^1.\label{aa2}
\end{equation}
Here, $\lambda^a = (\lambda^0\,,\, \lambda^1)$ are the worldsheet coordinates parameterizing the string trajectory \cite{Synge1960,JP}, $\mathcal{M}$ is the string tension (a positive constant with dimensions of energy per unit length), and $\gamma$ is the determinant of the induced metric on the string worldsheet:
\begin{equation}
\gamma_{ab} = g_{\mu\nu}\tfrac{\partial x^\mu}{\partial\lambda^a}\tfrac{\partial x^\nu}{\partial\lambda^b}.\label{aa3}
\end{equation}
The antisymmetric tensor $\Sigma^{\mu\nu}$ in Eq. (\ref{aa3}) represents the bi-vector density of the string cloud \cite{Letelier1979}, defined as:
\begin{equation}
\Sigma^{\mu\nu} = \epsilon^{ab}\tfrac{\partial x^\mu}{\partial\lambda^a}\tfrac{\partial x^\nu}{\partial\lambda^b},\label{aa4}
\end{equation}
where $\epsilon^{ab}$ is the two-dimensional Levi-Civita symbol. This formalism provides a relativistic description of string-like matter distributions in curved spacetime \cite{AV}. The corresponding energy-momentum tensor takes the form \cite{Letelier1979}:
\begin{equation}
T_{\mu\nu}^\text{CoS} = \tfrac{\rho \Sigma_{\mu\sigma}\Sigma_{\nu}^{\;\sigma}}{\sqrt{-\gamma}},\label{aa5}
\end{equation}
where the string bivector $\Sigma^{\mu\nu} = \epsilon^{ab}\partial_a x^\mu \partial_b x^\nu$ satisfies the conservation laws:
\begin{equation}
\Sigma^{\mu\beta}\nabla_\mu\left[\tfrac{\Sigma_\beta^{\;\nu}}{(-\gamma)^{1/2}}\right] = 0\quad,\quad \nabla_\mu(\rho\Sigma^{\mu\sigma})\Sigma_\sigma^{\;\nu} = 0.\label{aa6}
\end{equation}
Considering only the electric-like component ($\Sigma_{01}$) while maintaining $\gamma < 0$, the energy-momentum tensor for a cloud of strings is given by \cite{Letelier1979,YBZ}
\begin{equation}
T^{\mu}_{\nu}=\frac{\alpha}{8 \pi r^2}\,\mbox{diag}(-1,\,1,\,0,\,0),\label{aa11}
\end{equation}
where $\alpha$ represents a constant related to the string.

\begin{figure*}[tbhp]
    \includegraphics[width=0.32\linewidth]{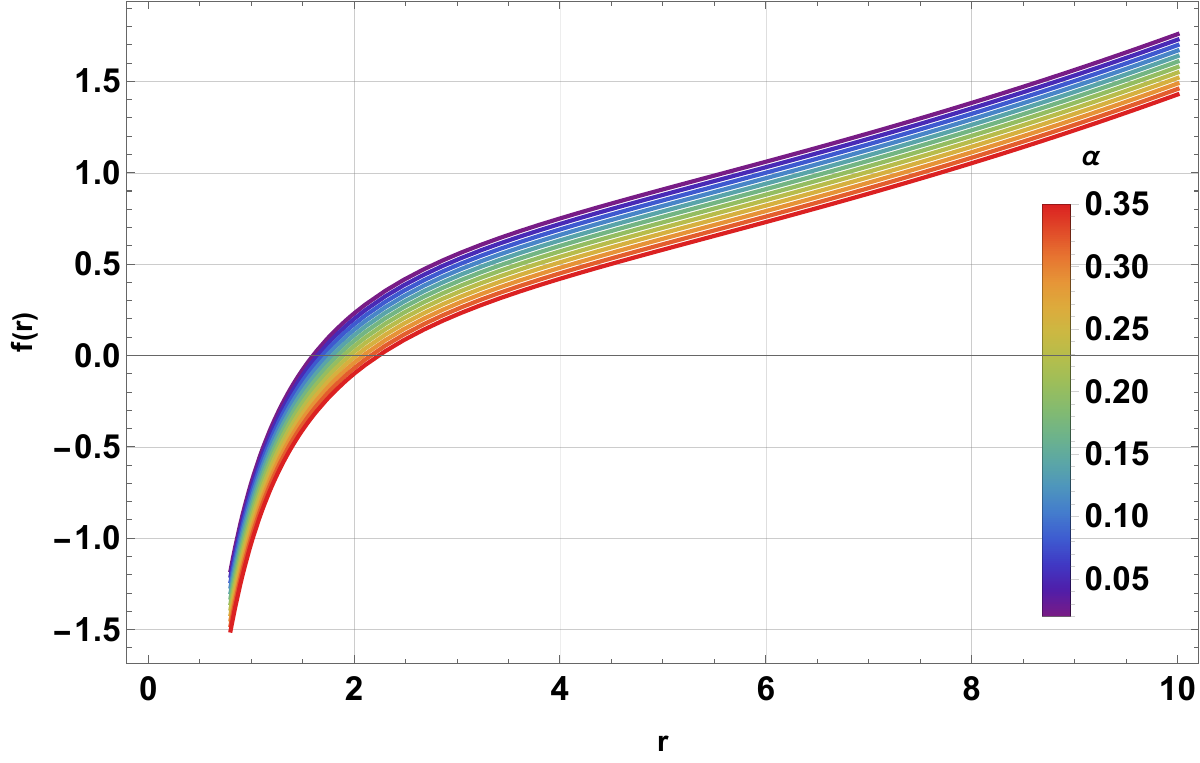}\quad
    \includegraphics[width=0.32\linewidth]{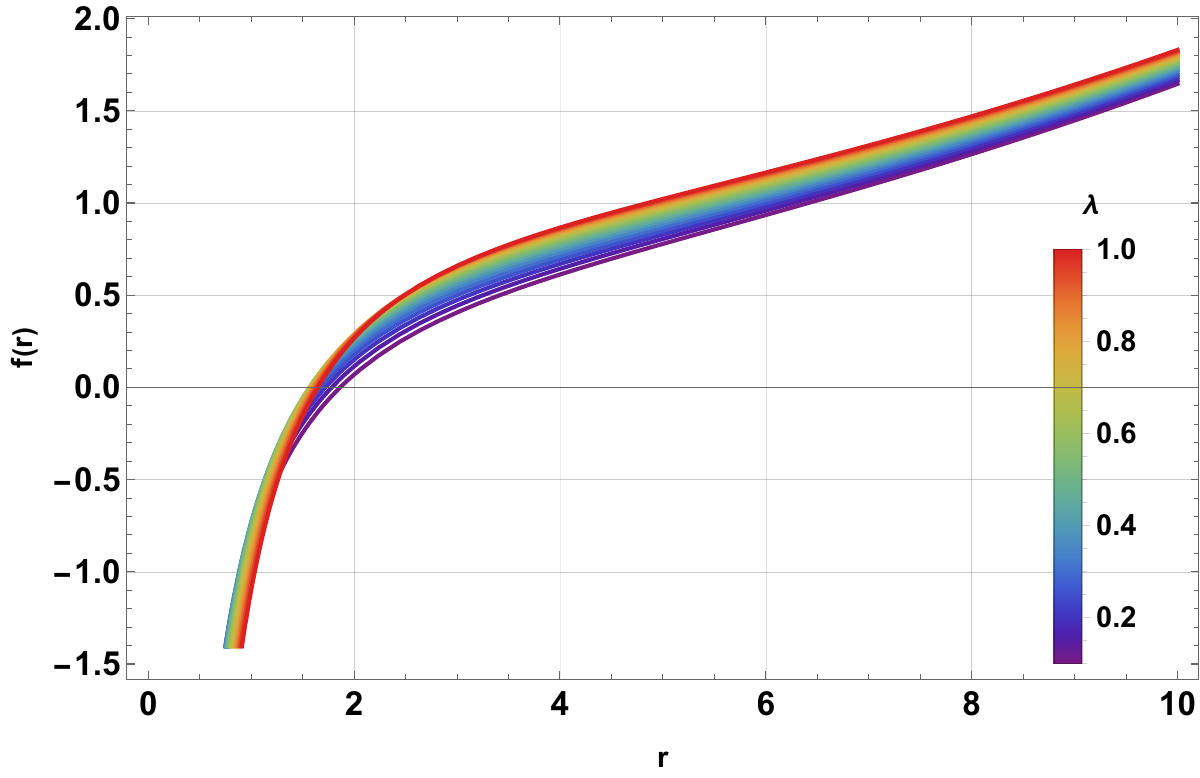}\quad
    \includegraphics[width=0.32\linewidth]{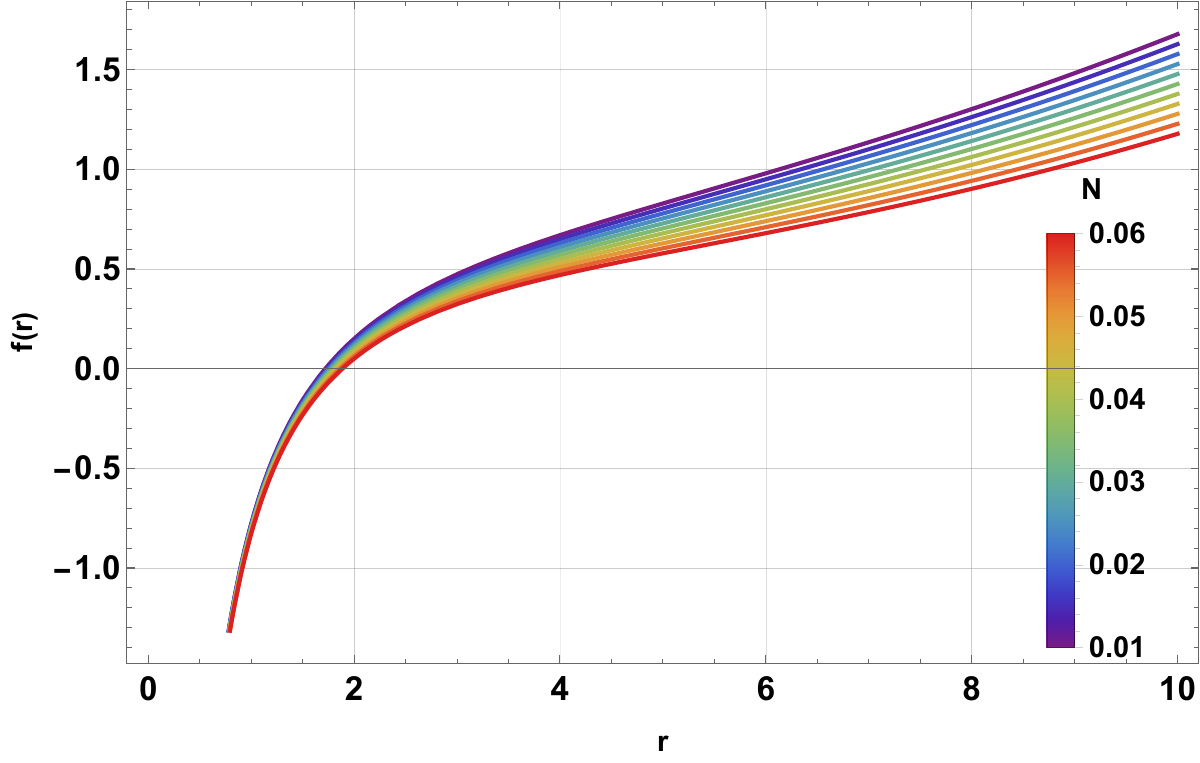}\\
     (i) $\lambda=0.2,\,N=0.01$ \hspace{4cm}  (ii) $\alpha=0.1,\,N=0.01$ \hspace{4cm} (iii) $\alpha=0.1,\,\lambda=0.2$\\
    \caption{\footnotesize Behavior of the metric function for different values of string parameter $\alpha$, perfect fluid dark matter parameter $\lambda$, and the normalization constant $N$ of quintessence-like field. Here $M=1,\,w=-2/3,\,\ell_p=10$.}
    \label{fig:metric}
\end{figure*}

The variation of the action \eqref{aa1} with metric $g_{\mu\nu}$ yields Einstein's equations with a nonzero cosmological constant.
\begin{equation}
R_{\mu\nu}-\frac{1}{2}\,g_{\mu\nu}\,R-\frac{3}{\ell^2_p}\,g_{\mu\nu}=8 \pi\,\left(T_{\mu\nu}^\mathrm{CoS}+T_{\mu\nu}^\mathrm{PFDM}-T_{\mu\nu}^\mathrm{QF}\right),\label{aa7}
\end{equation}
where $T_{\mu\nu}^\mathrm{CoS}$ is the energy-momentum tensor of string cloud, $T^{\mu}{}_{\nu}\big|_{\rm PFDM}=\mathrm{diag}(-\rho,\,p_r,\,p_t,\,p_t)$ is that for perfect fluid dark matter given by \cite{MHL}:
\begin{equation}
    -\rho^{\rm PFDM}=p^{\rm PFDM}_{r} =\frac{\lambda}{8\pi r^{3}}\quad, \quad p^{\rm PFDM}_{t} = -\frac{\lambda}{16\pi r^{3}}\,.\label{aa10}
\end{equation}
Here $\lambda$ denotes the perfect fluid dark matter parameter, which governs the modifications in the energy–momentum tensor influencing the spacetime geometry around the black hole. These modifications reflect the effects of the dark matter density profile in the vicinity of the black hole. 

And $T_{\mu\nu}^\mathrm{QF}$ is the energy-momentum tensor of a quintessence-like field given by \cite{Kiselev2003}
\begin{align}
& T^{t}{}_{t}\big|_{\rm QF}=T^{r}{}_{r}\big|_{\rm QF}= \rho^{\rm QF},\\
&T^{\theta}{}_{\theta}\big|_{\rm QF}=T^{\phi}{}_{\phi}\big|_{\rm QF}=-\tfrac{1}{2} \rho^{\rm QF}\,(3\, w+1). \label{qf}
\end{align}
The pressure and density of quintessence-like field obey the equation of state \(p^{\rm QF}=w\,\rho^{\rm QF}\), with
\begin{equation}
    \rho^{\rm QF}(r) = - \frac{N}{2}\, \frac{3\, w}{r^{3 (w +1)}} ,\label{qf2}
\end{equation}
where \(N>0\) and $w$ is the state parameter, lies in the ranges \(-1<w<-\tfrac{1}{3}\) for acceleration.

Considering a static and spherically symmetric asymptotically flat space-time described by
\begin{equation}
ds^2 = -f(r)\,dt^2 +\frac{1}{f(r)}\,dr^2 + r^2\,(d\theta^2 + \sin^2\theta d\phi^2),\label{final}
\end{equation}
where $f(r)$ is the metric function of the form
\begin{equation}
    f(r)=1-\frac{2\,m(r)}{r},\label{mass}
\end{equation}
with $m(r)$ is the mass function.

Incorporating the cloud of strings, perfect fluid dark matter, and quintessence-like field into the field equations (\ref{aa7}) and further simplification, one can obtain the metric function of the following form:
\begin{equation}
f(r) =1-\alpha- \frac{2\,M}{r} +\frac{\lambda}{r}\,\mbox{ln}\,\frac{r}{|\lambda|}-\frac{N}{r^{3\,w+1}}+\frac{r^2}{\ell^2_p},\label{function}
\end{equation}
where $2M$ and $\lambda$ appeared as integration constants, $N$ is the normalization constant of a quintessence-like field, and $\ell_p$ is the AdS radius. At large distance $r \to \infty$, the metric function behaves as $f (r) \to 1-\alpha+\frac{r^2}{\ell^2_p}$, indicating non-asymptotic flatness of the space-time.
    
Using the above metric function (\ref{function}), one can construct various well-known black hole solutions in the literature. For example, in the limit $N=0$, the metric function corresponds to the Letelier-AdS black hole solution surrounded by perfect fluid dark matter~\cite{PDU,AS,BH}. Furthermore, in the limit \( \lambda \to 0 \), the solution reduces to the Letelier-AdS black hole spacetime in the presence of quintessence-like field~\cite{MMDC}. On the other hand, when \( \alpha \to 0 \) and \(N=0\), the metric function corresponds to the Schwarzschild-AdS black hole surrounded by perfect fluid dark matter~\cite{MHL}. Finally, in the limit $\alpha \to 0$, the solution reduces to the AdS-black hole surrounded by perfect fluid dark matter in the presence of quintessence-like field, reported in \cite{BH2}.

Fig.~\ref{fig:metric} illustrates how each matter component alters the metric function \(f(r)\) (Eq.~\ref{function}), shifting horizon locations (zeros of \(f\)) and the near-horizon minimum that affects strong-field lensing and photon orbits.  
\emph{(i) Cloud of strings (\(\alpha\))}: Increasing \(\alpha\) reduces \(f\) by a constant \((1-\alpha)\), creating an angular deficit that lowers the curve, pushing horizons to larger radii and deepening the minimum, thereby expanding the photon sphere and critical impact parameter (panel a).  
\emph{(ii) perfect fluid dark matter (\(\lambda\))}: The radially varying perfect fluid dark matter term \(\frac{\lambda}{r}\ln(r/|\lambda|)\) changes sign, lowering \(f\) near the center and raising it at larger radii, which shifts the minimum and horizon position and modifies the photon potential barrier (panel b).  
\emph{(iii) Quintessence (\(N,w\))}: With \(w=-2/3\), \(-N/r^{3w+1} = -N r\) grows linearly, tilting \(f\) downward at large \(r\), pushing the outer horizon outward and deepening the intermediate minimum (panel c).  
Together, \(\alpha\), \(\lambda\), and \(N\) finely tune the spacetime’s causal and optical structure while maintaining the AdS behavior at large radii.

\section{Geodesic structure of BH}\label{sec:3}

Black holes exhibit remarkable optical properties due to the extreme curvature of spacetime near their event horizons. One of the most prominent effects is gravitational lensing, in which light from background sources is deflected around the black hole, producing multiple images or Einstein rings. The bending angle increases dramatically near the photon sphere, a region of unstable circular photon orbits \cite{Virbhadra2000,Claudel2001}. Another key feature is the black hole shadow, a dark region surrounded by a bright emission ring formed by photons just grazing the photon sphere. Its size and shape depend on the black hole's mass, spin, and spacetime geometry, as revealed by the Event Horizon Telescope's image of M87* \cite{Bardeen1973,EHT2019,Solanki2022, Perlick2022}.

\begin{figure*}[tbhp]
    \includegraphics[width=0.32\linewidth]{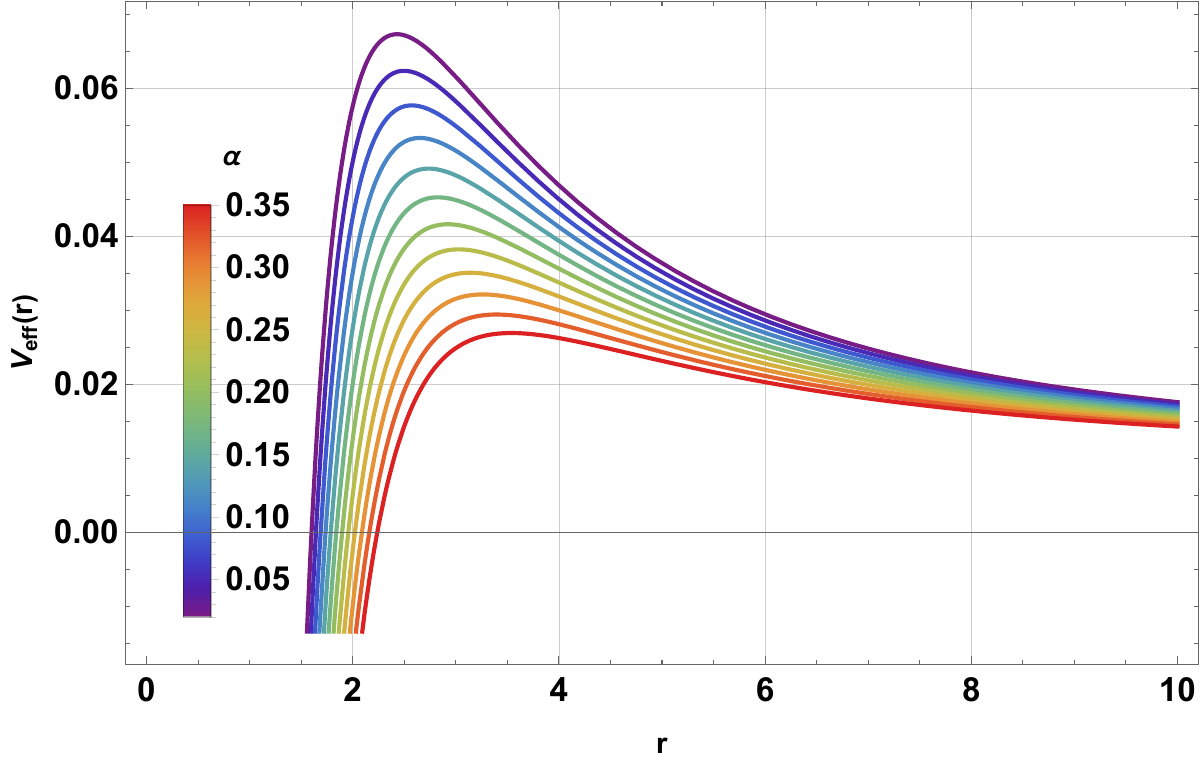}\quad
    \includegraphics[width=0.32\linewidth]{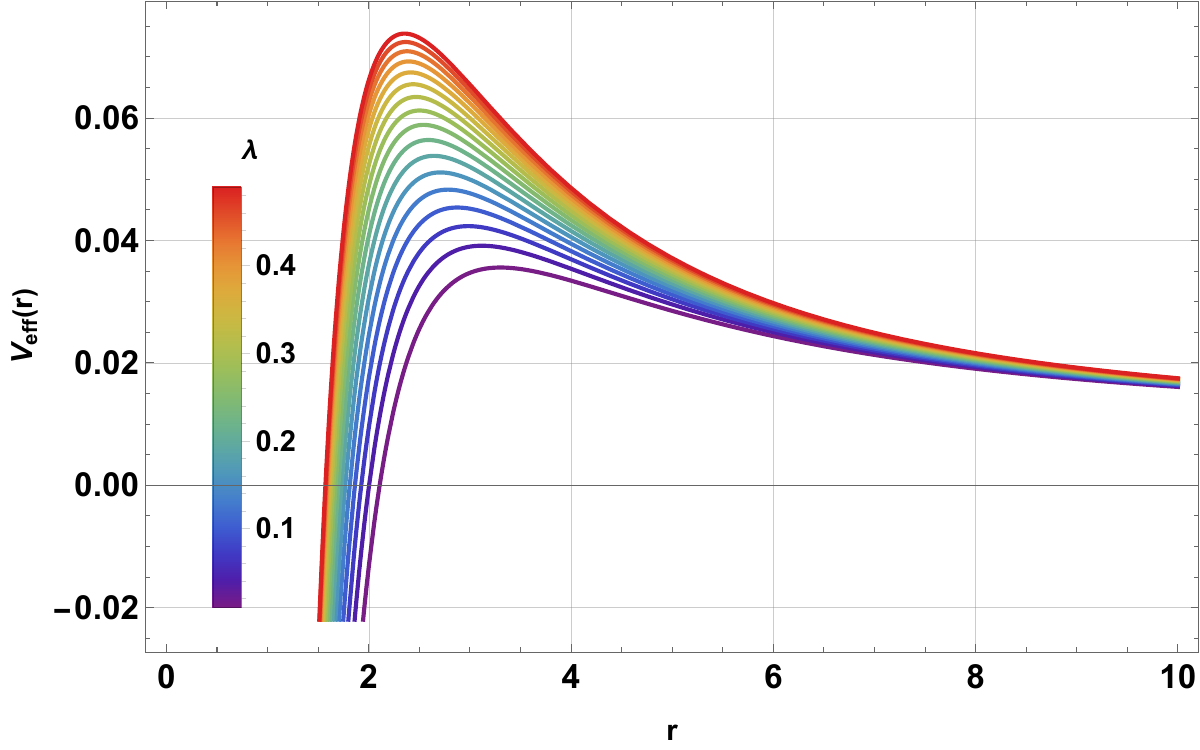}\quad
    \includegraphics[width=0.32\linewidth]{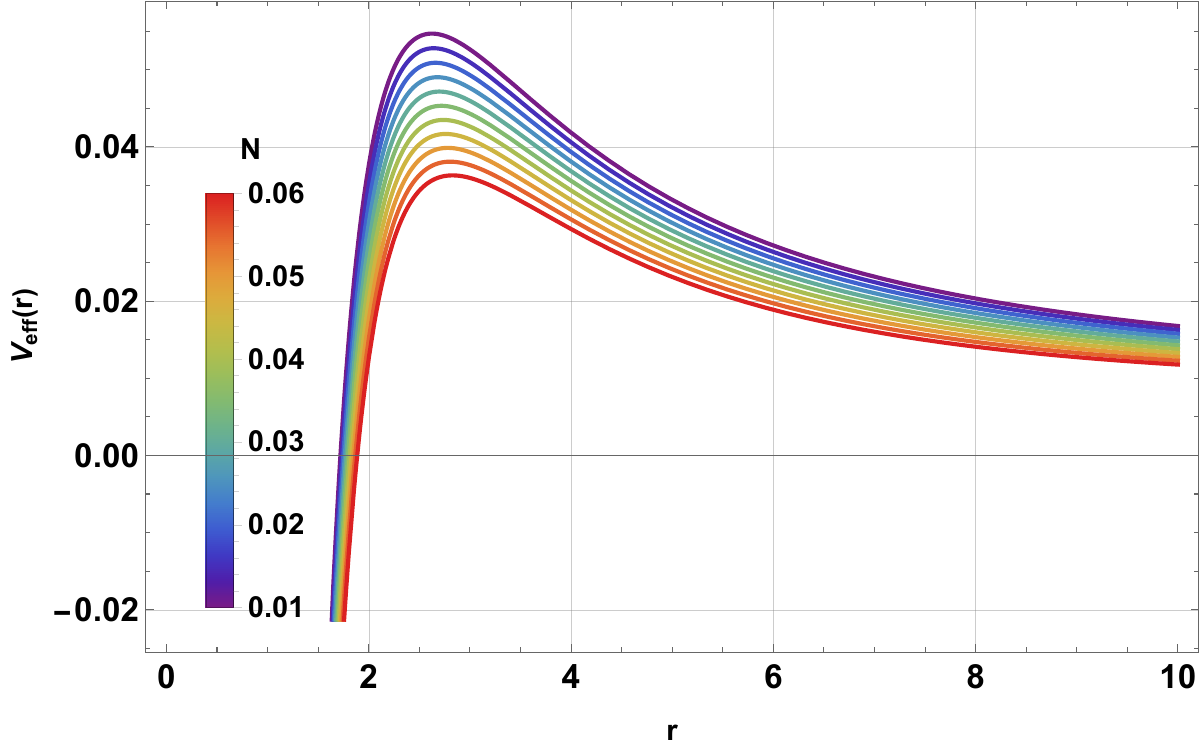}\\
    (i) $\lambda=0.2,\,N=0.01$ \hspace{4cm} (ii) $\alpha=0.1,\,N=0.01$ \hspace{4cm} (iii) $\alpha=0.1,\,\lambda=0.2$
    \caption{\footnotesize Behavior of the effective potential governs the dynamics of photon particles as a function of the radial coordinate $r$ for different values of $\alpha$, $\lambda$ and $N$. Here $M=1,\,w=-2/3,\,\ell_p=10,\,\mathrm{L}=1$.}
    \label{fig:null-potential}
\end{figure*}

Considering the geodesic motion in the equatorial plane defined by $\theta=\pi/2$ and $\dot{\theta}=0$. Expressing the chosen space-time (\ref{final}) in the form $ds^2=g_{\mu\nu}\,dx^{\mu}\,dx^{\nu}$, where $\mu, \nu=0,1,2,3$, the Lagrangian density function simplifies as \cite{chandrasekhar1998}: 
\begin{equation}
    \mathcal{L}=\frac{1}{2}\,\left[-f(r)\,\dot{t}^2+\frac{1}{f(r)}\,\dot{r}^2+r^2\,\dot{\phi}^2\right].\label{bb}
\end{equation}
As the space-time is static and spherically symmetric, there exist two Killing vector fields associated with the temporal coordinate $t$ denoted by $\xi_{(t)} \equiv \partial_{t}$ and the rotational symmetry associated with the azimuthal coordinate $\phi$ denoted by $\xi_{(\phi)} \equiv \partial_{\phi}$. The corresponding conserved quantities associated with these cyclic coordinates are the conserved energy defined by $-\mathrm{E}=g_{t\nu}\,\dot{x}^{\nu}$ and the conserved angular momentum $\mathrm{L}=g_{\phi \nu}\,\dot{x}^{\nu}$. Moreover, the conjugate momentum can be obtained using the relation $p_{\mu}=\frac{\partial \mathcal{L}}{\partial \dot{x}^{\mu}}$. In our case, these are explicitly given as,
\begin{align}
p_t&=\frac{\partial \mathcal{L}}{\partial \dot{t}}=-f(r)\,\dot{t}=-\mathrm{E},\label{bb1}\\
p_r&=\frac{\partial \mathcal{L}}{\partial \dot{r}}=\frac{1}{f(r)}\,\dot{r},\label{bb2}\\
p_{\phi}&=\frac{\partial \mathcal{L}}{\partial \dot{\phi}}=r^2\,\dot{\phi}=\mathrm{L}.\label{bb3}
\end{align}

\begin{center}
    \large{\bf A.\, Photon dynamics}
\end{center}
Eliminating $\dot{t}$ and $\dot{\phi}$ using Eqs.~(\ref{bb1}) and (\ref{bb3}) into Eq. (\ref{bb}), we find the equation of motion for photon particles as,
\begin{equation}
    \dot{r}^2+V_\text{eff}(r)=\mathrm{E}^2\label{bb4}
\end{equation}
which is equivalent to the one-dimensional equation of motion of unit mass particles having energy $\mathrm{E}^2$ and the effective potential $V_\text{eff}(r)$, which governs the dynamics of massless and massive particles. This effective potential is given by
\begin{equation}
    V_\text{eff}(r)=\frac{\mathrm{L}^2}{r^2}\,f(r).\label{bb5}
\end{equation}

From expression (\ref{bb5}), the effective potential for null geodesics is controlled by the spacetime function \(f(r)\); hence it depends on the parameters \(\alpha\) (string cloud), \(\lambda\) (perfect fluid dark matter), \((N,w)\) (quintessence-like field), as well as on \(M\) entering \(f(r)\). Collectively, these parameters reshape the curvature and, in turn, the potential experienced by the test particles.

Figure \ref{fig:null-potential} shows the behavior of the effective potential that governs the dynamics of photon particles by varying the string parameter $\alpha$, the perfect fluid dark matter parameter $\lambda$, and the normalization constant $N$ of the quintessence-like field for a fixed state parameter. In panels (i) and (iii), we observe that as the parameters $\alpha$ and $N$ increase, the effective potential decreases. In contrast, panel (ii) shows an increasing trend as the parameter $\lambda$ value increases.

\begin{center}
    \large{\bf I.\,\,Effective Radial Force and Trajectory}
\end{center}

In this part, we will discuss the photon's trajectory and show how combined matter fields, including string clouds, alter its path in a gravitational field.

Using the given metric function, we find $(2\,f(r)-r\,f'(r))$ which is useful throughout the study as follows:
\begin{align}
   \tfrac{2\,f(r)-r\,f'(r)}{2}=1-\alpha - \tfrac{3\,M}{r}+\tfrac{\lambda}{2\,r}\left(3\,\ln\!\tfrac{r}{|\lambda|} - 1\right)-\tfrac{N\,(3\,w+3)}{2\,r^{3\,w+1}}.\label{special} 
\end{align}

The effective radial force acting on the photon particles in the gravitational field can be obtained from the effective potential discussed earlier. This force can be determined by $\mathcal{F}=-\frac{1}{2}\,\frac{dV_\text{eff}(r)}{dr}$. Using the effective potential given in Eq.~(\ref{bb5}), we find
\begin{align}
\mathcal{F}_\text{rad}=\tfrac{\mathrm{L}^2}{r^3}\,\Bigg[1-\alpha - \tfrac{3\,M}{r}+\tfrac{\lambda}{2r}\left(3\ln\!\tfrac{r}{|\lambda|}- 1\right)-\tfrac{3 N (w+1)}{2 r^{3\,w+1}}\Bigg].\label{cc12}
\end{align}

\begin{figure*}[tbhp]
    \includegraphics[width=0.32\linewidth]{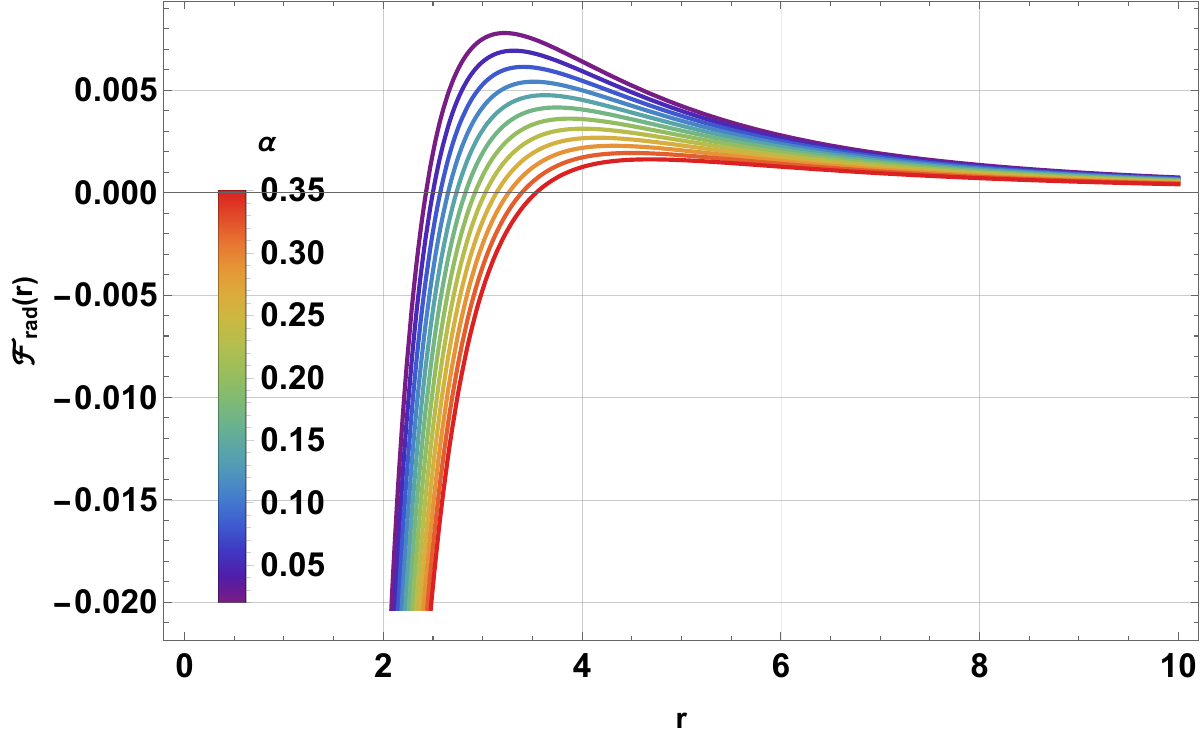}\quad
    \includegraphics[width=0.32\linewidth]{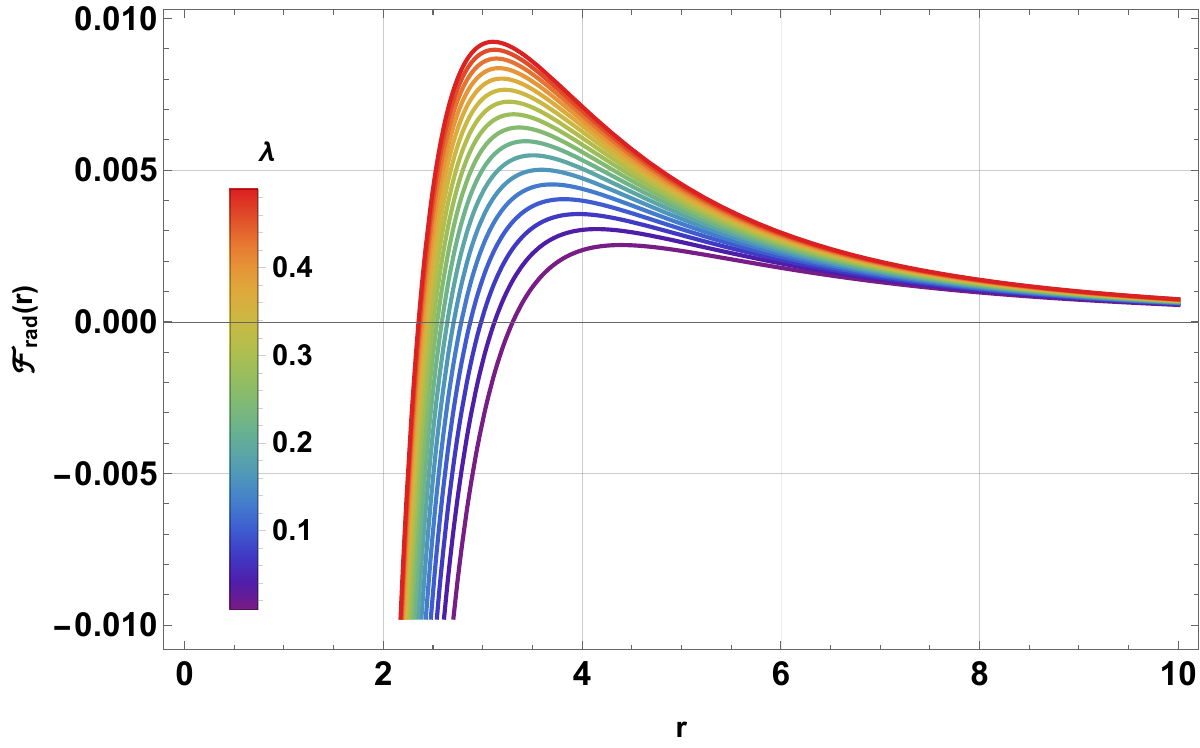}\quad
    \includegraphics[width=0.32\linewidth]{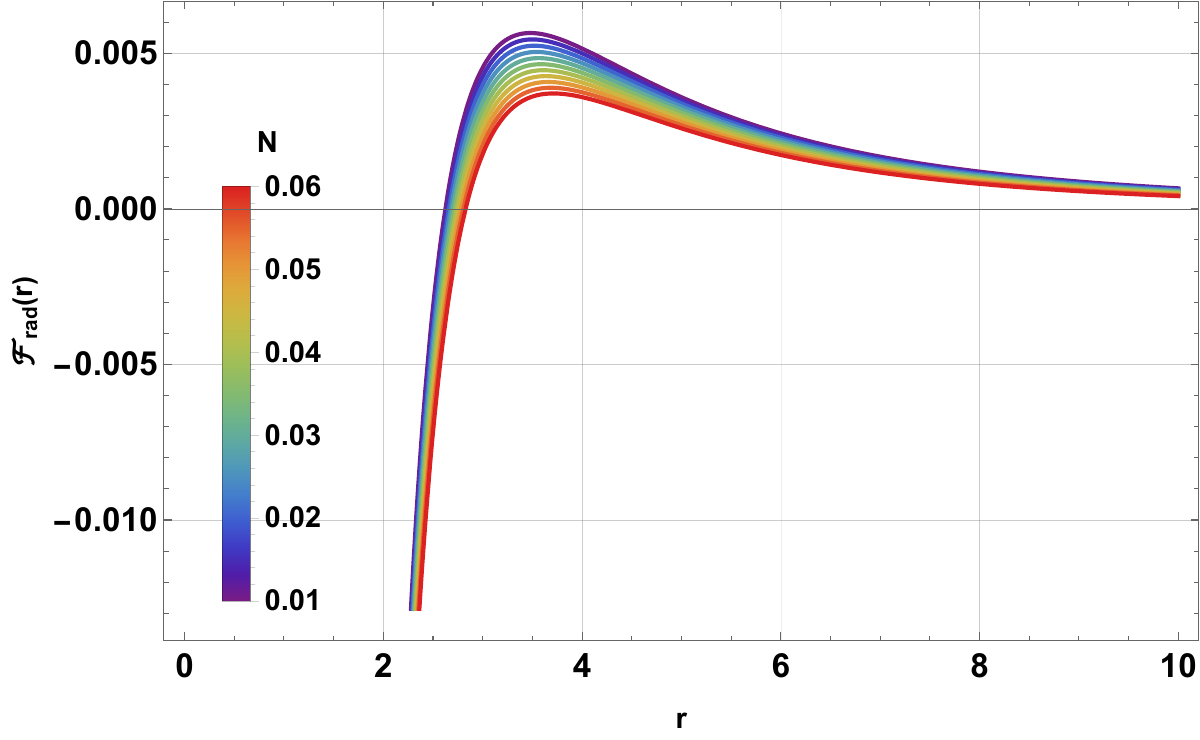}\\
    (i) $\lambda=0.2,\,N=0.01$ \hspace{4cm}  (ii) $\alpha=0.1,\,N=0.01$   \hspace{4cm} (iii) $\alpha=0.1,\,\lambda=0.2$
    \caption{\footnotesize Behavior of the effective radial force for different values of $\alpha$, $\lambda$ and $N$. Here $M=1,\,w=-2/3,\,\ell_p=10,\,\mathrm{L}=1$.}
    \label{fig:radial-force}
\end{figure*}

From Eq.~(\ref{cc12}), we observed that the effective radial force experienced by the photons depends on key geometric parameters: (\(\alpha,\,\lambda,\,N,\,w,\,M\)) solely through \(f(r)\). 

Figure \ref{fig:radial-force} shows the behavior of the effective radial force experienced by photon particles by varying the string parameter $\alpha$, the perfect fluid dark matter parameter $\lambda$, and the normalization constant $N$ of the quintessence-like field for a fixed state parameter $w=-2/3$. In panels (i) and (iii), we observe that as the parameters $\alpha$ and $N$ increase, the effective radial force decreases. In contrast, panel (ii) shows an increasing trend as the parameter $\lambda$ value increases.

The photon trajectory can be obtained using the Eqs. (\ref{bb3}) to (\ref{bb5}) as follows. The equation of the orbit
\begin{equation}
    \left(\tfrac{1}{r^2}\,\tfrac{dr}{d\phi}\right)^2=\tfrac{1}{\beta^2}-\tfrac{1}{\ell^2_p}-\tfrac{1-\alpha}{r^2}+\tfrac{2 M }{r^3}-\tfrac{\lambda}{r^3}\,\ln\!\tfrac{r}{|\lambda|}+\tfrac{N}{r^{3 w+3}},\label{tra1}
\end{equation}
where $\beta=\mathrm{L}/\mathrm{E}$ is the impact parameter for photon particles.

Transforming to a new variable via $u(\phi)=\frac{1}{r(\phi)}$ into the above equation results
\begin{align}
    \left(\tfrac{du}{d\phi}\right)^2=\tfrac{1}{\beta^2}-(1-\alpha) u^2+2 M u^3+\lambda u^3 \ln(|\lambda| u)+N u^{3 w+3}.\label{tra2}
\end{align}
Differentiating both sides w. r. to $\phi$ and after simplification results
\begin{align}
\tfrac{d^2u}{d\phi^2}&+(1-\alpha) u\notag\\ &=3 M u^2+\tfrac{\lambda u^2}{2} (3  \ln(|\lambda| u)+1)+\tfrac{N (3 w+3)}{2} u^{3 w+2}.\label{tra3}
\end{align}

\begin{figure*}[tbhp]
    \centering
    \includegraphics[width=0.5\linewidth]{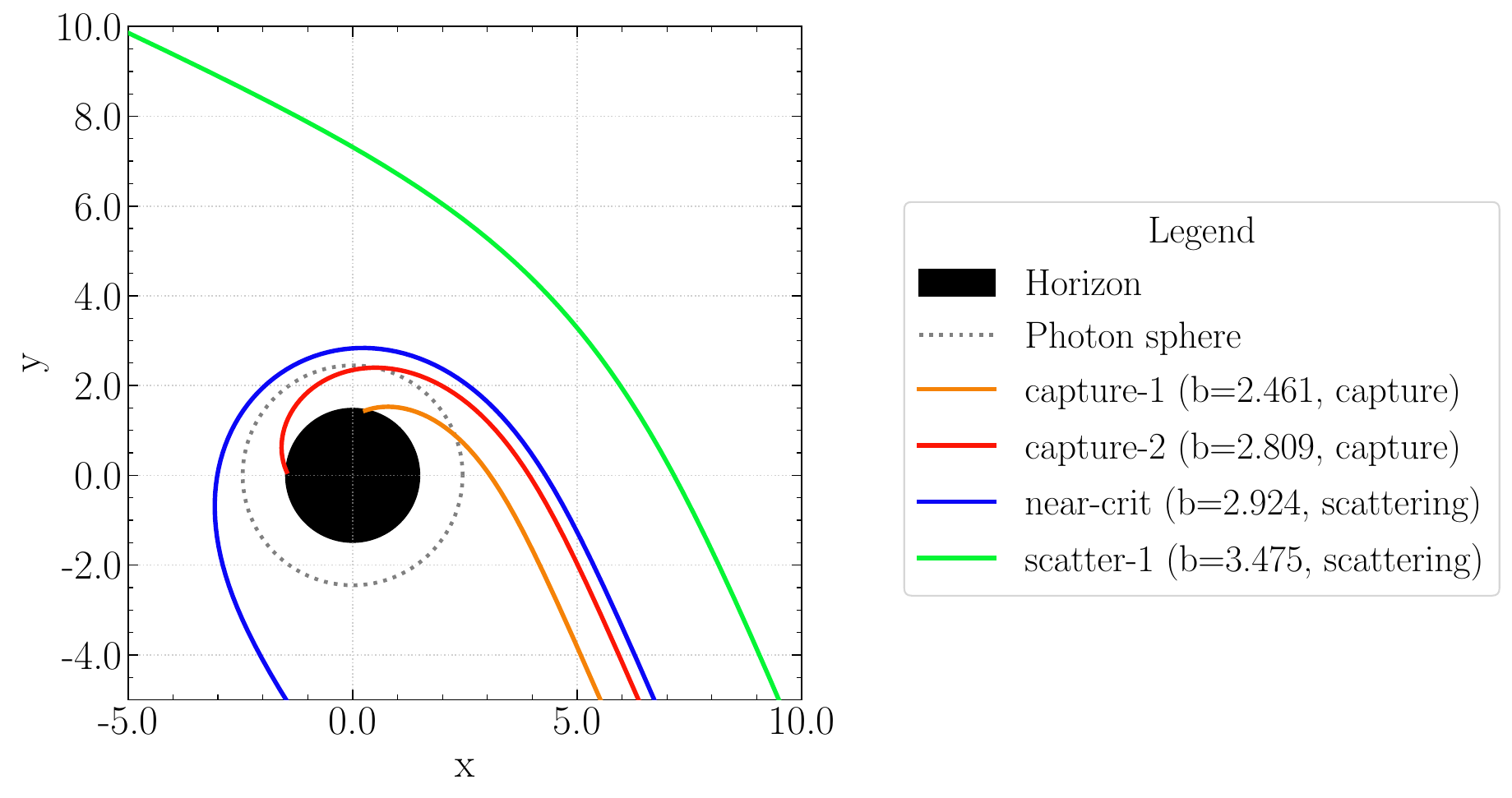}
    \caption{\footnotesize  Null geodesics in the equatorial plane for the AdS black hole with a cloud of strings and perfect fluid dark matter in the presence of a quintessence-like field. We fix $M=1$, $w_q=-2/3$, $\alpha=0.10$, $\lambda=0.30$ and choose $(N,\ell_p)$ such that a photon sphere exists outside the event horizon (in the example shown, the auto-tuned configuration gives $N=0$ and $\ell_p=4$, which yields an event-horizon radius $r_h\simeq 1.471$, a photon-sphere radius $r_{\rm ph}\simeq 2.450$, and the critical impact parameter $b_c\simeq 2.895$). The solid black disk depicts the horizon, the dotted gray circle shows the photon sphere, and the colored curves are null orbits with impact parameters $b=f\,b_c$ for representative factors $f\in\{0.85,0.97,1.01,1.20,1.60,2.20\}$. Orbits with $b<b_c$ are captured (red/orange), the near-critical one ($b\approx b_c$) asymptotically approaches the photon sphere (blue), while for $b>b_c$ the trajectories undergo scattering (green) according to the AdS ``return/escape'' criterion described in the text.}
    \label{fig:null-orbits-vary-b}
\end{figure*}
We integrate null geodesics confined to the equatorial plane $\theta=\pi/2$ of the line element \eqref{final}. Introducing $u(\phi)=1/r(\phi)$, the first integral controlling photon motion is Eq.~\eqref{tra2}, while the second--order equation advanced by a fourth--order Runge--Kutta scheme is Eq.~\eqref{tra3}. Circular null orbits (photon sphere) are selected by the invariant condition $2f-r f'=0$ [Eq.~\eqref{cc6}], which gives Eq.~\eqref{cc7} for the present lapse $f(r)$; the associated critical impact parameter is $b_c=r_{\rm ph}/\sqrt{f(r_{\rm ph})}$. In asymptotically AdS spacetimes, light is not required to escape to infinity to be classified as scattering. Operationally, after the periapsis we monitor the radial growth and label an orbit as ``scattering'' if $r(\phi)$ exceeds a fixed multiple of its initial value (here $1.5\,r_{\rm in}$); otherwise, if the trajectory crosses the horizon, it is labeled as capture, and near $b\!\approx\! b_c$, we observe the usual whirl behavior.

\smallskip
\noindent
\textit{Role of each parameter.}
The impact of the matter sources and of the AdS curvature can be understood from the structure of
\begin{equation*}
f(r)=1-\tfrac{2M}{r}-\underbrace{\alpha}_{\text{CoS}}
+\underbrace{\tfrac{\lambda}{r}\,\ln\!\tfrac{r}{|\lambda|}}_{\text{PFDM}}
\;-\;\underbrace{\tfrac{N}{r^{3w+1}}}_{\text{QF}}\;+\;\underbrace{\tfrac{r^2}{\ell^2_p}}_{\text{AdS}}\, .
\end{equation*}
\emph{(i) Cloud of strings, $\alpha$.} The string cloud subtracts a constant from $f$, producing an angular deficit. In the simple limit where the other sectors are weak near $r_{\rm ph}$, the photon-sphere condition yields $r_{\rm ph}\simeq 3M/(1-\alpha)$; hence increasing $\alpha$ (with $\alpha<1$) moves the photon sphere outward, enlarges the critical impact parameter $b_c$, and strengthens capture/whirl effects. Geodesically, the reduced ``effective $g_{tt}$'' enhances deflection at fixed $M$.

\emph{(ii) perfect fluid dark matter, $\lambda$.} The perfect fluid dark matter term $\frac{\lambda}{r}\ln(r/|\lambda|)$ is radially varying and changes sign with $\ln(r/|\lambda|)$. For $\lambda>0$ it typically raises $f$ at large radii (where $\ln r>0$) but can lower $f$ in the inner region ($r\!\lesssim\!|\lambda|$). In practice (and for the ranges we explore), a positive $\lambda$ deepens the effective potential well around the photon region, pushing $r_{\rm ph}$ and $b_c$ upward and producing longer whirl phases; $\lambda<0$ tends to have the opposite qualitative trend. This non-monotonic contribution is also responsible for modest shifts of the turning points of scattering rays.

\emph{(iii) Quintessence, $N,w$.} The quintessence-like fluid contributes $-N/r^{3w+1}$. For the accelerating range $-1<w<-\tfrac{1}{3}$ we have $3w+1\in(-2,0)$, so this term grows with $r$ (e.g. for $w=-2/3$, it scales as $-N\,r$). It therefore lowers $f$ outside the near-horizon region and competes with the AdS term. Increasing $N$ (at fixed $w$) enlarges the horizon radius and typically raises $b_c$, favoring capture and enhancing the ``confinement'' felt by near-critical photons. Varying $w$ at fixed $N$ changes how quickly the quintessence term turns on with radius, thereby shifting both $r_{\rm ph}$ and the location/strength of radial turning points.

\emph{(iv) AdS curvature, $\ell_p$.} The term $+r^2/\ell_p^2$ is a genuine confining piece. Smaller $\ell_p$ (stronger AdS curvature) increases this upward quadratic contribution, producing additional turning points and allowing photons to move away from the black hole after periapsis even though the spacetime is not asymptotically flat. This is precisely why we adopt an operational scattering criterion based on the post-periapsis growth of $r$.

\emph{(iv) Mass, $M$.} As usual, $M$ sets the length scale of the strong-field region; all characteristic radii (horizon, photon sphere, near-critical whirl radius) scale with $M$ once the dimensionless combinations $(\alpha,\lambda/M,\,N/M^{3w+1})$ are fixed.

\smallskip
\noindent
\textit{Capture vs. scattering.}
For $b<b_c$, the centrifugal barrier is insufficient, and null geodesics plunge (capture). For $b\gtrsim b_c$, the effective potential develops a sharp maximum at $r\simeq r_{\rm ph}$, generating long-lived whirl segments before the photon either falls in or is reflected by the large-$r$ AdS barrier; in the latter case, the trajectory satisfies our operational scattering criterion. Increasing $\alpha$ or $N$ (at fixed $w$) and, to a lesser extent, taking $\lambda>0$ tend to raise $b_c$ and widen the capture cone; reducing the AdS curvature (larger $\ell_p$) moves the behavior toward the asymptotically flat limit and weakens the confining reflection.

\begin{figure*}[tbhp]
    \includegraphics[width=0.32\linewidth]{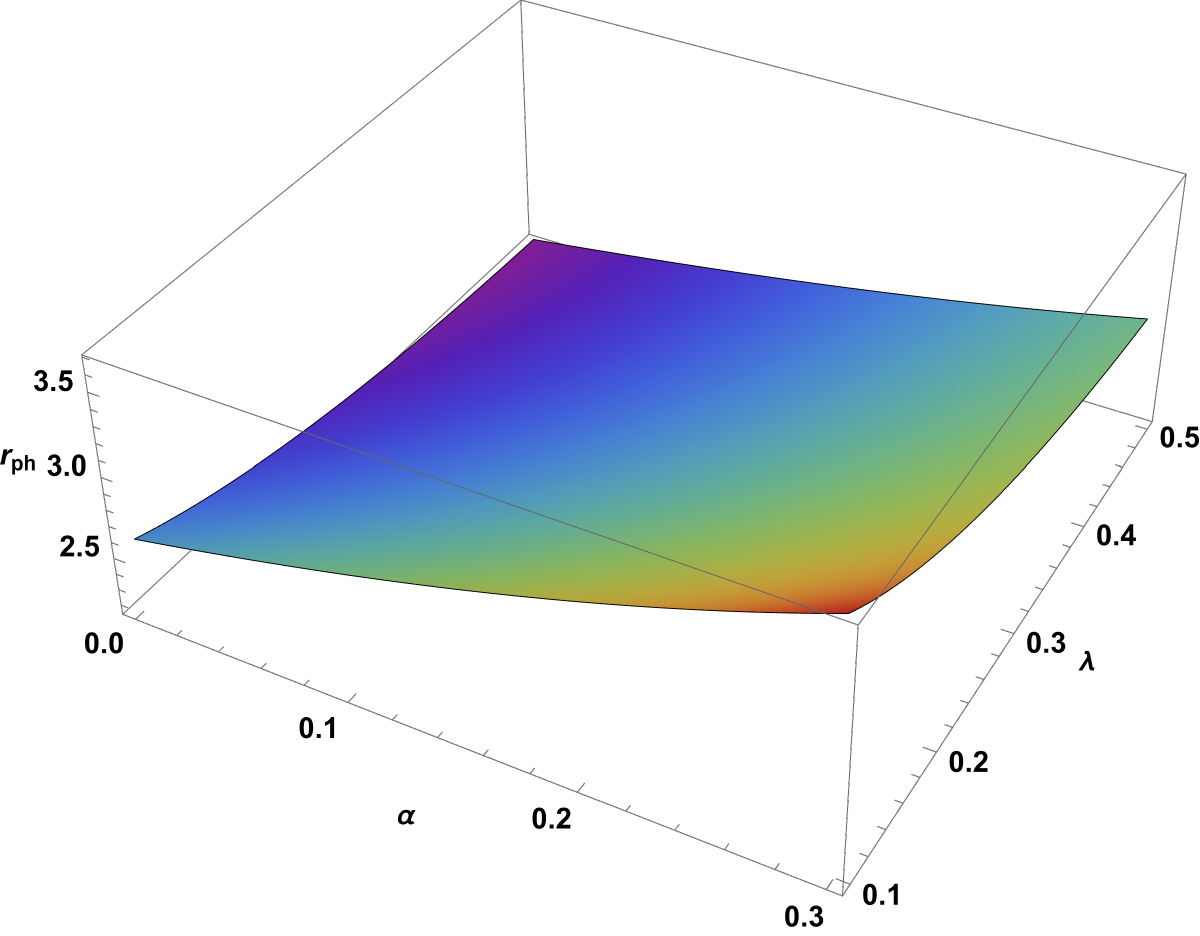}\quad
    \includegraphics[width=0.32\linewidth]{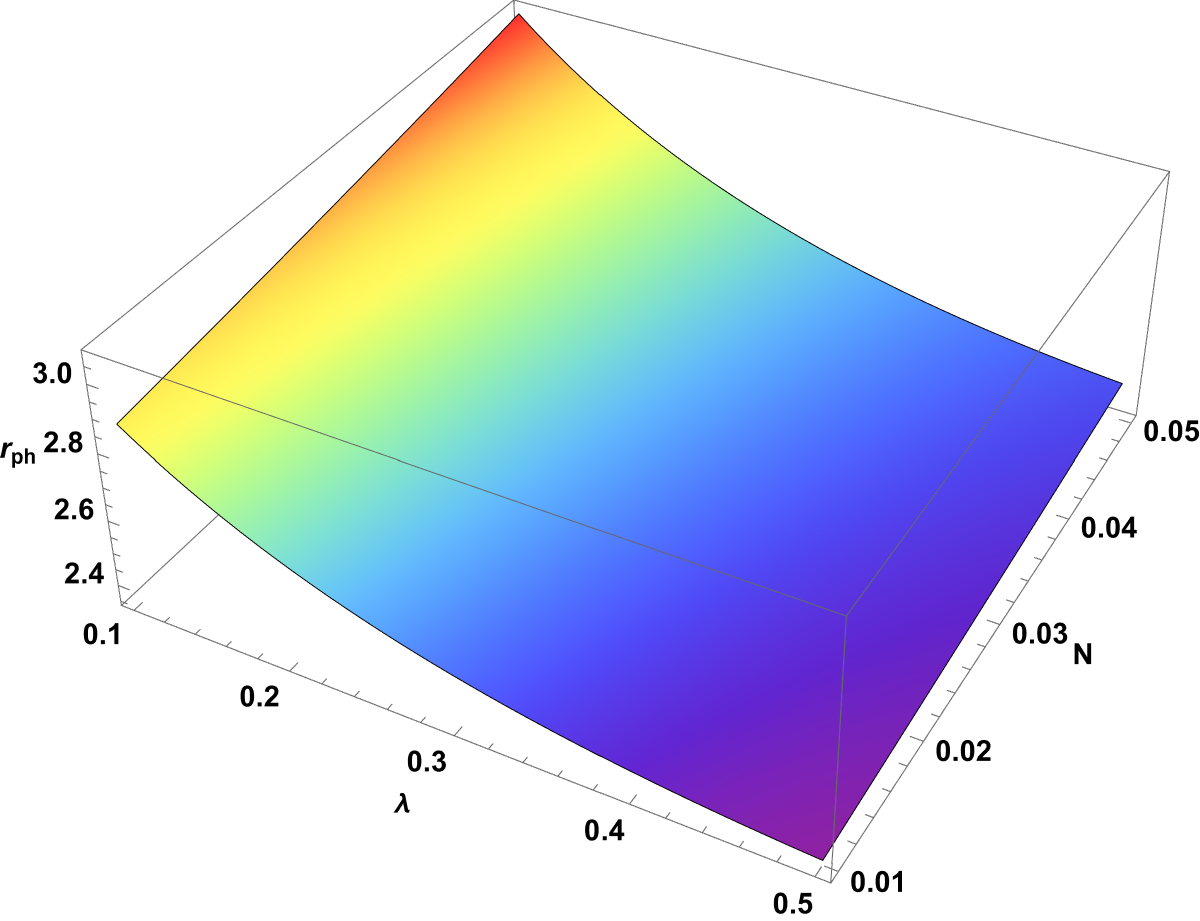}\quad
    \includegraphics[width=0.32\linewidth]{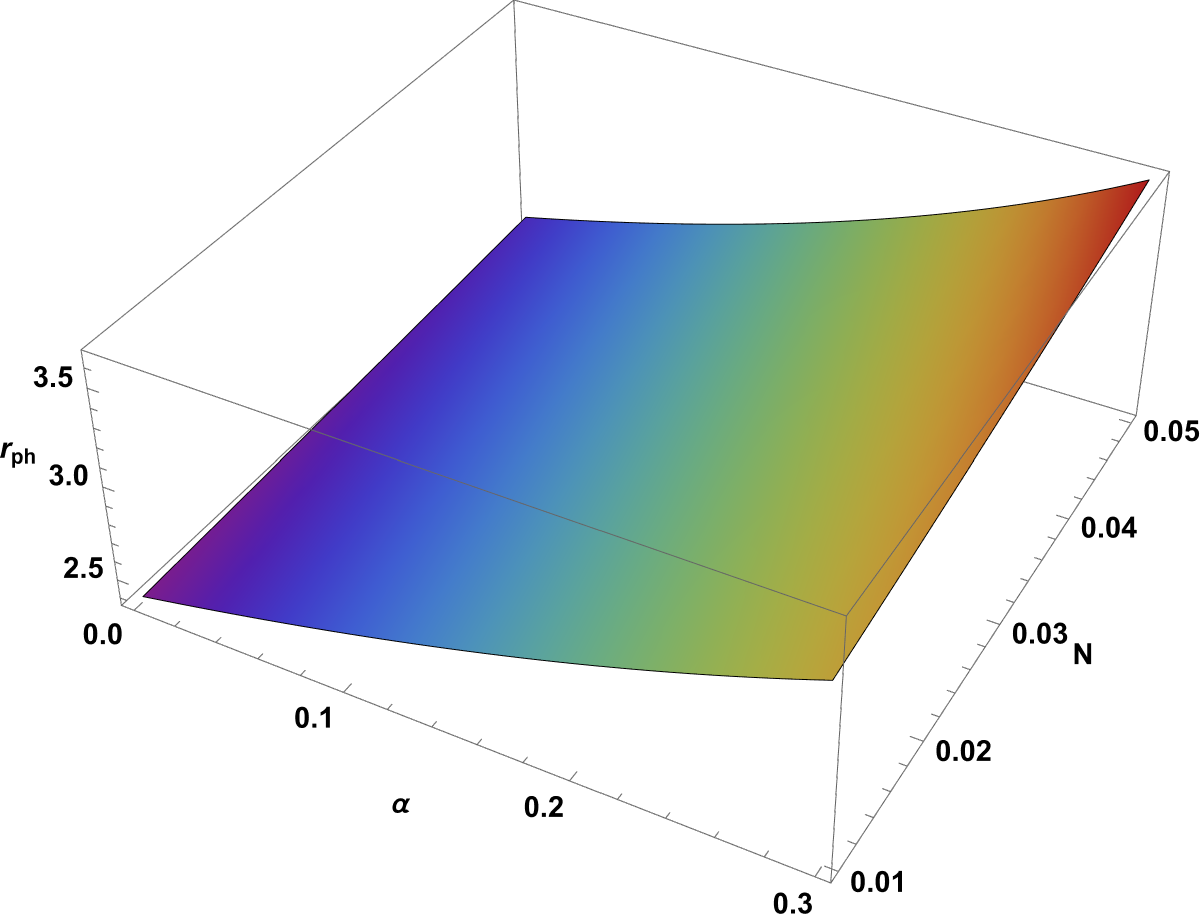}\\
     (i) $N=0.01$ \hspace{4cm} (ii)$\alpha=0.1$ \hspace{4cm} (iii) $\lambda=0.2$
    \caption{\footnotesize Three-dimensional plot of the photon sphere radius \(r_{\rm ph}\) as a function of combination of $\alpha$, $\lambda$ and $N$. Here $M=1,\,w=-2/3$.}
    \label{fig:photon-sphere}
\end{figure*}

\begin{center}
    \large{\bf II.\,\,Photon sphere and BH shadow }
\end{center}

For circular null orbits, the conditions $\dot{r}=0$ and $\ddot{r}=0$ must be satisfied \cite{Virbhadra2000}. Using Eq. (\ref{bb4}), these conditions simplify to the following relations
\begin{equation}
    \mathrm{E}^2=V_\text{eff}(r)=\frac{\mathrm{L}^2}{r^2}\,f(r).\label{cc2}
\end{equation}
And
\begin{equation}
    V'_\text{eff}(r)=0,\label{cc3}
\end{equation}
where prime denotes partial derivative w. r. to $r$. 

The second relation (\ref{cc3}) using (\ref{bb5}) simplifies as,
\begin{equation}
    2\,f(r)-r\,f'(r)=0.\label{cc6}
\end{equation}
Substituting Eq.~(\ref{special}) into the above Eq.~(\ref{cc6}) results
\begin{align}
   1 -\alpha- \tfrac{3\,M}{r_{\rm ph}}+\tfrac{\lambda}{2\,r_{\rm ph}}\left(3\,\ln\!\tfrac{r_{\rm ph}}{|\lambda|} - 1\right)-\tfrac{N\,(3\,w+3)}{2\,r_{\rm ph}^{3\,w+1}}=0.\label{cc7} 
\end{align}
An exact analytical solution of the above equation yields the photon sphere radius \( r_{\rm ph} \), but it is quite challenging due to the presence of the logarithmic function. However, one can obtain numerical values of \( r_{\rm ph} \) by choosing suitable geometric parameters \( (\alpha, M, N, w, \lambda) \).

Figure \ref{fig:photon-sphere} shows the photon sphere radius $r_{\rm ph}$ as a function of combined parameters, such as string parameter $\alpha$, perfect fluid dark matter parameter $\lambda$, and normalization constant $N$ of quintessence-like field for a fixed state parameter $w=-2/3$.

\begin{figure*}[tbhp]
    \includegraphics[width=0.32\linewidth]{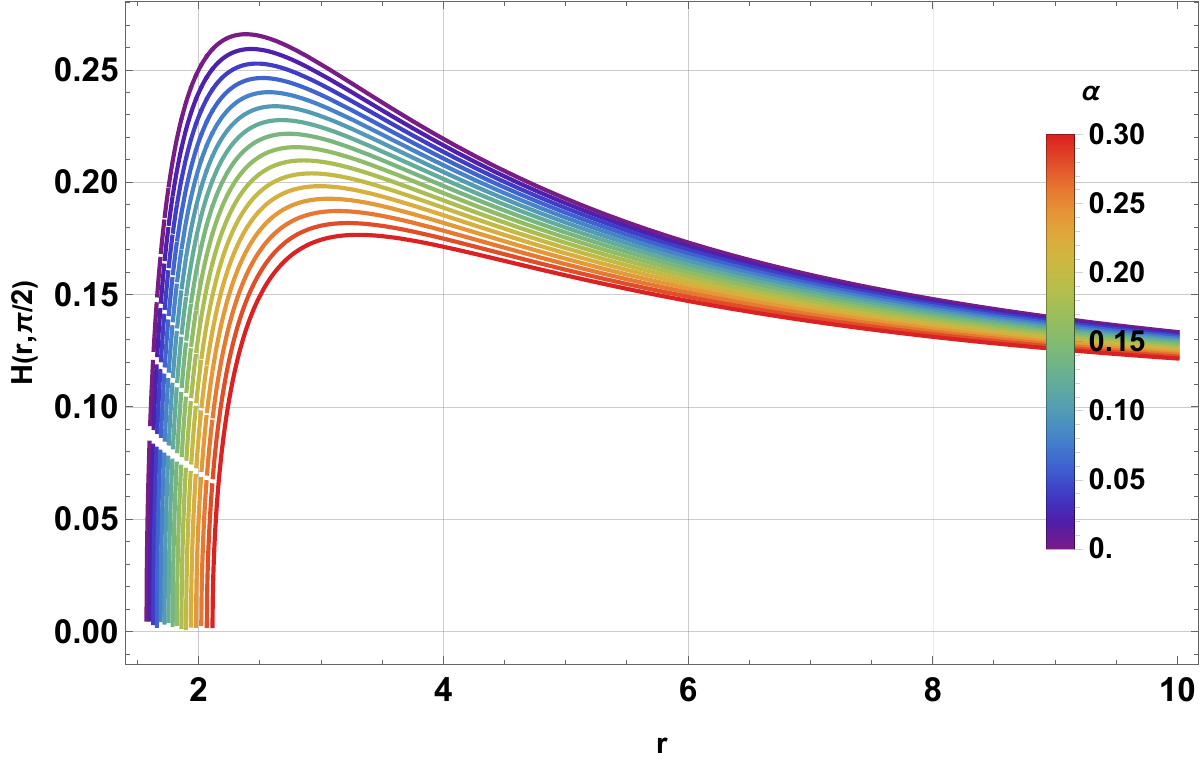}\quad
    \includegraphics[width=0.32\linewidth]{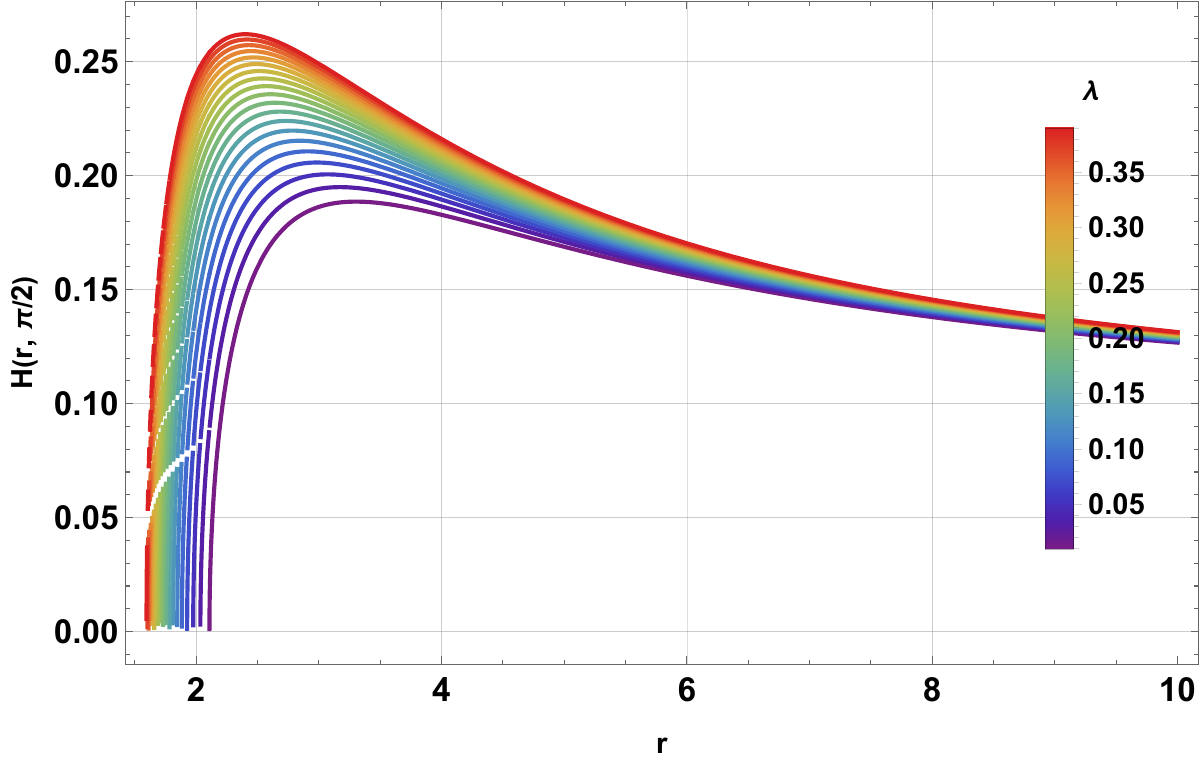}\quad
    \includegraphics[width=0.32\linewidth]{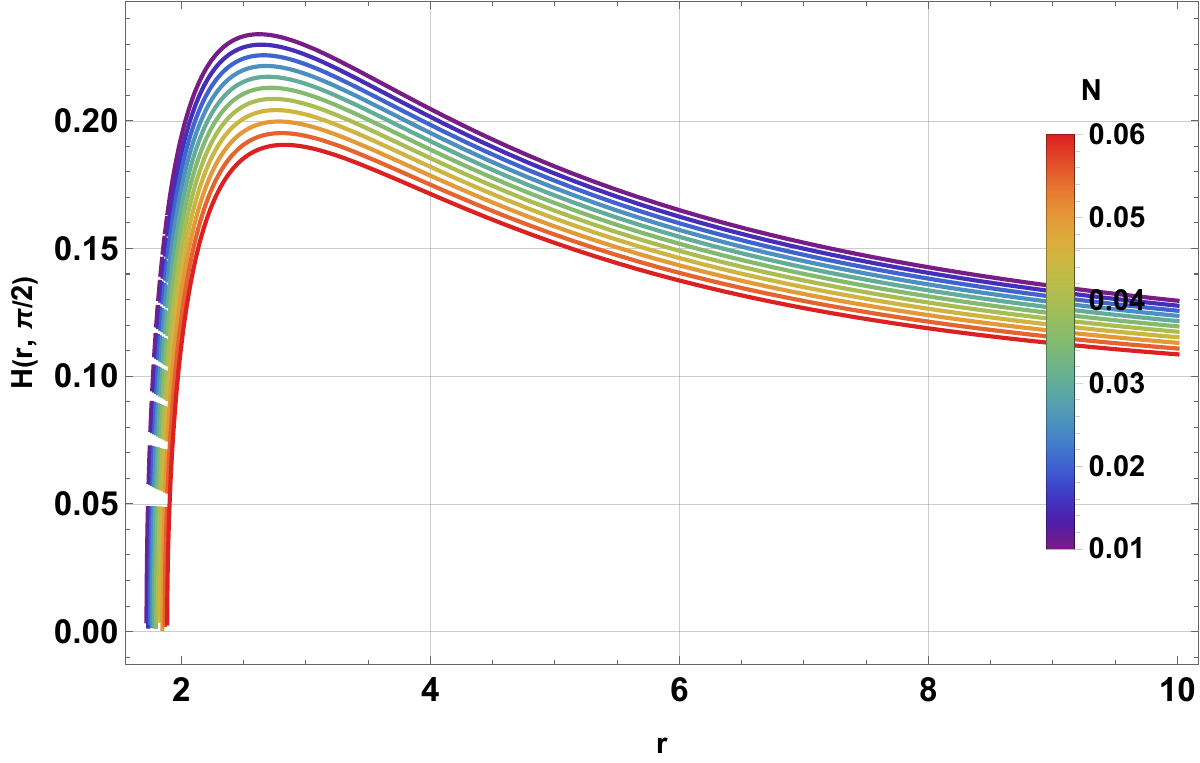}\\
    (i) $\lambda=0.2,\,N=0.01$ \hspace{4cm}  (ii) $\alpha=0.1,\,N=0.01$ \hspace{4cm} (iii) $\alpha=0.1,\,\lambda=0.2$
    \caption{\footnotesize Behavior of the potential function $H(r, \pi/2)$ for different values of $\alpha$, $\lambda$ and $N$. Here $M=1,\,w=-2/3$.}
    \label{fig:potential-function}
\end{figure*}

Next, we focus on the black hole shadows cast by the black hole solution and examine how the geometric and physical parameters influence their size. The shadow size $R_s$ of a non-asymptotically flat metric can be determined using the geometric method, as discussed in detail in \cite{Perlick2022}. The angular size of the shadow cast by the black hole is given by \cite{Perlick2022}
\begin{equation}
    \sin^2\vartheta_{\rm sh}=\left(\tfrac{r_{\rm ph}}{r_{\rm O}}\right)^2\,\tfrac{f(r_{\rm O})}{f(r_{\rm ph})},\label{cc10}
\end{equation}
where $r_{\rm O}$ is the position of a distant observer from the black hole.

The black hole shadow size is therefore given by
\begin{equation}
    R_{\rm sh}=r_{\rm O}\,\sin \vartheta_{\rm sh}=r_{\rm ph}\,\sqrt{\tfrac{f(r_{\rm O})}{f(r_{\rm ph})}}.\label{cc11}
\end{equation}

\section{Topological properties of Photon Sphere }\label{sec:4}

The topological study of photon spheres in black holes is a compelling area of theoretical physics. The photon sphere, where gravity is strong enough to force photons into circular orbits, plays a key role in phenomena such as black hole shadows and light bending. Topological methods provide insights into the stability, structure, and dynamics of photon spheres, as well as their interactions with surrounding physical processes, offering a deeper understanding of black hole behavior \cite{Cardoso2014, Wei2020}.

Beyond the classical approach in determining the photon sphere radius, a topological viewpoint offers additional insight by assigning a topological charge to each photon orbit, encapsulating its stability traits, as discussed in detail in \cite{Cunha2020, Wei2020, Cunha2017}. This approach involves defining the scalar potential function as,
\begin{align}\label{p6}
H(r, \theta)=\sqrt{\tfrac{-g_{tt}}{g_{\phi\phi}}}=\tfrac{\sqrt{1-\alpha - \frac{2\,M}{r}+\frac{\lambda}{r}\,\mbox{ln}\,\frac{r}{|\lambda|}-\frac{N}{r^{3\,w+1}}+\frac{r^2}{\ell^2_p}}}{r\,\sin \theta},
\end{align}
where $g_{tt}$ and $g_{\phi\phi}$ are metric components.

Figure \ref{fig:potential-function} shows the behavior of the potential function $H(r, \theta)$ by varying string parameter $\alpha$, perfect fluid dark matter parameter $\lambda$, and normalization constant $N$ of quintessence-like field for a fixed state parameter $w=-2/3$. In panels (i) and (iii), we observe that as parameters $\alpha$ and $N$ increase, the function $H(r, \pi/2)$ decreases. In contrast, panel (ii) shows an increasing trend as the parameter $\lambda$ value increases.

The radii of photon spheres correspond to critical points of $H$ with respect to $r$, identified by the condition
\begin{equation}\label{p7}
\frac{dH}{dr} = 0.
\end{equation}
To further explore the topological structure, one constructs a vector field $\boldsymbol{\varphi}$ on the two-dimensional manifold spanned by $(r, \theta)$ coordinates, with components defined as \cite{Wei2020}
\begin{equation}\label{p8}
\varphi^{r} =\sqrt{f(r)}\, \tfrac{dH}{dr}\quad, \quad \varphi^{\theta} = \tfrac{1}{r}\,\tfrac{d H}{d\theta}.
\end{equation}
This vector field can be expressed in complex or polar form as
\begin{equation}\label{p9}
\varphi = ||\varphi||\,e^{i\,\Theta} = \varphi^{r} + i\,\varphi^{\theta},
\end{equation}
with magnitude
\begin{equation}\label{p10}
||\varphi|| = \sqrt{(\varphi^{r})^2 + (\varphi^{\theta})^2}.
\end{equation}
Normalizing $\boldsymbol{\varphi}$ yields a unit vector field
\begin{equation}\label{p11}
n^{a} = \tfrac{\varphi^{a}}{||\varphi||},
\end{equation}
where the index $a = 1,2$ corresponds to directions along $r$ and $\theta$, respectively. What is more, a topological current can be established through the utilization of Duan’s theory~\cite{Duan1, Duan2} on $\phi$-mapping topological currents as follows:
\begin{equation}
j^{\mu} = \tfrac{1}{2\pi} \epsilon^{\mu\nu\rho} \epsilon_{ab} \partial_{\nu} n^{a} \partial_{\rho} n^{b}
\quad (\mu, \nu, \rho = 0, 1, 2),
\label{p12}
\end{equation}
where $\partial_{\nu} = \partial / \partial x^{\nu}$ and $x^{\nu} = (t, r, \theta)$. According to Noether's theorem, this topological current obeys the conservation law:
\begin{equation}
\partial_{\mu} j^{\mu} = 0.
\label{p13}
\end{equation}

\begin{figure*}[tbhp]
    \includegraphics[width=0.29\linewidth]{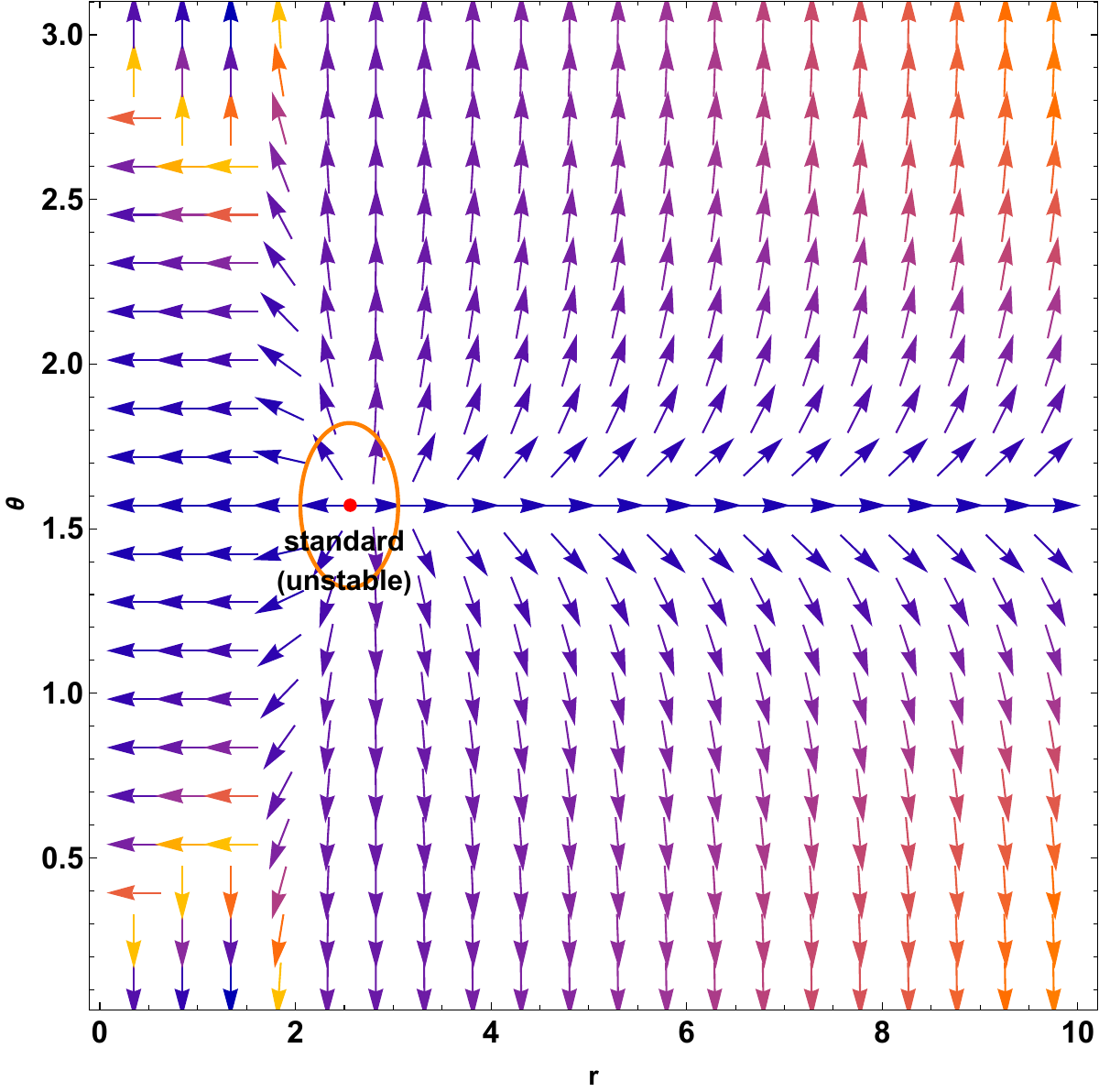}\quad
    \includegraphics[width=0.29\linewidth]{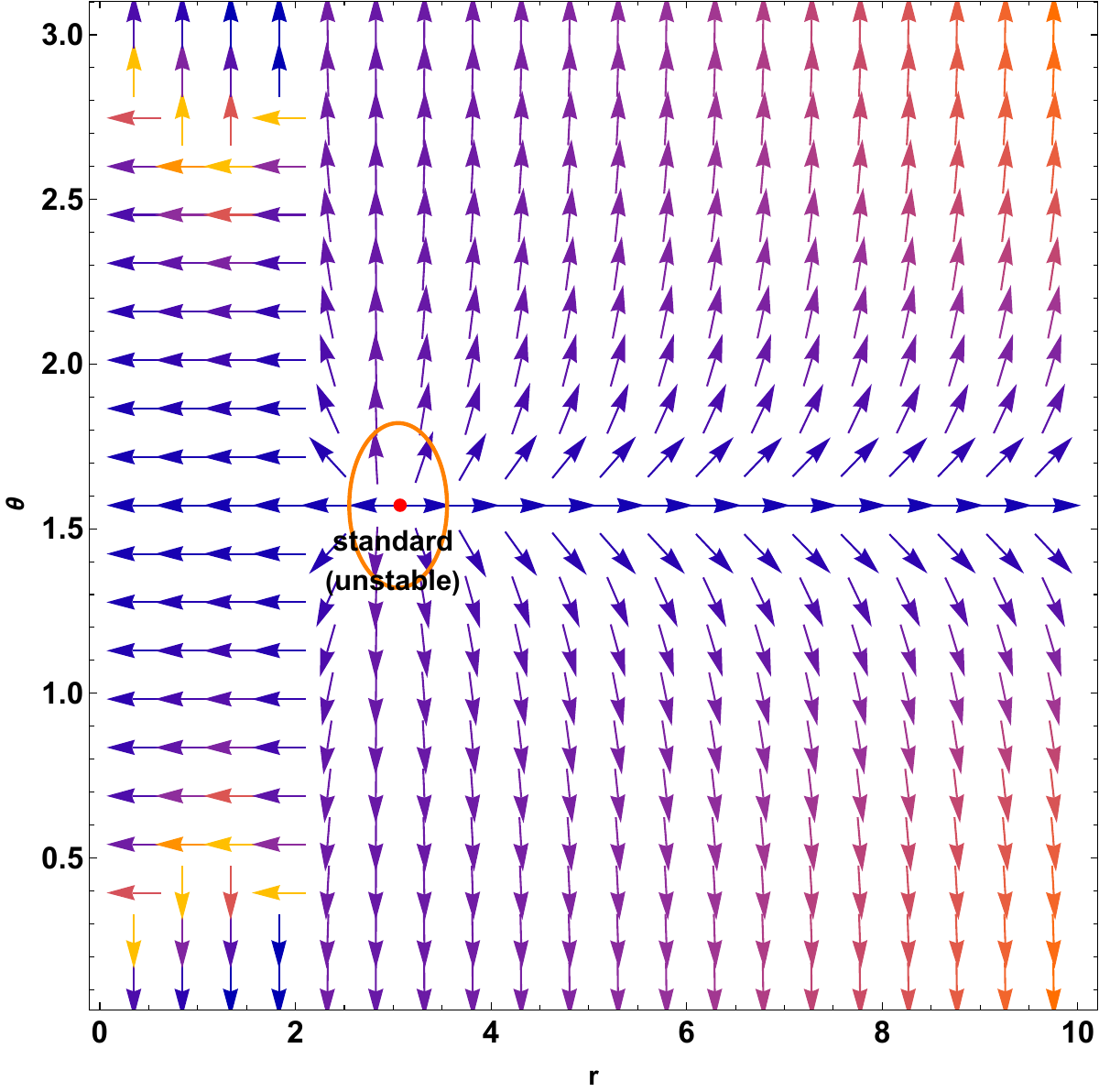}\quad
    \includegraphics[width=0.29\linewidth]{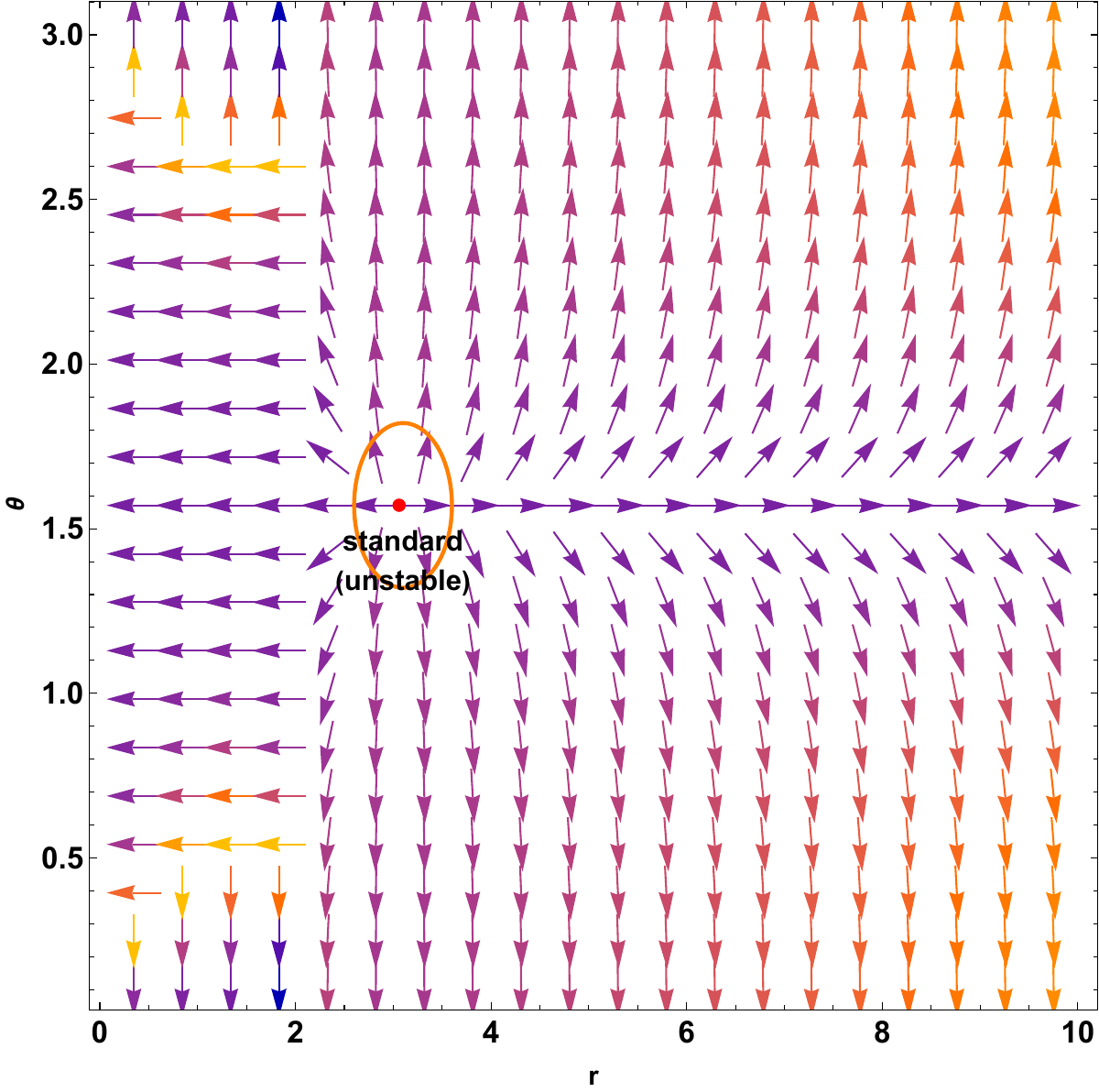}\\
    (i) $\alpha=0.05$ \hspace{6cm}  (ii) $\alpha=0.10$ \hspace{6cm} (iii) $\alpha=0.15$
    \caption{\footnotesize The arrows represent the normalized vector field $\boldsymbol{n}$ on a portion of the $r$–$\theta$ plane for a static black hole with parameters $M = 1$, $N = 0.01$, $w = -\tfrac{2}{3}$, $\lambda = 0.1$, and $\ell_p = 10$. The standard (or zero) points are indicated by the red dots located at $(r, \theta) = (r_0, \pi/2)$, where (i) $r_0 = 2.56$ for $\alpha = 0.05$, (ii) $r_0 = 3.07$ for $\alpha = 0.1$, and (iii) $r_0 = 3.06$ for $\alpha = 0.15$. The photon sphere is enclosed by the orange color closed loop. Evidently, the topological charge of the photon sphere is $Q = +1$.
} \label{fig:unit-vector-1}
\end{figure*}

\begin{figure*}[tbhp]
    \includegraphics[width=0.29\linewidth]{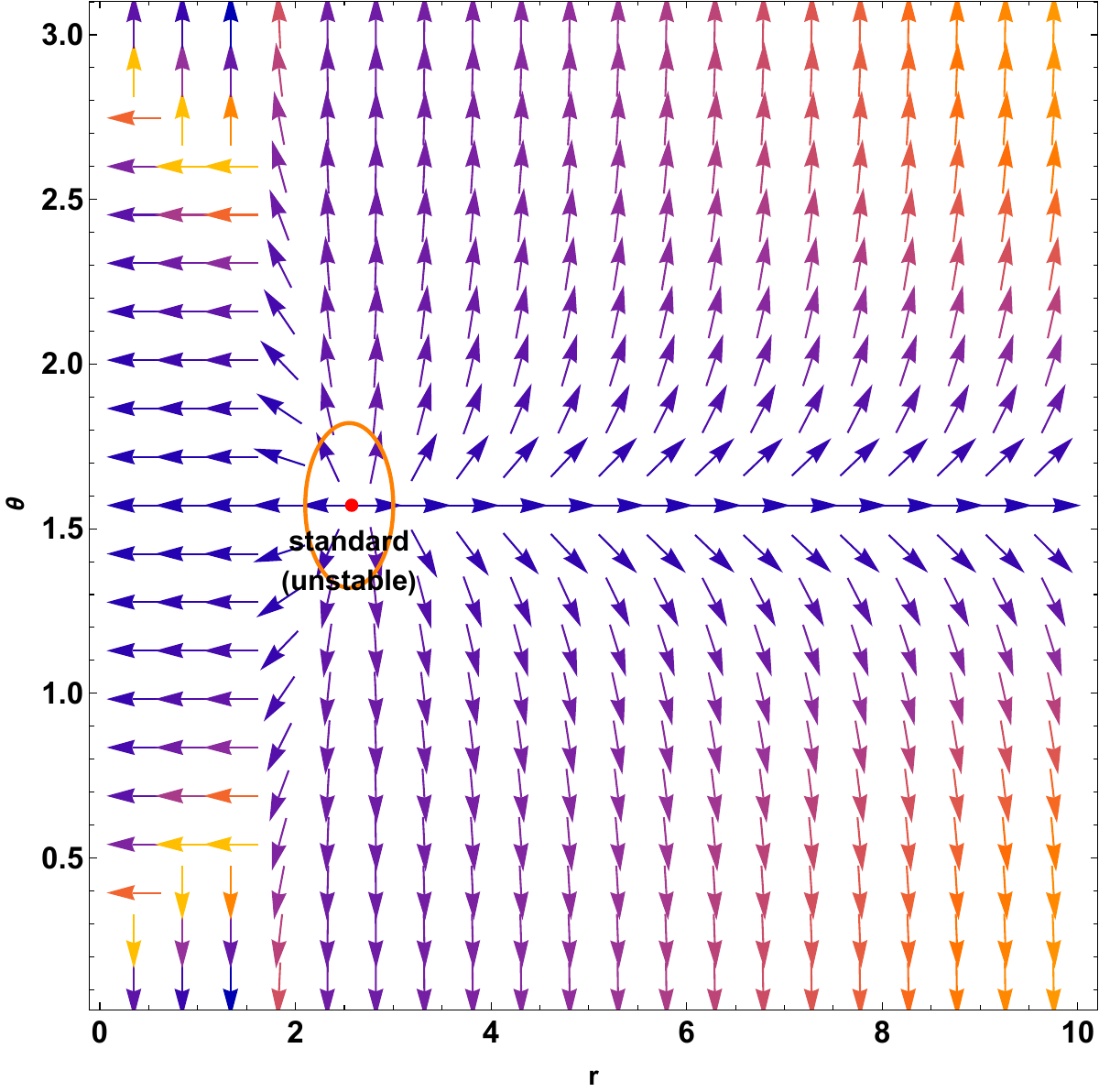}\quad
    \includegraphics[width=0.29\linewidth]{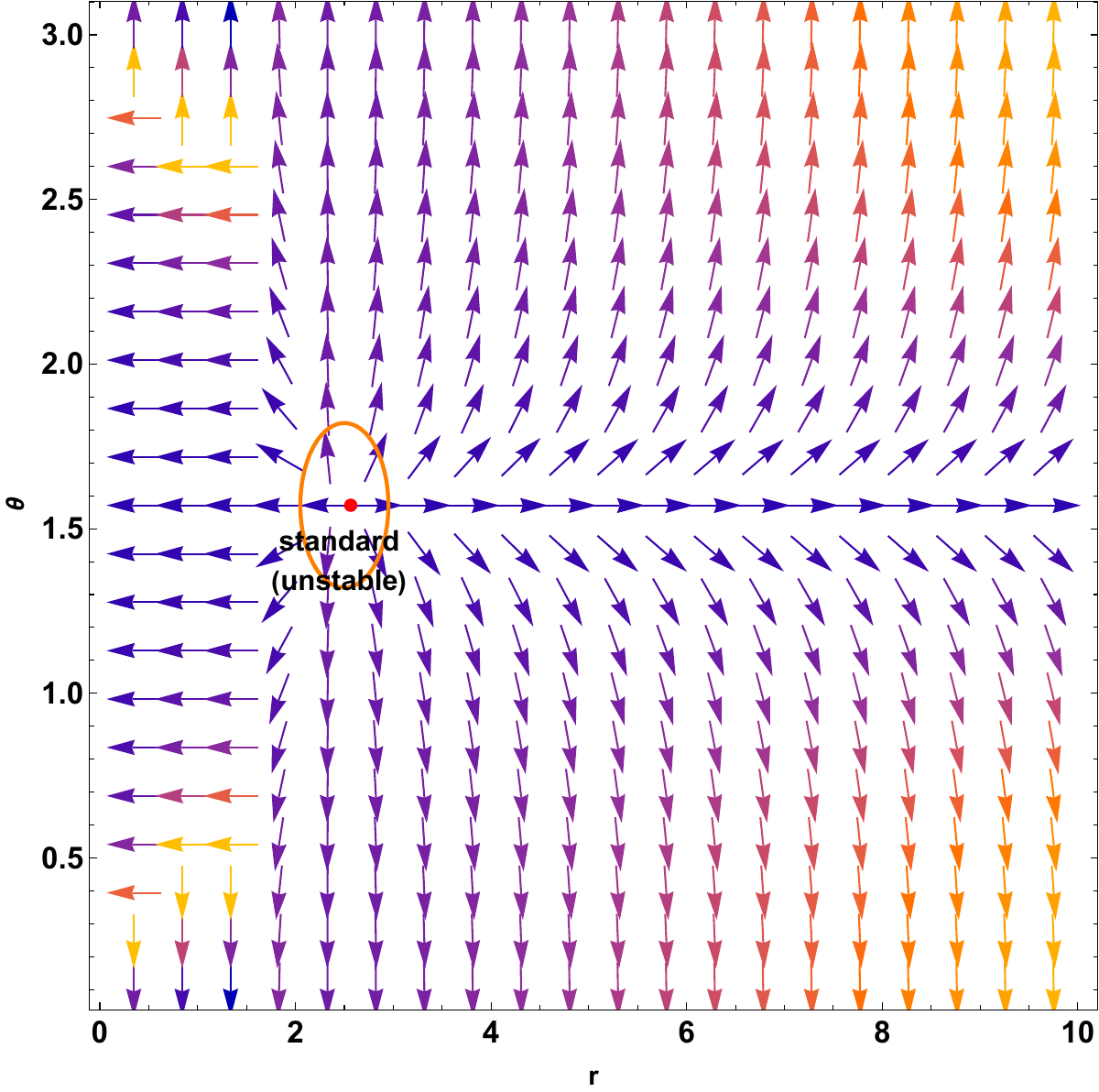}\quad
    \includegraphics[width=0.29\linewidth]{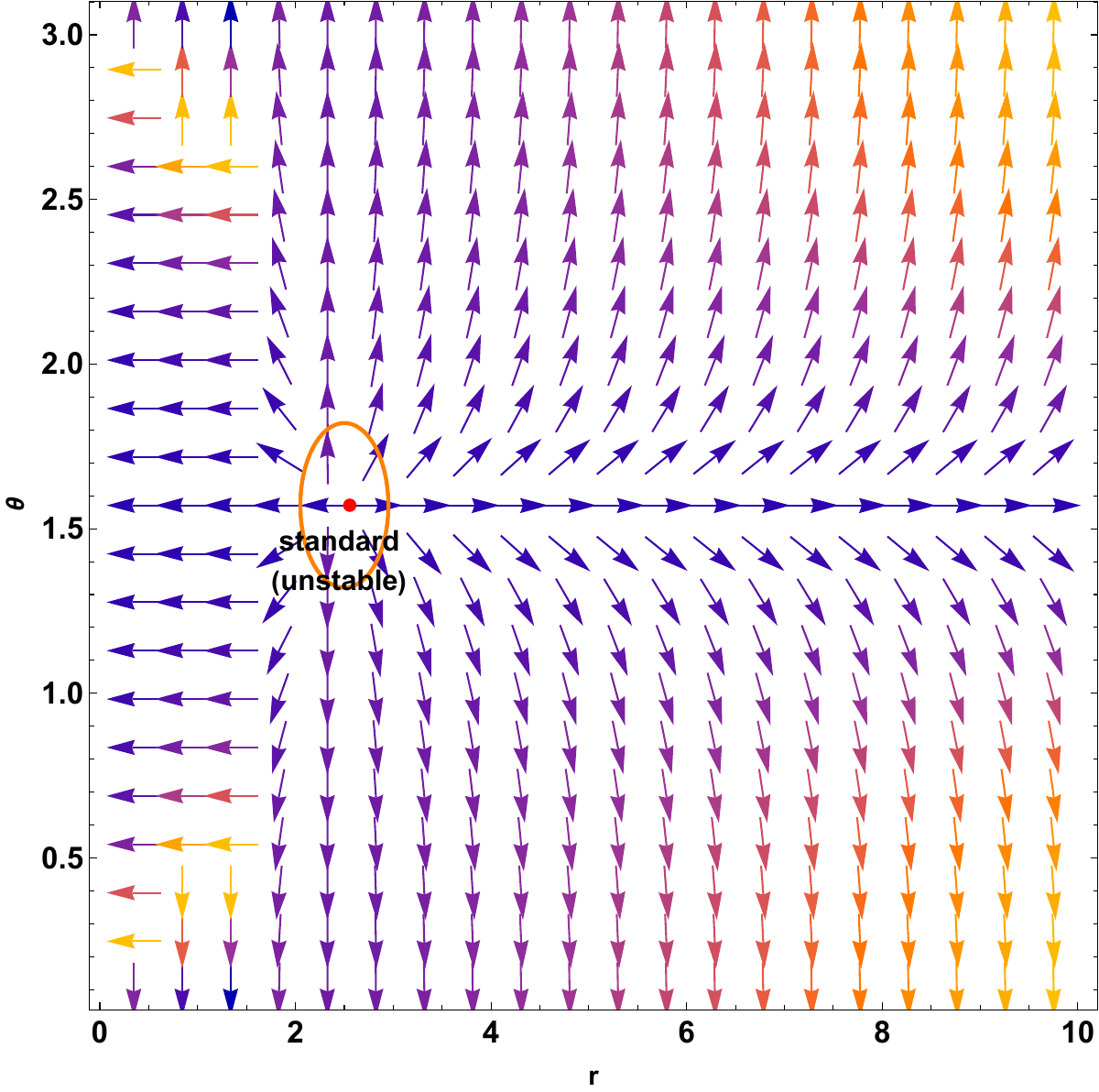}\\
    (i) $\lambda=0.2$ \hspace{6cm}  (ii) $\lambda=0.4$ \hspace{6cm} (iii) $\lambda=0.6$
    \caption{\footnotesize The arrows represent the normalized vector field $\boldsymbol{n}$ on a portion of the $r$–$\theta$ plane for a static black hole with parameters $M = 1$, $N = 0.01$, $w = -\tfrac{2}{3}$, $\alpha = 0.1$, and $\ell_p = 10$. The standard (or zero) points are indicated by the red dots located at $(r, \theta) = (r_0, \pi/2)$, where (i) $r_0 = 2.575$ for $\lambda = 0.2$, (ii) $r_0 = 2.565$ for $\lambda = 0.4$, and (iii) $r_0 = 2.555$ for $\lambda = 0.6$. The photon sphere is enclosed by the orange color closed loop. Evidently, the topological charge of the photon sphere is $Q = +1$.
}
\label{fig:unit-vector-2}
\end{figure*}

This argument obviously indicates that $j^{\mu}$ is nonzero only at the zero points of $\phi^{a}(x_{i})$, namely, $\phi^{a}(x_{i}) = 0$. Therefore, the topological charge at the given parameter region $\Sigma$ can be determined by using the following formula:
\begin{equation}
Q = \int_{\Sigma} j^{0} \, d^{2}x = \sum_{i=1}^{N} \beta_{i} \eta_{i} = \sum_{i=1}^{N} w_{i}.
\label{p14}
\end{equation}
When a closed curve includes a zero point, the charge $Q$ is exactly equal to the winding number. If the curve encloses multiple zero points, $Q$ is the sum of the winding numbers at each zero point. If it encloses no zero points, the charge is zero. To summarize how to calculate topological charges and total charges for each case: (i) If the vector field lines diverge around the zero point, the topological charge is $+1$. This corresponds to a naked singularity. (ii) If the field lines converge around the zero point, the topological charge is $-1$. This corresponds to a black hole with an unstable photon sphere. (iii) If the contour does not include a winding, the charge is zero. The total charges for each shape are the algebraic sum of the charges, which can be visualized by drawing a contour that includes all the zero points. From these findings, we can conclude that each black hole may have a photon sphere. The topological current is non-zero only at the zero point of the vector field, determining the photon sphere's location. Thus, a single topological charge can be assigned to each photon sphere.

Here, the positive Hopf index $\beta_{i}$ represents the number of loops made by $\phi^{a}$ in the vector $\phi$-space as $x^{\mu}$ moves around the zero point $z_{i}$; the Brouwer degree 
\begin{equation}
    \eta_{i} = \text{sign}\big(J^{0}(\phi/x)_{z_{i}}\big) = \pm 1.\label{p15}
\end{equation}
For a closed, smooth loop $C_{i}$ that encloses the $i$th zero point of $\phi$ while excluding other zero points, the winding number of the vector is then given by
\begin{equation}
w_{i} = \frac{1}{2\pi} \oint_{C_{i}} d\Omega,
\label{p16}
\end{equation}
where \(\Omega = \arctan(\phi^{\vartheta}/\varphi^{r}).\)

In our case, the normalized vector field components $\varphi^{r}$ and $\varphi^{\theta}$ are given by
\begin{align}
\varphi^{r}&=- \tfrac{\left(1-\alpha - \frac{3\,M}{r}+\frac{\lambda}{2\,r}\left(3\,\ln\!\frac{r}{|\lambda|} - 1\right)\notag-\frac{N\,(3\,w+3)}{2\,r^{3\,w+1}}\right)}{r^{2}\,\sin\theta},\\
\varphi^{\theta}&= -\tfrac{\sqrt{1-\alpha - \frac{2\,M}{r}+\frac{\lambda}{r}\,\mbox{ln}\,\frac{r}{|\lambda|}-\frac{N}{r^{3\,w+1}}+\frac{r^2}{\ell^2_p}}}{r^2}\,\frac{\cot\theta}{\sin\theta}.\label{p17}
\end{align}
Noted that zero points occur at $r=r_{\rm ph}$ satisfying the relation given in Eq.~(\ref{cc7}) and $\theta=\pi/2$, the equatorial plane.

From the above vector field components, we observe that the geometric and physical key parameters \((\alpha, \lambda, N, w)\), as well as the black hole mass $M$ and curvature radius $\ell_p$, later determine the topological characteristics of the photon sphere. 

For the static black hole solution considered in the present study, the normalized vector field $\boldsymbol{n}$ is plotted on a portion of the $r$–$\theta$ plane, as shown in Figure~\ref{fig:unit-vector-1}, with parameters $M = 1$, $N = 0.01$, $w = -2/3$, $\lambda = 0.1$, and $\ell_p = 10$. In this figure, the photon sphere is located at the point, indicated by the red dot at $(r, \theta) = (r_0, \pi/2)$, where (i) $r_0 = 2.56$ for $\alpha = 0.05$, (ii) $r_0 = 3.07$ for $\alpha = 0.1$, and (iii) $r_0 = 3.06$ for $\alpha = 0.15$. Similarly, Figure~\ref{fig:unit-vector-2} shows the normalized vector field $\boldsymbol{n}$ on the $r$–$\theta$ plane for $M = 1$, $N = 0.01$, $w = -2/3$, $\alpha = 0.1$, and $\ell_p = 10$. In this case, the photon sphere is again located at the point, marked by the red dot at $(r, \theta) = (r_0, \pi/2)$, where (i) $r_0 = 2.575$ for $\lambda = 0.2$, (ii) $r_0 = 2.565$ for $\lambda = 0.4$, and (iii) $r_0 = 2.555$ for $\lambda = 0.6$. 

From both figures, it is evident that the topological charge of the photon sphere is $Q = +1$, as the vector field diverges outward and the photon sphere is enclosed by the orange color closed loop. Based on the classification of photon spheres (or light rings), this configuration is of the \emph{standard} type and is dynamically \emph{unstable}~\cite{Cunha2020,Wei2020}.

\begin{center}
\large{\bf B.\,Test Particle Dynamics}
\end{center}

In this part, we study the test-particle dynamics, specifically focusing on the ISCO radius and showing how several geometric and physical parameters influence it.

\begin{figure*}[tbhp]
  \includegraphics[width=0.32\linewidth]{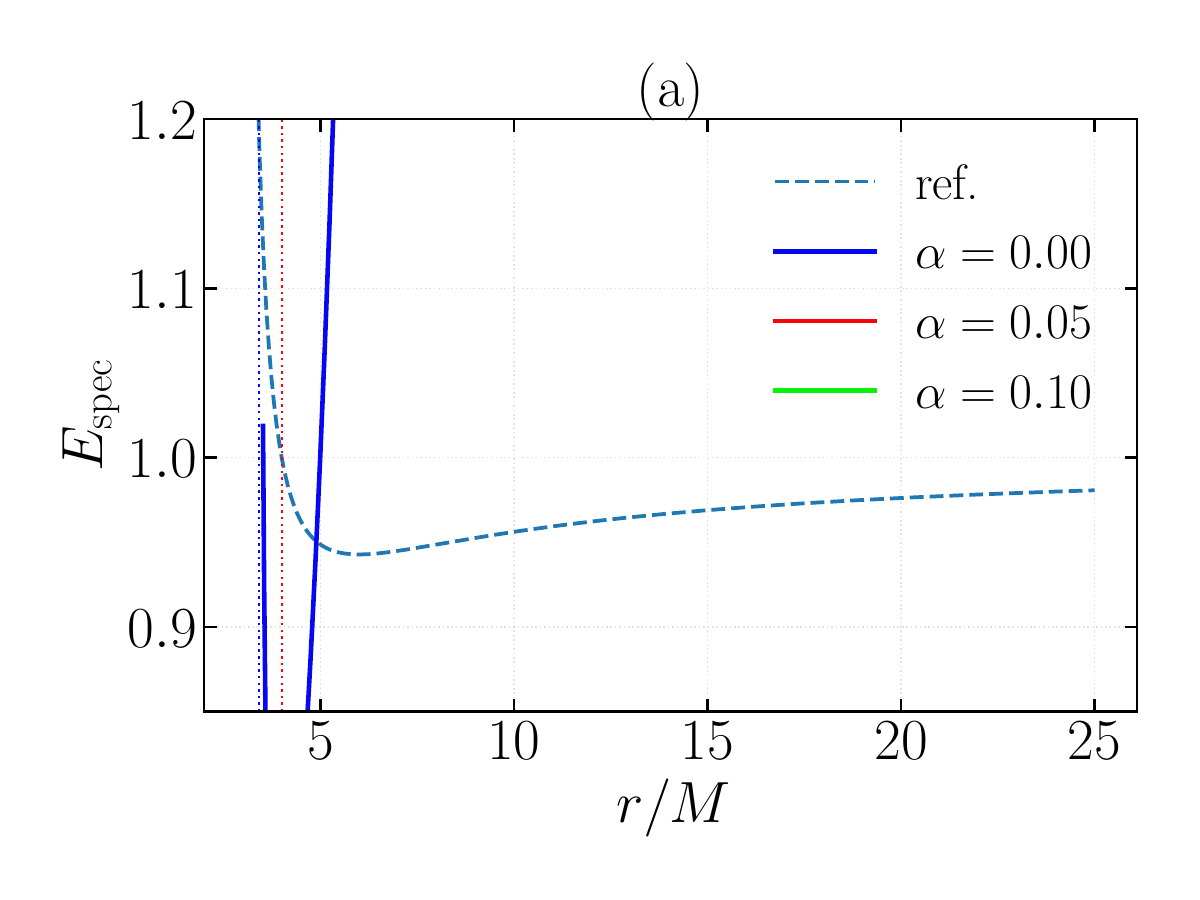}
  \includegraphics[width=0.32\linewidth]{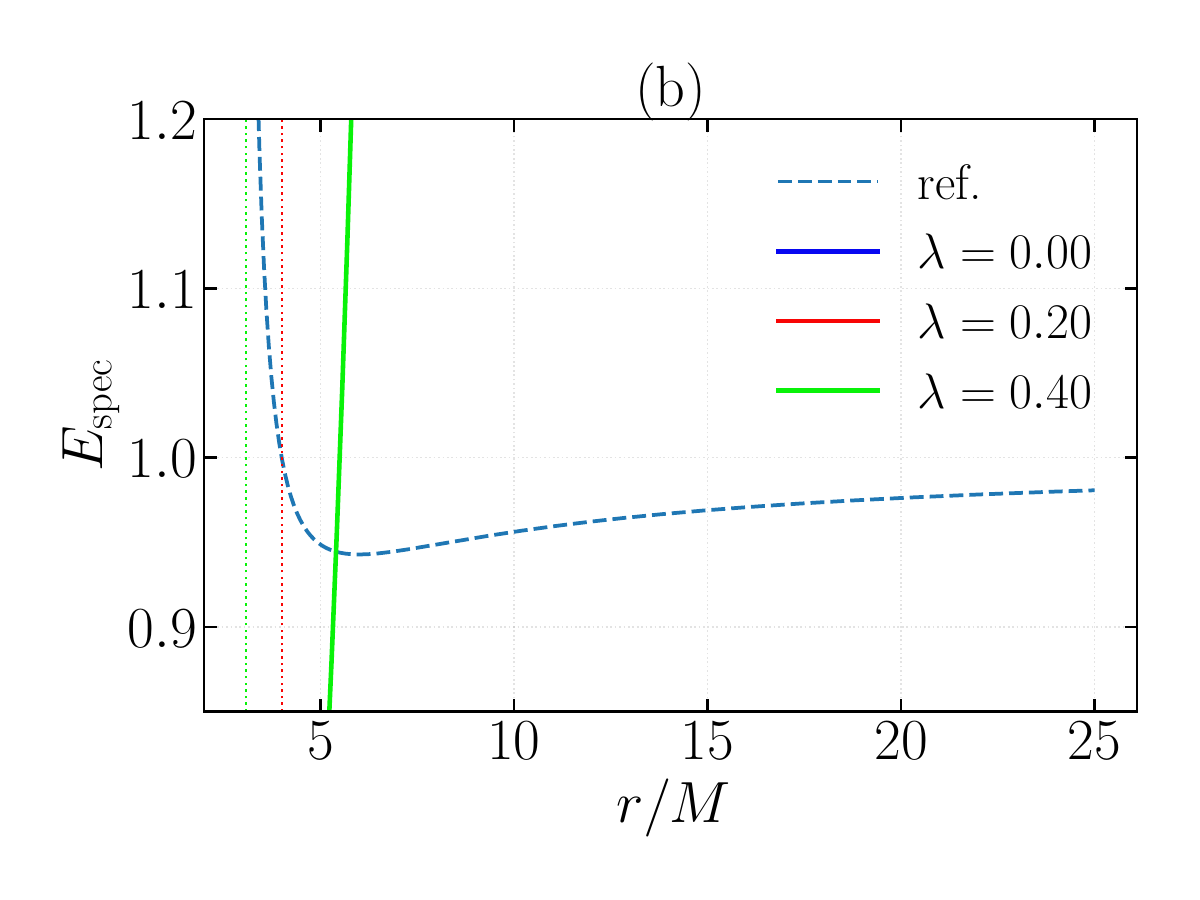}
  \includegraphics[width=0.32\linewidth]{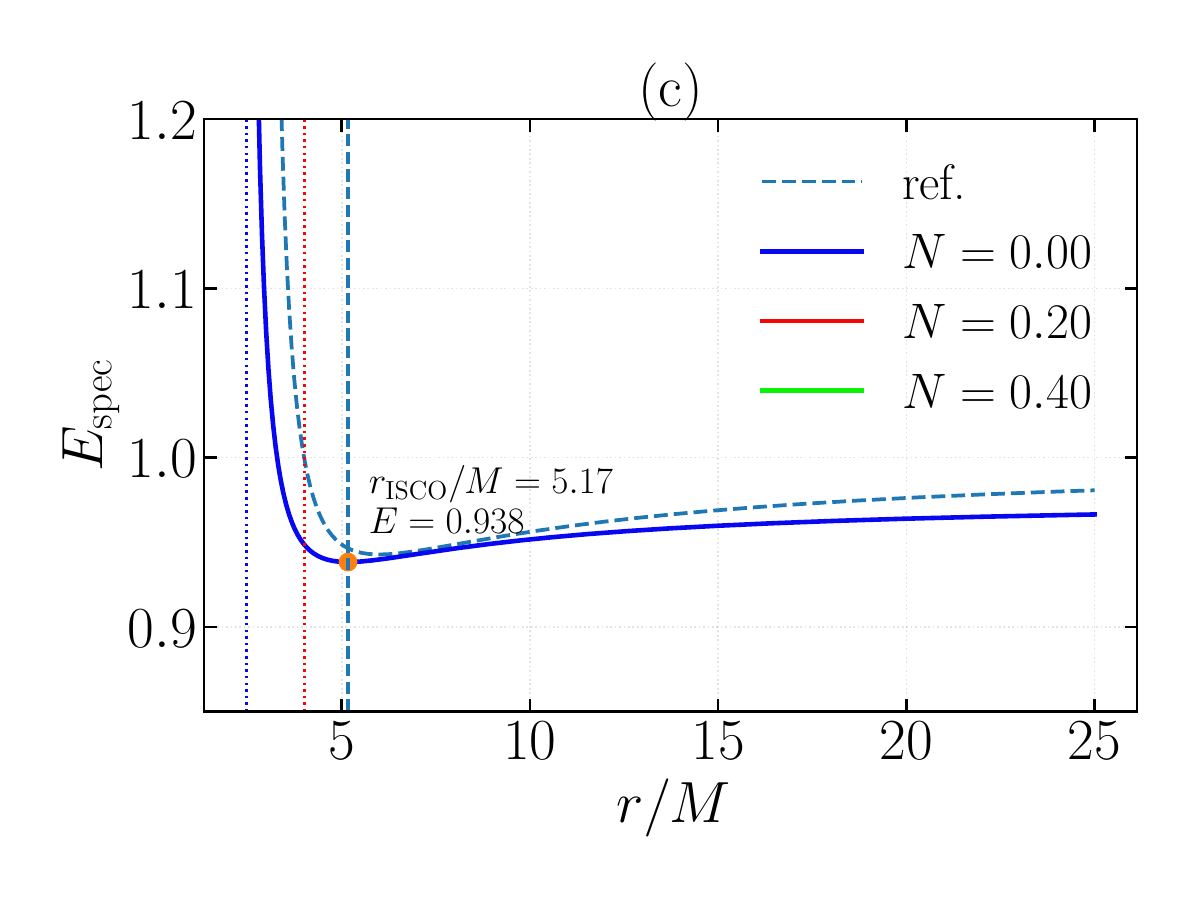}
  \caption{\footnotesize
  Specific energy and ISCO trends.
  Figures (a)-(c) show the specific energy of circular timelike orbits, $E_{\rm spec}(r)$, for controlled sweeps of the external sectors in our spacetime.
  \textbf{(a)} Sweep in the cloud-of-strings parameter $\alpha$ (CoS);
  increasing $\alpha$ effectively subtracts a constant from $f(r)$, shifting both the photon sphere and the ISCO outward and slightly raising the minimum of $E_{\rm spec}(r)$.
  \textbf{(b)} Sweep in the perfect fluid dark matter parameter $\lambda$; the $(\lambda/r)\ln(r/|\lambda|)$ contribution produces modest, non-monotonic changes that are small over the ranges explored.
  \textbf{(c)} Sweep in the quintessence normalization $N$ at fixed $w=-2/3$; larger $N$ lowers $f(r)$ at intermediate/large radii, shifting the photon sphere and the ISCO outward and increasing the value at the minimum.
  In all panels, $M=1$ and the other parameters are held fixed at the reference values quoted in the text.
  The dashed curve is the Schwarzschild reference ($\alpha=\lambda=N=0$).
  Vertical dotted and dashed lines mark, respectively, the photon-sphere radius $r_{\rm ph}$ and the ISCO radius $r_{\rm ISCO}$ obtained from $3 f f' - 2 r (f')^2 + r f f''=0$.}
  \label{fig:EspecificEnergy}
\end{figure*}

For massive test particles, following the same procedure as before, the equation for the radial component is given by 
\begin{equation}
    \dot{r}^2+U_{\rm eff}=\mathbb{E}^2\label{dd0}
\end{equation}
with the effective potential defined as:
\begin{equation}
V_\text{eff}(r)=\left(1+\frac{\mathbb{L}^2}{r^2}\right)\,f(r).\label{dd1}
\end{equation}

For circular orbits, the following conditions must be satisfied \cite{Shaymatov2021b}
\begin{align}
    &\mathcal{E}^2=U_\text{eff}(r)=\left(1+\frac{\mathbb{L}^2}{r^2}\right)\,f(r),\label{dd2}\\
    &\partial_{r} U_\text{eff}(r)=0,\label{dd3}\\
    &\partial_{rr} U_\text{eff}(r) \geq 0.\label{dd4}
\end{align}
Here $\mathbb{E}$ and $\mathbb{L}$ respectively are the energy and angular momentum per unit mass of the test particles.

Substituting the effective potential in condition (\ref{dd3}), we find the specific angular momentum of test particles as
\begin{align}
\mathrm{L}_\text{sp}=\tfrac{r\,\sqrt{\frac{M}{r}-\frac{\lambda}{2\,r}\,\left(\ln\!\frac{r}{|\lambda|}-1\right) + \frac{\mathrm{N}\,(3\, w + 1)}{2\,r^{3\,w + 1}}+\frac{r^2}{\ell^2_p}}}{\sqrt{1-\alpha - \frac{3\,M}{r}+\frac{\lambda}{2r}\left(3\ln\!\frac{r}{|\lambda|}- 1\right)-\frac{3 N (w+1)}{2 r^{3\,w+1}}}}.\label{dd5}
\end{align}
Thereby, using (\ref{dd5}) in the condition (\ref{dd2}), we find the specific energy of test particles as
\begin{equation}
\mathrm{E}_\text{sp}=\pm\,\tfrac{1 -\alpha- \frac{2\,M}{r}+\frac{\lambda}{r}\,\ln\!\frac{r}{|\lambda|}-\frac{\mathrm{N}}{r^{3\,w+1}}+\frac{r^2}{\ell^2_p}}{\sqrt{\,1-\alpha - \frac{3\,M}{r}+\frac{\lambda}{2r}\left(3\ln\!\frac{r}{|\lambda|}- 1\right)-\frac{3 N (w+1)}{2 r^{3\,w+1}}\,}}.\label{dd6}
\end{equation}   

\begin{figure*}[thbp]
  \includegraphics[width=0.31\linewidth]{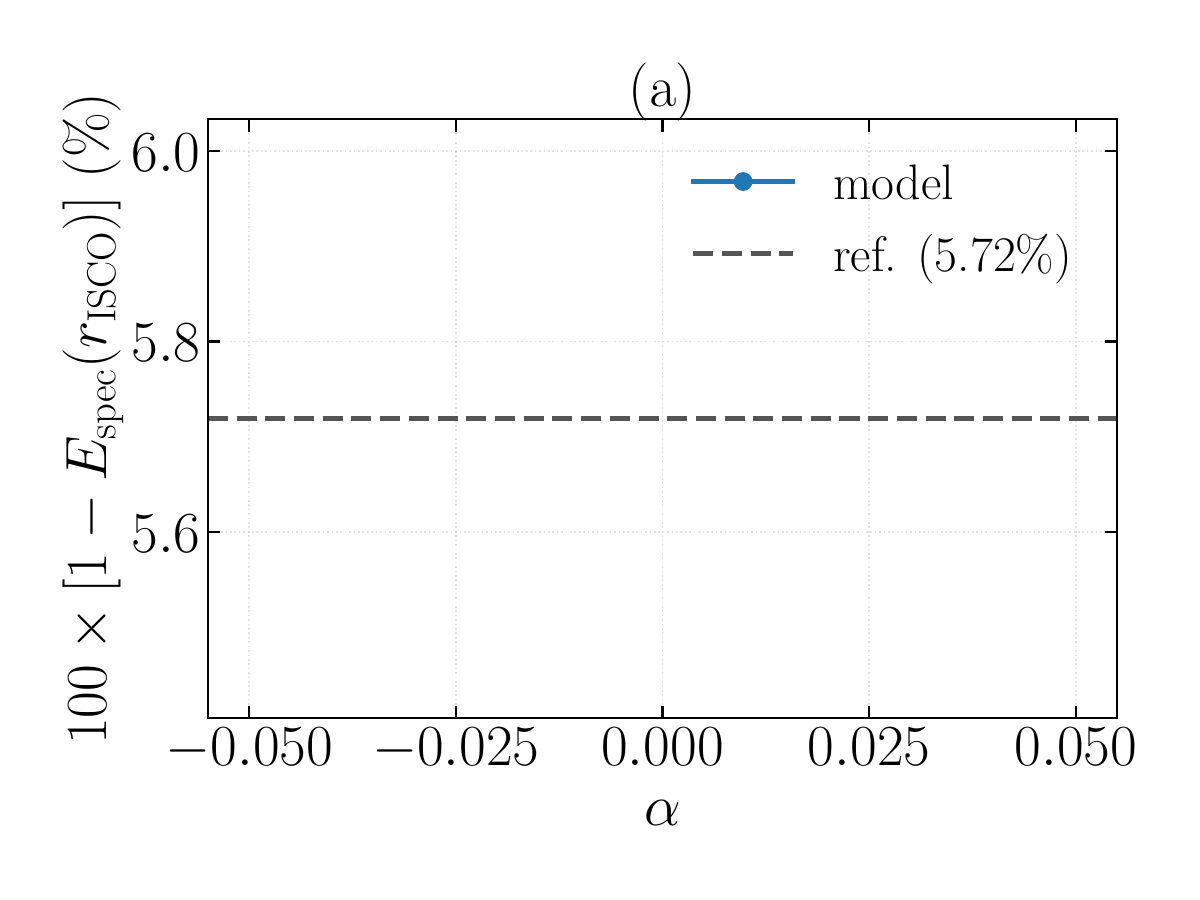}
  \includegraphics[width=0.31\linewidth]{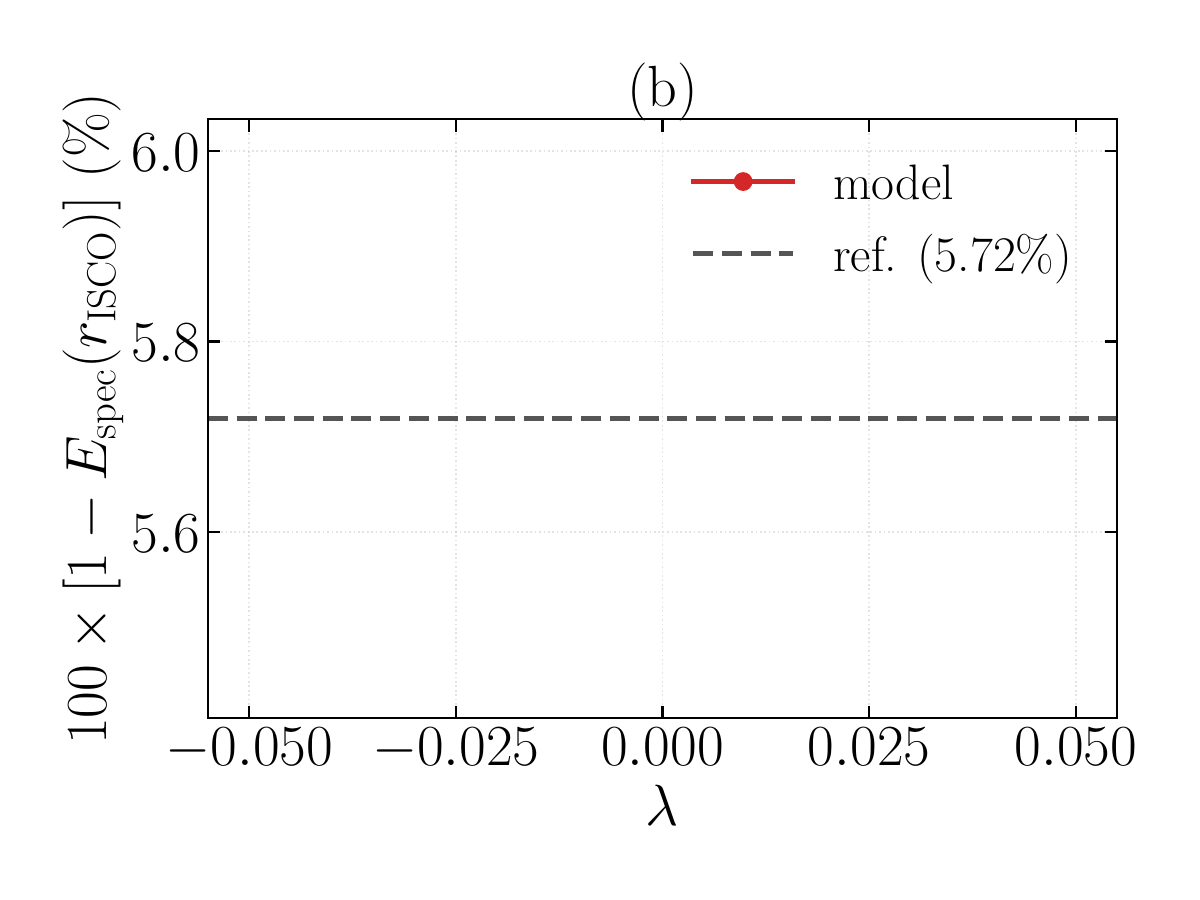}
  \includegraphics[width=0.31\linewidth]{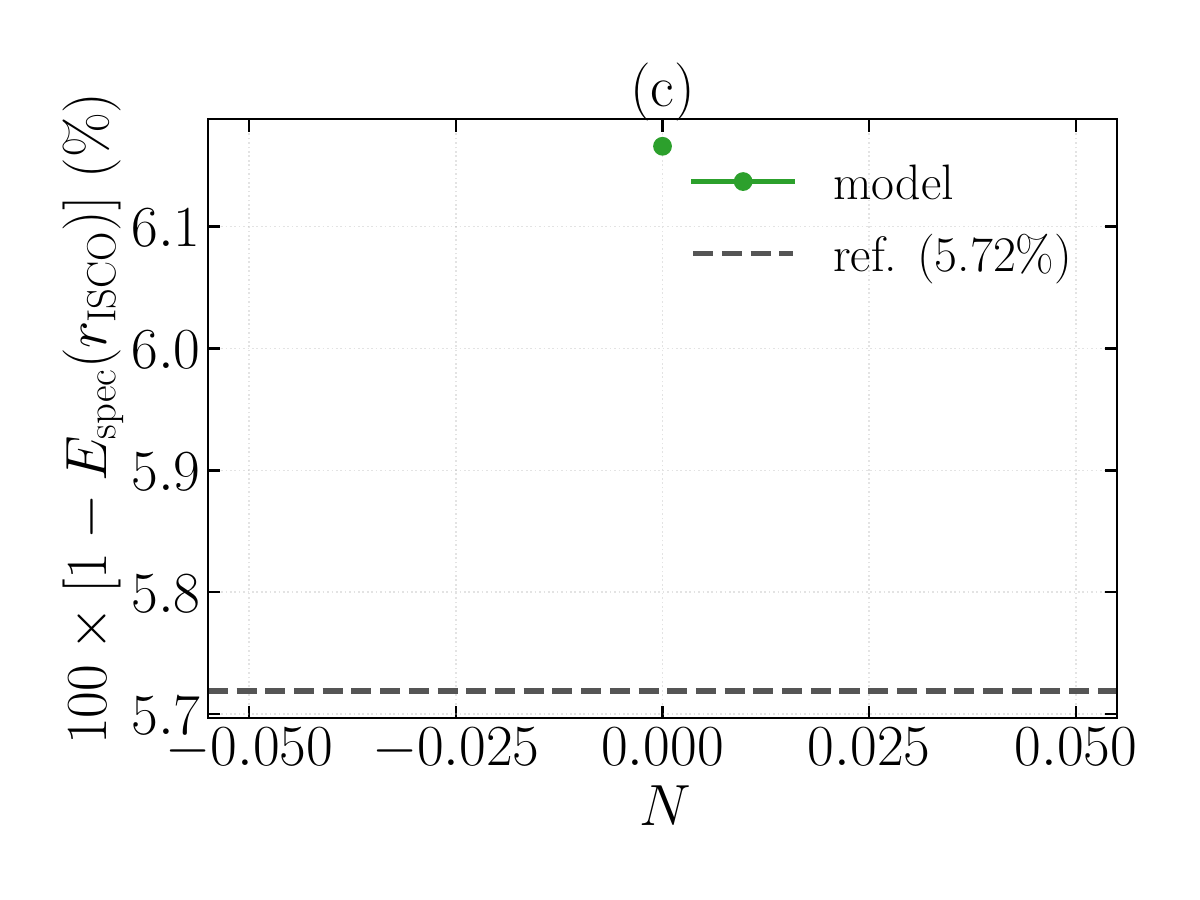}
  \caption{\footnotesize
  Radiative efficiency at the ISCO.
  Panels (a)–(c) plot $100\times\big[1 - E_{\rm spec}(r_{\rm ISCO})\big]$ (percent) for sweeps in each sector:
  \textbf{(a)} $\alpha$ (CoS), \textbf{(b)} $\lambda$ (perfect fluid dark matter), and \textbf{(c)} $N$ (quintessence, with $w=-2/3$).
  The horizontal dashed line shows the Schwarzschild benchmark.
  Across the explored windows, the efficiency is remarkably stable—varying at the sub-percent level with $\alpha$ and $\lambda$, and only modestly with $N$—indicating that the external sectors mainly shift characteristic radii (photon sphere and ISCO) while leaving the thin-disk efficiency close to the Schwarzschild value.}
  \label{fig:BindingISCO}
\end{figure*}

Equations \eqref{dd5}-\eqref{dd6} show that the specific angular momentum and energy depend on key parameters \((\alpha,\,\lambda,\,N,\,w,\,M)\) consistently through \(f(r)\).

Substituting the effective potential given in Eq.~(\ref{dd1}) into the  condition (\ref{dd4}) and using (\ref{dd3}), we find the following equation satisfying the ISCO radius $r=r_\text{ISCO}$ as (setting $w=-2/3$)
\begin{align}
& 3\,f(r)\,f'(r)-2\,r\,(f'(r))^2+r\,f(r)\,f''(r)= 0.\label{dd8}
\end{align}

From Eq.~(\ref{dd8}), it becomes clear that the ISCO radius depends on several geometric and physical parameters. No extra CoS parameter beyond \(\alpha\) is assumed. An analytic closed form for \(r_{\rm ISCO}\) is typically not available; numerical evaluation is straightforward once \((\alpha,\,\lambda,\,N,\,w,\,M)\) are fixed.

\paragraph*{Specific energy and ISCO trends.}

Figure~\ref{fig:EspecificEnergy} (panels (a)–(c)) displays the specific energy of circular timelike orbits, $E_{\rm spec}(r)$, for controlled sweeps of the external sectors.
In all figures, the curves asymptote to the Schwarzschild reference (dashed) at large radii and develop a shallow minimum that locates the ISCO.
In Fig. \ref{fig:EspecificEnergy}(a), increasing the cloud–of–strings parameter $\alpha$ effectively subtracts a constant from $f(r)$, pushing both the photon sphere and the ISCO outward; correspondingly, the minimum of $E_{\rm spec}(r)$ shifts to larger $r/M$ and its value rises slightly.
Figure \ref{fig:EspecificEnergy}(b) shows that the perfect fluid dark matter term $(\lambda/r)\ln(r/|\lambda|)$ alters the profile more subtly: the changes are modest over the ranges explored and reflect the non-monotonic radial dependence of this contribution.
In Fig. \ref{fig:EspecificEnergy}(c), increasing the quintessence normalization $N$ (with $w=-2/3$) lowers $f(r)$ at intermediate/large radii, shifting both the photon sphere and the ISCO outward and lifting the ISCO minimum of $E_{\rm spec}$.

\paragraph*{Radiative efficiency (binding at the ISCO).}

The corresponding efficiencies $100\times[1-E_{\rm spec}(r_{\rm ISCO})]$ are plotted in Fig.~\ref{fig:BindingISCO}(a)–(c).
Figure \ref{fig:BindingISCO}(a) indicates that the efficiency varies only at the sub-percent level with $\alpha$ across the considered range.
A similarly small sensitivity is seen for $\lambda$ in Fig.~\ref {fig:BindingISCO}(b), while Fig.~\ref{fig:BindingISCO}(c) shows a slightly more visible, but still modest, variation with $N$. Overall, the external sectors chiefly shift characteristic radii (photon sphere and ISCO) while leaving the accretion efficiency close to the Schwarzschild benchmark, complementing the null-geodesic analysis. 

\begin{figure*}[tbhp]
    \includegraphics[width=0.32\linewidth]{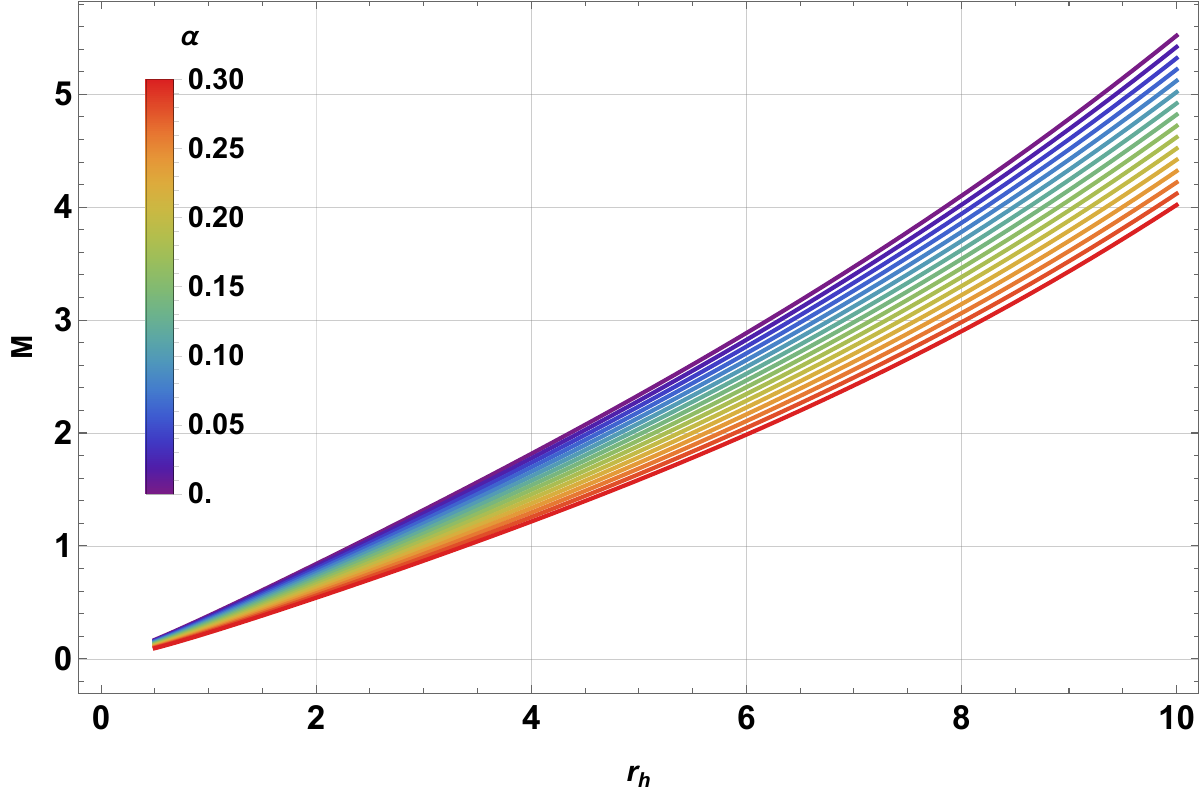}
    \includegraphics[width=0.32\linewidth]{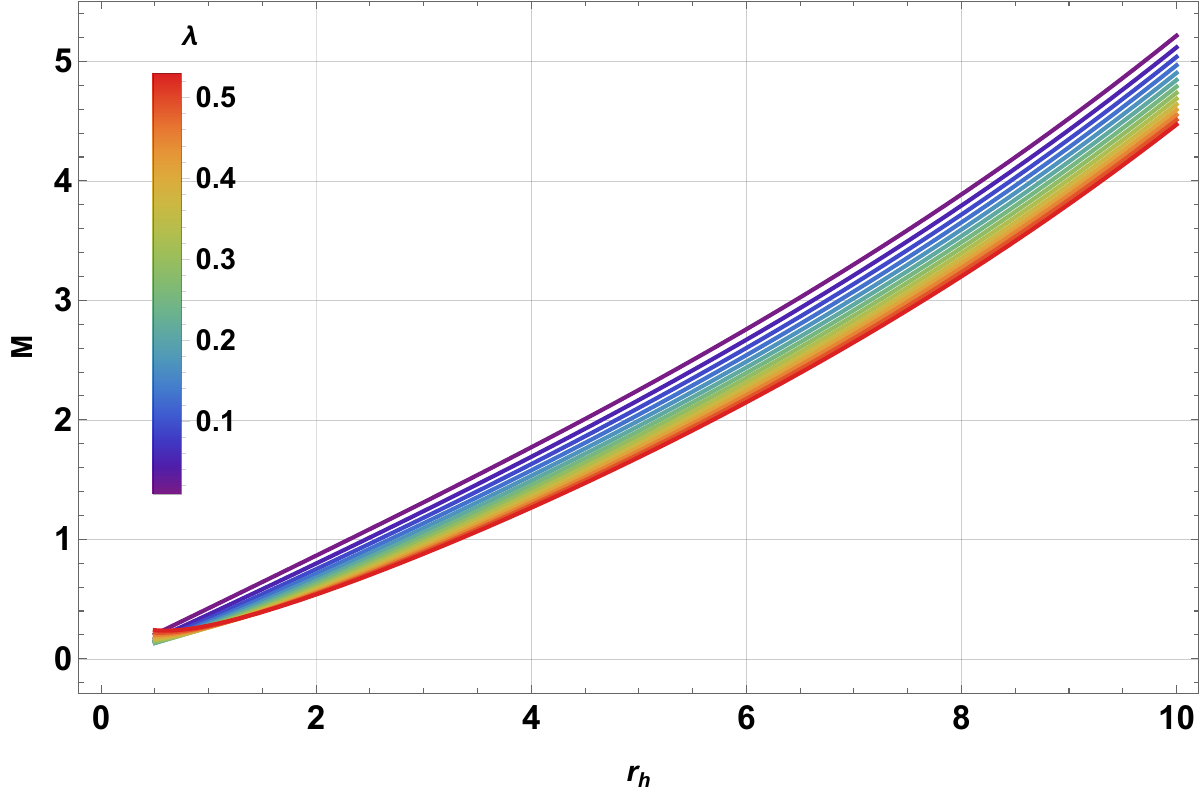}
    \includegraphics[width=0.32\linewidth]{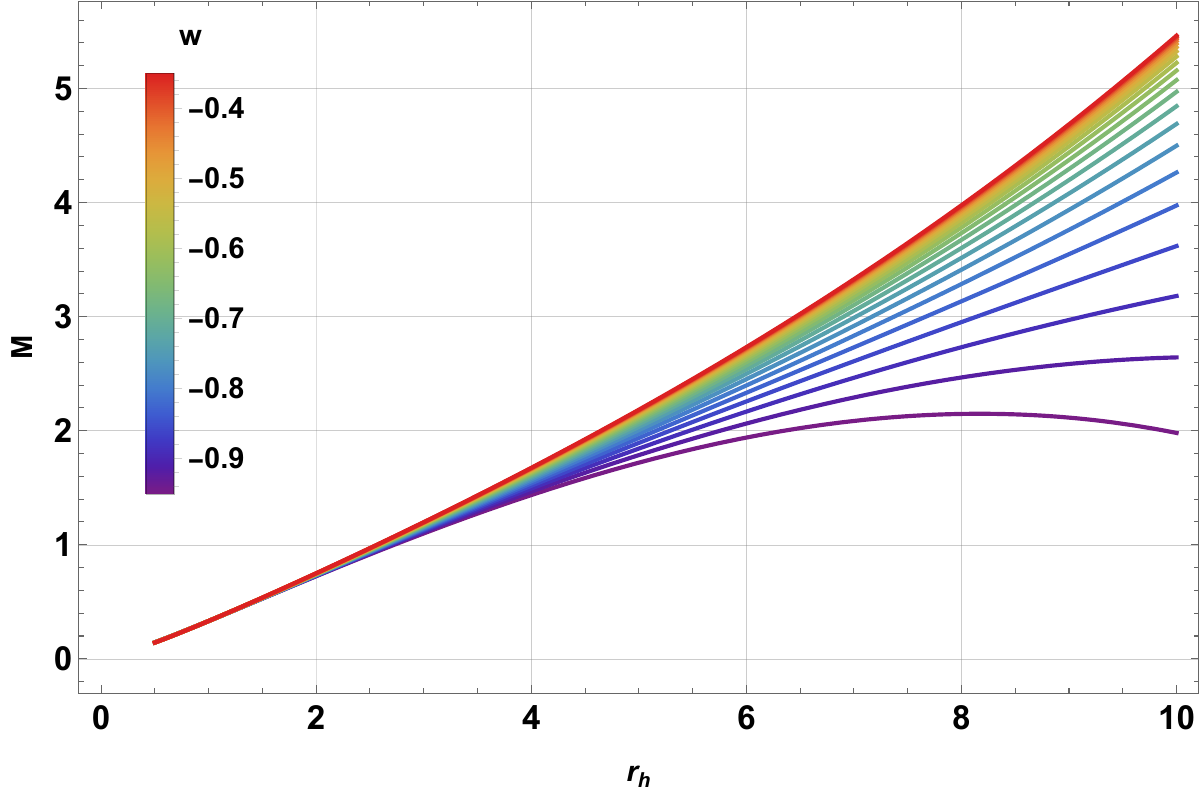}\\
    (i) $\lambda=0.1,\,w=-2/3$ \hspace{4cm} (ii) $\alpha=0.1,\,w=-2/3$ \hspace{4cm} (iii) $\alpha=0.1,\,\lambda=0.1$
    \caption{\footnotesize Behavior of the ADM as a function of the horizon radius. Each panel shows $M(r_h)$ while sweeping a single parameter with a colorbar: (a) string-cloud strength $\alpha$, (b) perfect fluid dark matter parameter $\lambda$, and (c) state parameter $w$. Here $N=0.01,\,\ell_p=\sqrt{\tfrac{3}{8\pi P}}=20$.}
    \label{fig:ADM-mass}
\end{figure*}

\begin{figure*}[tbhp]
    \includegraphics[width=0.32\linewidth]{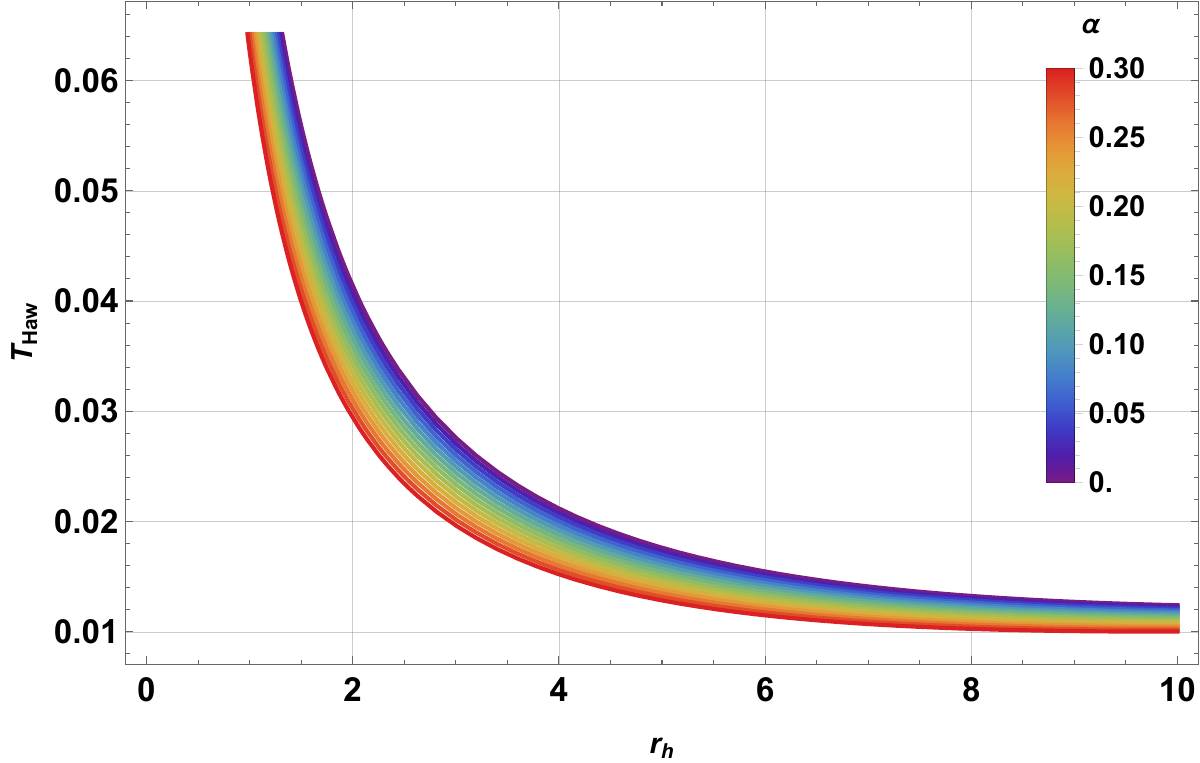}
    \includegraphics[width=0.32\linewidth]{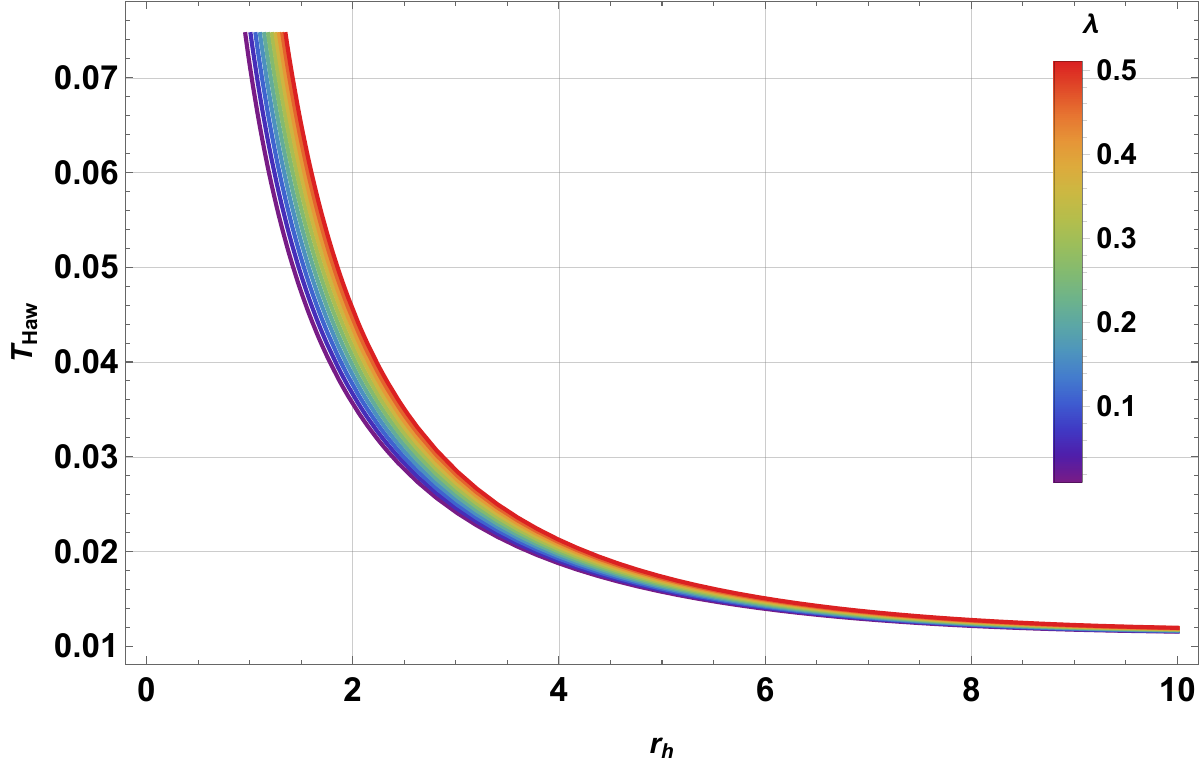}
    \includegraphics[width=0.32\linewidth]{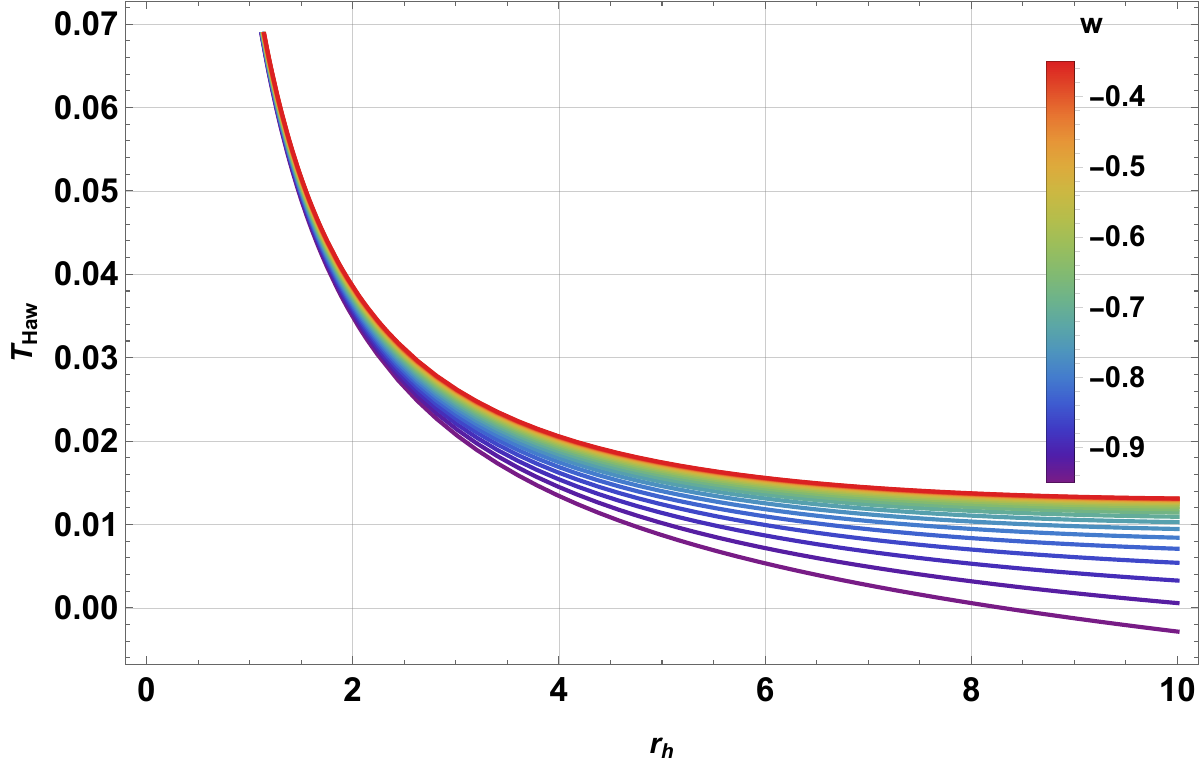}\\
    (i) $\lambda=0.1,\,w=-2/3$ \hspace{4cm} (ii) $\alpha=0.1,\,w=-2/3$ \hspace{4cm} (iii) $\alpha=0.1,\,\lambda=0.1$
    \caption{\footnotesize Behavior of the 
    Hawking temperature as a function of the horizon radius. Each panel shows $T_{\rm H}(r_h)$ while sweeping a single parameter with a colorbar: (a) string-cloud strength $\alpha$, (b) perfect fluid dark matter parameter $\lambda$, and (c) state parameter $w$. Here $N=0.01,\,\ell_p=\sqrt{\tfrac{3}{8\pi P}}=20$.}
    \label{fig:Hawking-temperature}
\end{figure*}

\begin{figure*}[tbhp]
    \includegraphics[width=0.32\linewidth]{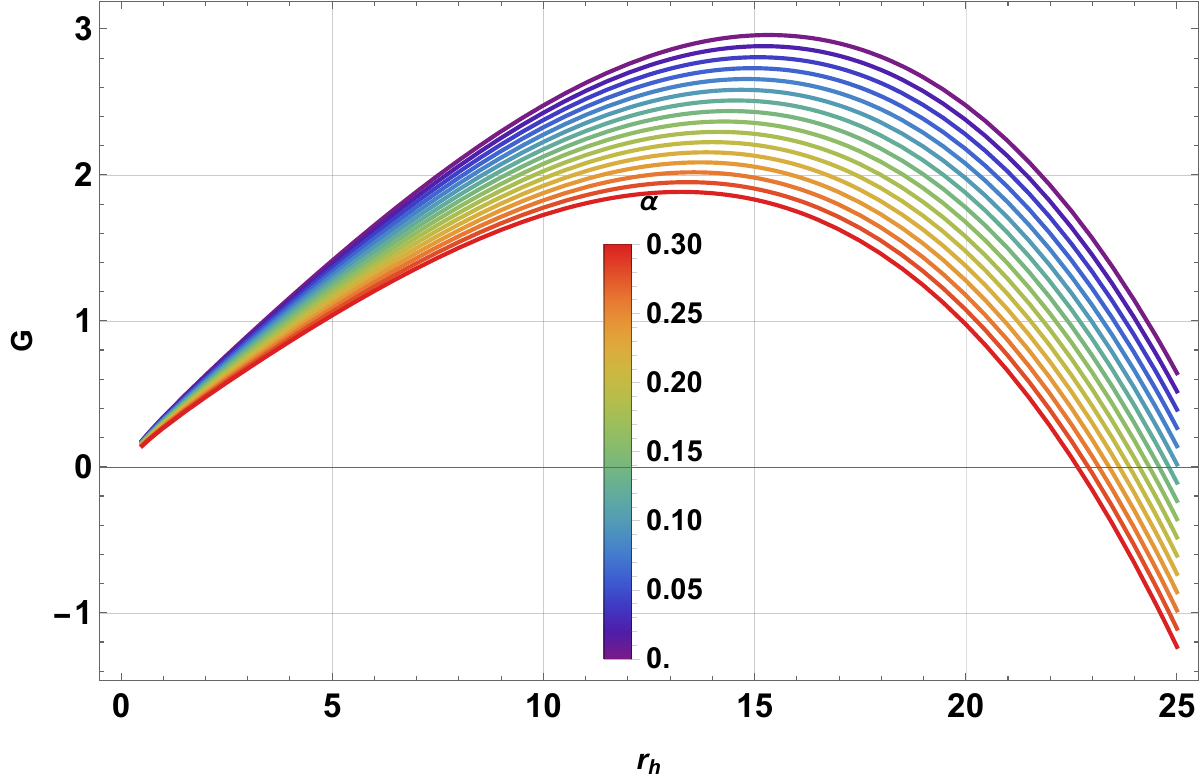}
    \includegraphics[width=0.32\linewidth]{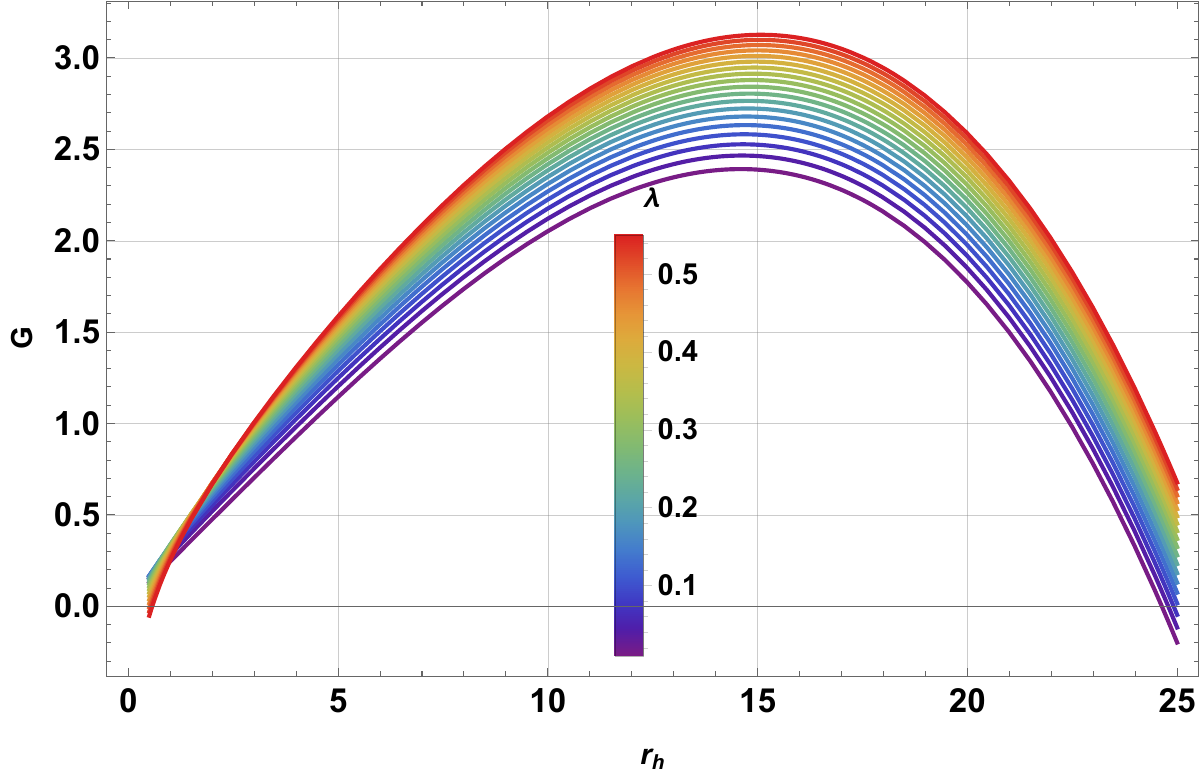}
    \includegraphics[width=0.32\linewidth]{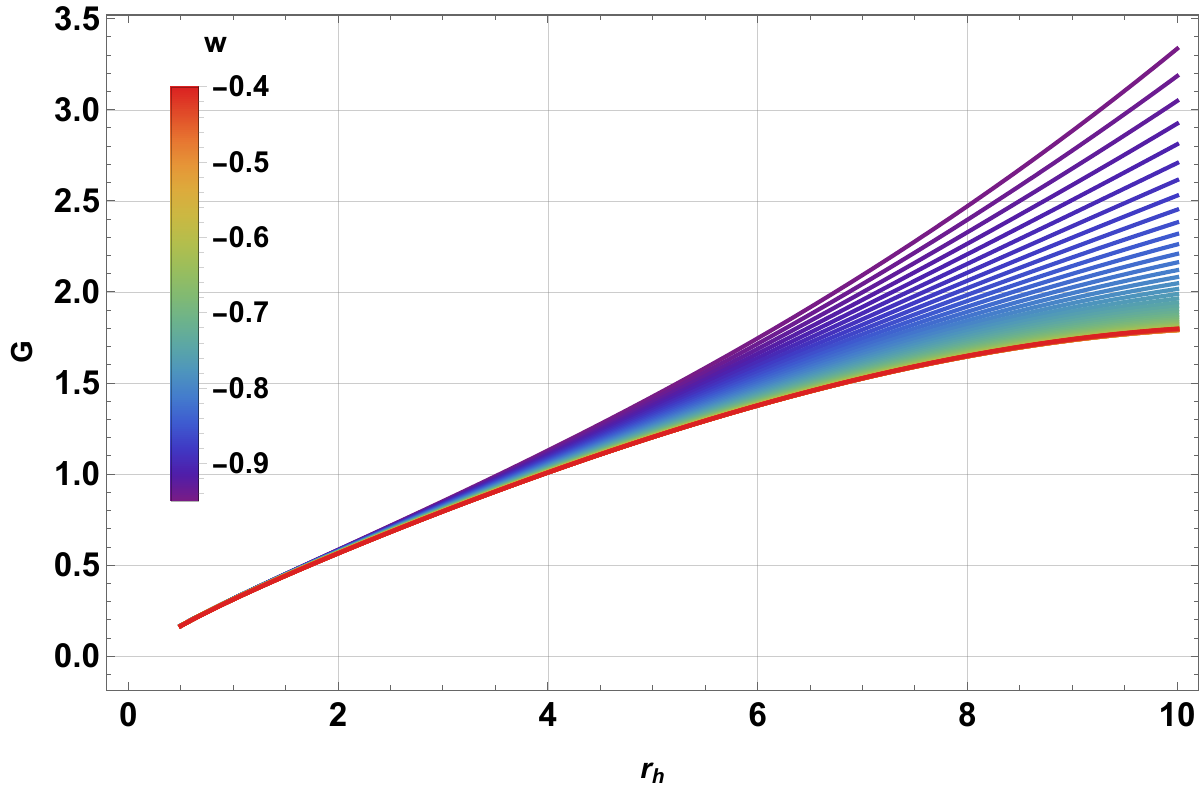}\\
    (i) $\lambda=0.1,\,w=-5/6$ \hspace{4cm} (ii) $\alpha=0.1,\,w=-5/6$ \hspace{4cm} (iii) $\alpha=0.1,\,\lambda=0.1$
    \caption{\footnotesize Behavior of the Gibbs free energy as a function of the horizon radius. Each panel shows $G$ while sweeping a single parameter with a colorbar: (a) string-cloud strength $\alpha$, (b) perfect fluid dark matter parameter $\lambda$, and (c) state parameter $w$. Here $N=0.01,\,\ell_p=\sqrt{\tfrac{3}{8\pi P}}=20$.}
    \label{fig:Gibb-energy}
\end{figure*}

\begin{figure*}[tbhp]
    \includegraphics[width=0.32\linewidth]{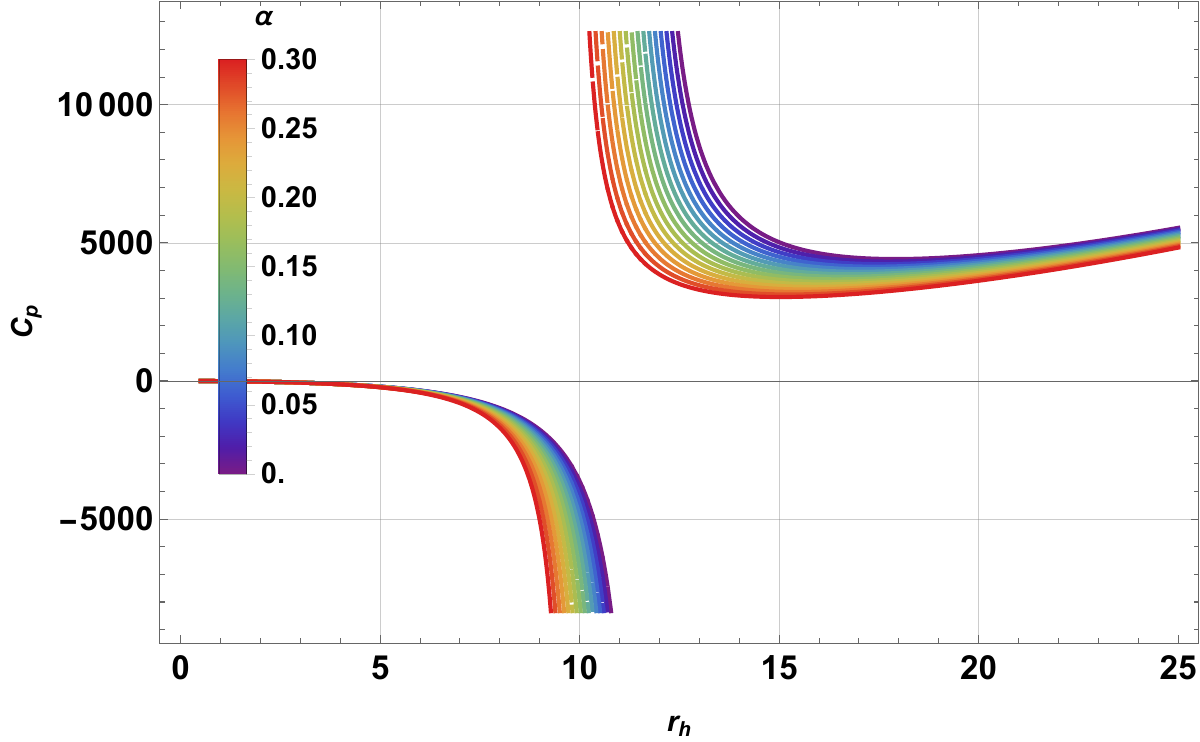}
    \includegraphics[width=0.32\linewidth]{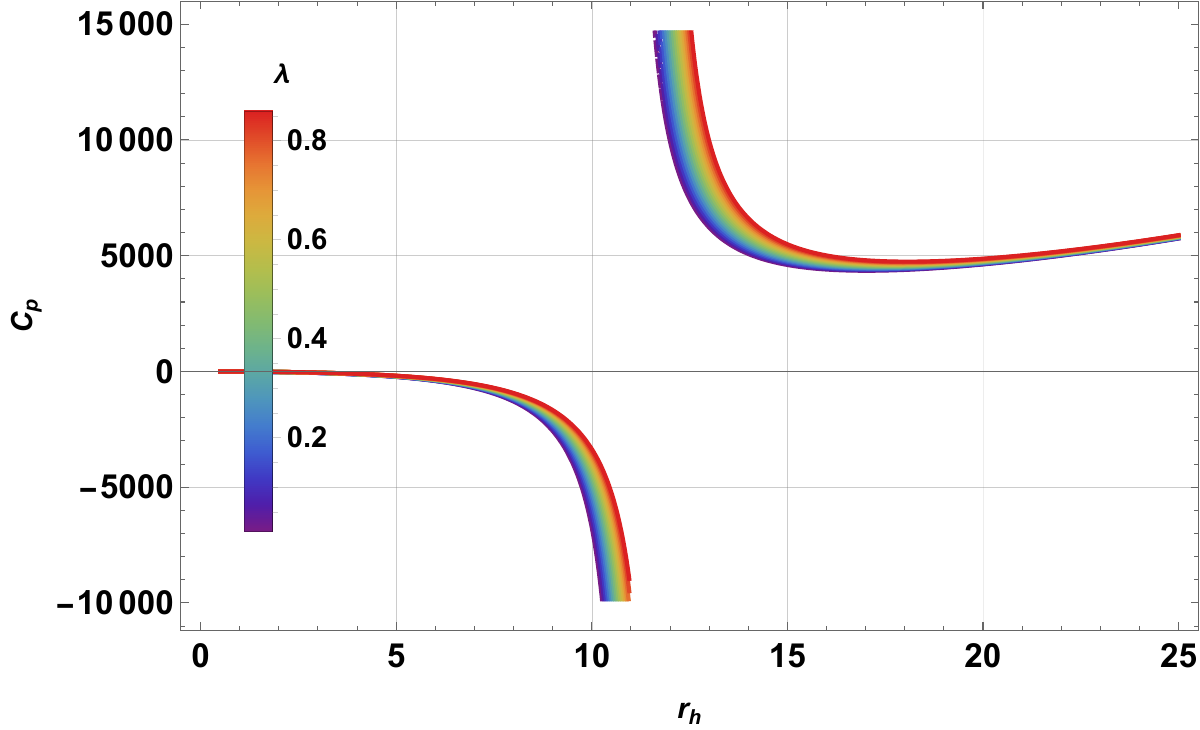}
    \includegraphics[width=0.32\linewidth]{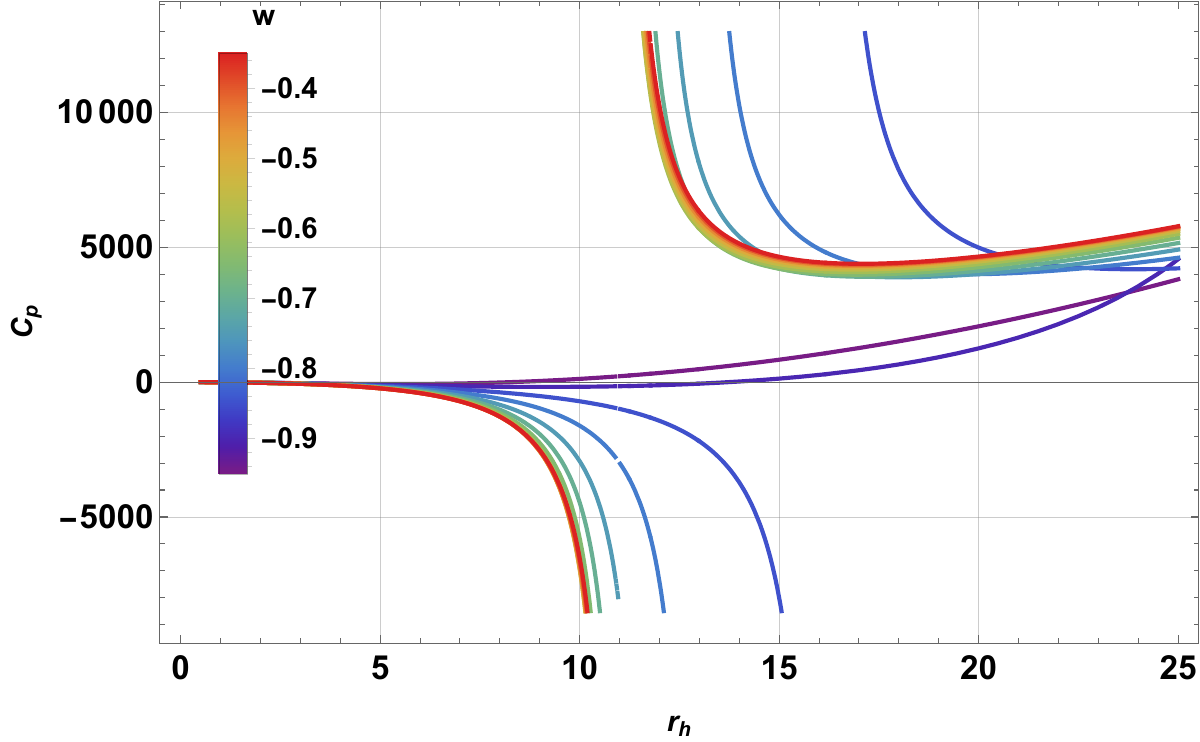}\\
    (i) $\lambda=0.1,\,w=-2/3$ \hspace{4cm} (ii) $\alpha=0.1,\,w=-2/3$ \hspace{4cm} (iii) $\alpha=0.1,\,\lambda=0.1$
    \caption{\footnotesize Behavior of the specific heat capacity $C_p$ as a function of the horizon radius. Each panel shows $C_p$ while sweeping a single parameter with a colorbar: (a) string-cloud strength $\alpha$, (b) perfect fluid dark matter parameter $\lambda$, and (c) state parameter $w$. Here $N=0.01,\,\ell_p=\sqrt{\tfrac{3}{8\pi P}}=20$.}
    \label{fig:specific-heat}
\end{figure*}

\section{Thermodynamics of BH }\label{sec:5}

In this section, we investigate the thermodynamics of the selected black hole and analyze the combined effects of the cloud of strings, perfect-fluid dark matter, and a quintessence-like field. The thermodynamics of black holes explores the deep connection between gravity, quantum mechanics, and statistical physics. In this framework, black holes behave like thermodynamic systems with well-defined quantities such as temperature, entropy, and heat capacity. The Hawking temperature arises from quantum effects near the event horizon, where particle–antiparticle pairs lead to thermal radiation known as Hawking radiation, implying that black holes can lose mass over time.

The ADM mass can be determined from the condition $f(r)=0$ at the radius $r=r_h$, called the event horizon. Therefore, the ADM mass is given by
\begin{equation}
M=\frac{r_h}{2}\left[1-\alpha+\frac{\lambda}{r_h}\,\mbox{ln}\,\frac{r_h}{|\lambda|}-\frac{N}{r^{3\,w+1}_h}+\frac{r^2_h}{\ell^2_p}\right].\label{ff1}
\end{equation}

Figure~\ref{fig:ADM-mass} illustrates the behavior of the ADM mass as a function of horizon for different values of $\alpha$, $\lambda$, and $w$. Panels (i) and (ii) show that increasing $\alpha$ and $\lambda$ lowers $M$, as expected from the term $(1-\alpha)$ and the perfect fluid dark matter term $(\lambda/r_h)\bigl(1-\ln|\lambda|\bigr)$. In contrast, panel (iii) indicates that increasing $w$ raises ADM mass.

The Hawking temperature is given by \cite{Hawking1974,Hawking1975,Hawking1983}
\begin{equation}
T_H=\frac{f'(r)}{4\pi}\Big{|}_{r=r_h}=\frac{1}{4\pi r_h}\Bigg[1-\alpha+\frac{\lambda}{r_h}+\frac{3w N}{r_h^{3w+1}}+\frac{3}{\ell^2_p}\,r^2_h\Bigg].
\label{ff2}
\end{equation}

Figure~\ref{fig:Hawking-temperature} illustrates the behavior of the Hawking temperature as a function of horizon for different values of $\alpha$, $\lambda$, and $w$. Panel (i) shows that increasing $\alpha$ systematically lowers $T_{\rm H}$, as expected from the $(1-\alpha)$ contribution in the numerator of $T_{\rm H}(r_h)$. In panel (ii), the perfect fluid dark matter term $(\lambda/r_h)\bigl(1-\ln|\lambda|\bigr)$ induces a non-monotonic dependence on $\lambda$: within the ranges shown, positive $\lambda$ typically raises the temperature while negative $\lambda$ lowers it, with the overall effect diminishing at larger $r_h$. Panel (iii) indicates that increasing $w$ decreases $T_{\rm H}$ and slightly shrinks the physically admissible $r_h$ domain, consistent with the $+(3wN)/r_h^{3w+1}$ term (negative for $w<0$). 
Across all panels, we restrict to $M(r_h)>0$ and $T_{\rm H}\ge 0$ to exclude non-outer or non-physical branches; the dashed $T_{\rm H}=0$ baseline marks the extremal limit.

In an extended phase space, the cosmological constant and hence the curvature radius $\ell_p$ is related to the thermodynamic pressure $P$ as \cite{Hawking1983},
\begin{equation}
    \Lambda=-\frac{3}{\ell^2_p}=-8\pi P.\label{ff3}
\end{equation}
Moreover, the entropy of the system is given by
\begin{equation}
    S=\int \frac{dM}{T_H}=\pi r^2_h\label{ff5}
\end{equation}
which is similar to the Bekenstein-Hawking entropy.

Thereby, in this extended phase space, the ADM mass and the Hawking temperature can be rewritten as 
\begin{align}
    M&=\frac{r_h}{2}\left[1-\alpha+\frac{\lambda}{r_h}\,\mbox{ln}\,\frac{r_h}{|\lambda|}-\frac{N}{r^{3\,w+1}_h}+\frac{8\pi P}{3}\,r^2_h\right],\label{ff4a}\\
    T_H&=\frac{1}{4\pi r_h}\Bigg[1-\alpha+\frac{\lambda}{r_h}+\frac{3w N}{r_h^{3w+1}}+8\pi P\,r^2_h\Bigg].\label{ff4}
\end{align}

Using Hawking temperature and ADM mass, which are considered the enthalpy of the thermodynamic system, we can now determine other thermodynamic variables and analyze the results. The Gibbs free energy is given by
\begin{align}
    G&=M-T_H\,S\nonumber\\
    &=\tfrac{r_h}{4}\left[
1 - \alpha
+ \tfrac{\lambda}{r_h}\left(2\ln\tfrac{r_h}{|\lambda|} - 1\right)
-\tfrac{N (2 + 3 w)}{r_h^{3w+1}}
- \tfrac{8\pi P}{3}\,r_h^2
\right].\label{ff6}
\end{align}

Figure~\ref{fig:Gibb-energy} illustrates the Gibbs free energy as a function of horizon for different values of $\alpha$, $\lambda$, and $w$. Figures ~\ref{fig:Gibb-energy}(i) and (iii) show that increasing $\alpha$ and $w$ lowers $G$, as expected from the term $(1-\alpha)$ and quintessence term $-(2+3w)N/r_h^{3w+1}$. In contrast, panel (ii) indicates an increasing $G$ with higher values of $\lambda$. 

Finally, the specific heat capacity at constant pressure is given by
\begin{align}
    C_p=T_H\,\left(\frac{\partial S}{\partial T_H}\right)_{P}=-\tfrac{
2\pi r_h^2 \left(1-\alpha+\frac{\lambda}{r_h}+\frac{3wN}{r_h^{3w+1}}+8\pi P r_h^2\right)
}{
\left(1-\alpha+\frac{2\lambda}{r_h} + \frac{3w(3w+2)N}{r_h^{3w+1}} - 8\pi P r_h^2
\right)}.\label{ff7}
\end{align}

\begin{figure*}[tbhp]
    \includegraphics[width=0.33\linewidth]{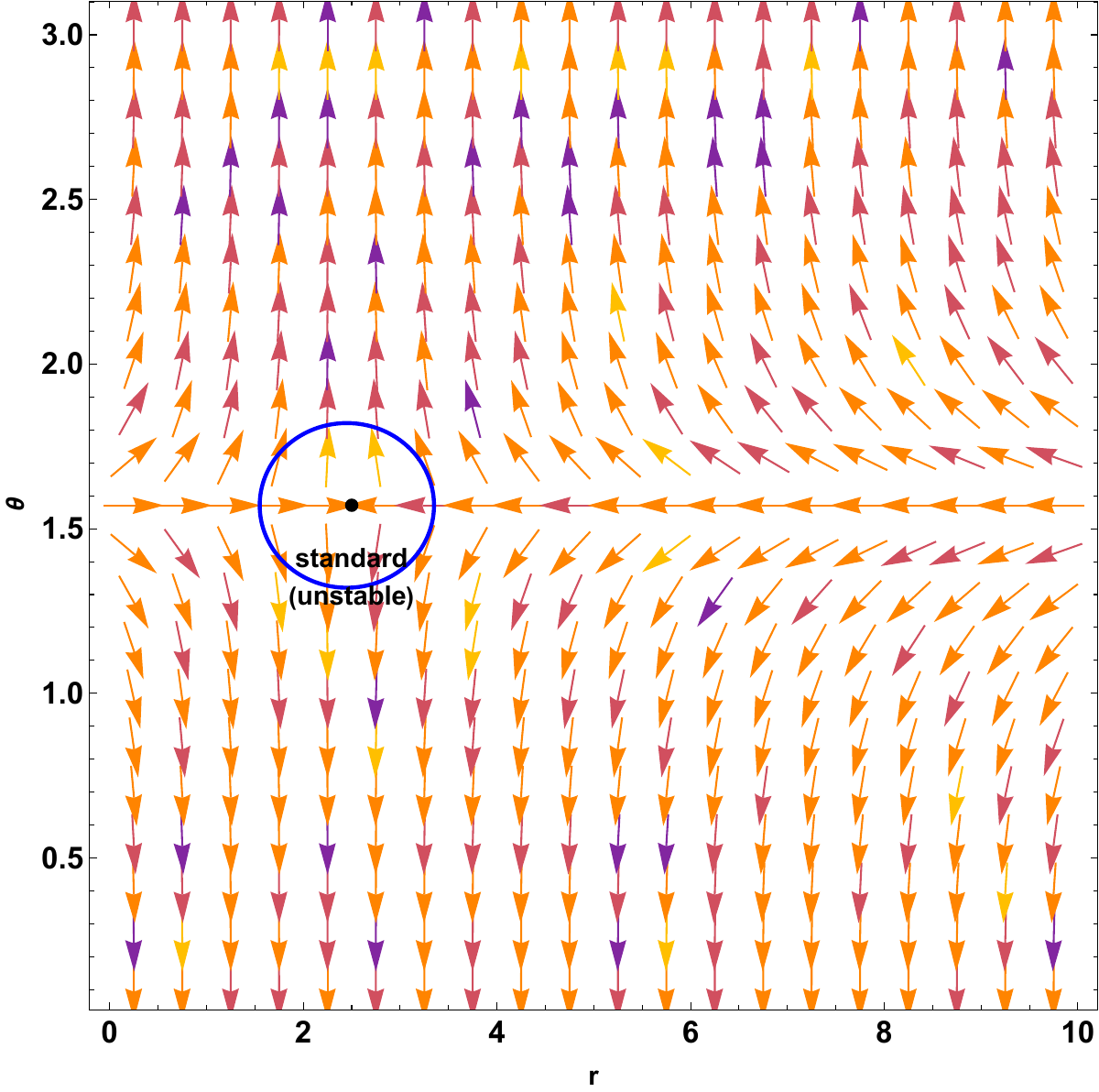}\qquad
    \includegraphics[width=0.33\linewidth]{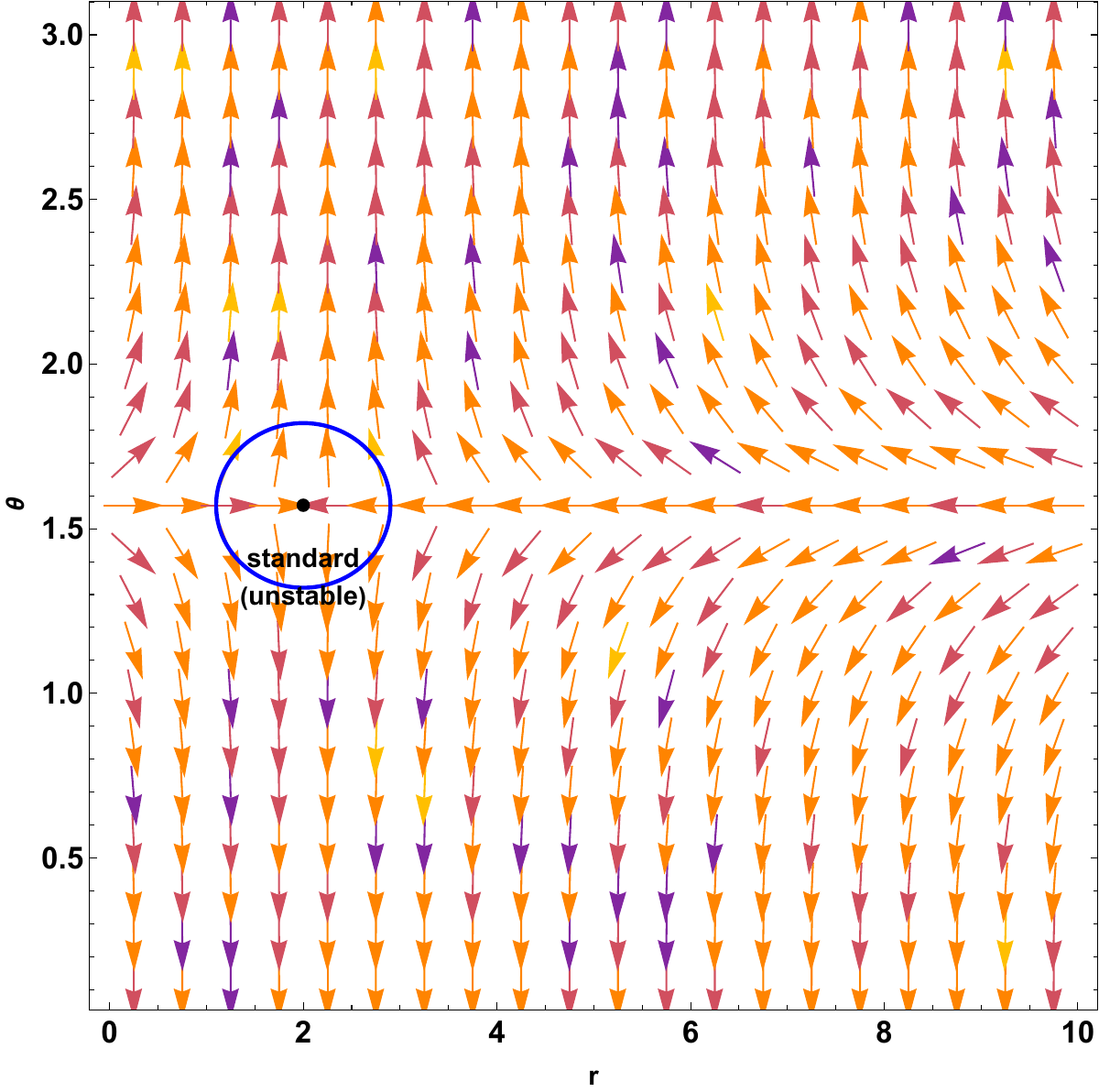}\\
    (i) $\alpha=0.1$ \hspace{6cm}  (ii) $\alpha=0.2$
    \caption{\footnotesize The arrows represent the normalized vector field $\boldsymbol{n}_{\mathcal{F}}$ of the energy $_{\mathcal{F}}$ on a portion of the $r$–$\theta$ plane for a static black hole with parameters $N = 0.1$, $w = -\tfrac{2}{3}$, $\lambda = 0.1$, $\ell_p =\sqrt{\tfrac{3}{8\pi P}} =50$, and $\tau=20 \pi$. The standard (or zero) point is indicated by the black dot located at $(r, \theta) = (r_h, \pi/2)$, where (i) $r_h = 2.5$ for $\alpha= 0.1$, and (ii) $r_h = 2.0$ for $\alpha = 0.2$. The horizon radius is enclosed by the blue color closed loop. Evidently, the topological charge is $W = -1$.}
    \label{fig:unit-vector-3}
\end{figure*}

\begin{figure*}[t!]
    \includegraphics[width=0.33\linewidth]{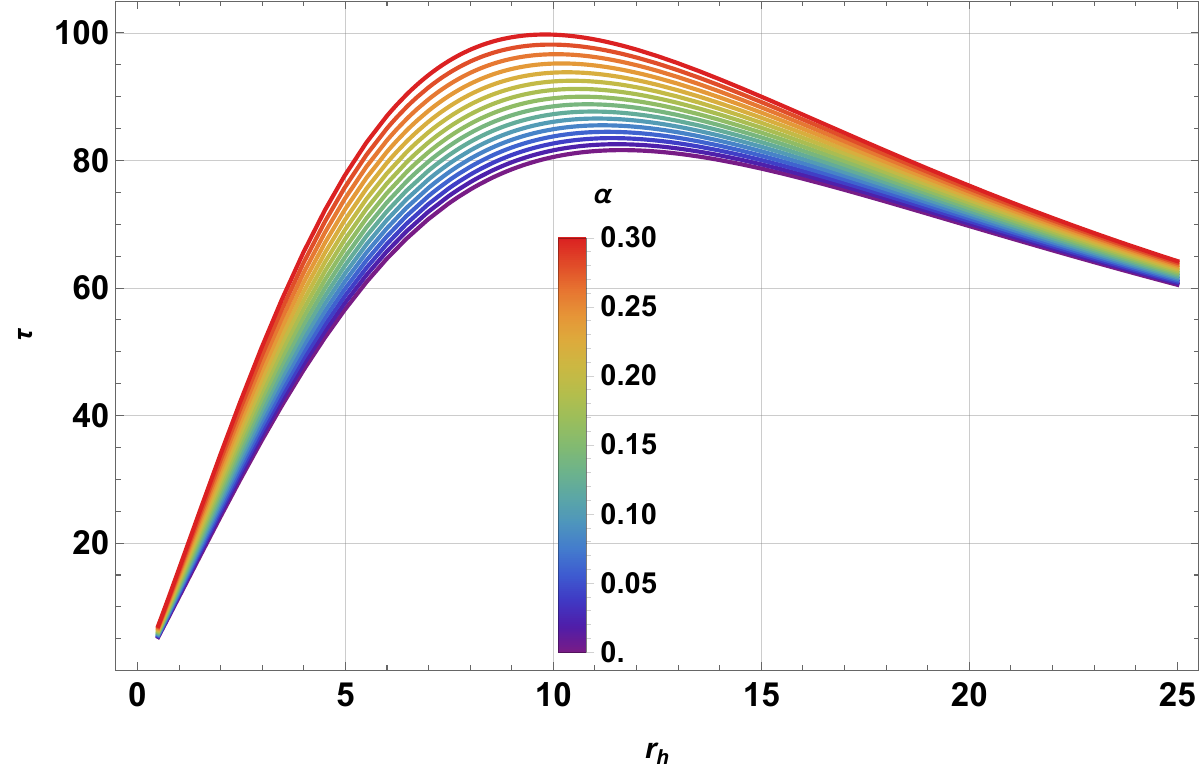}\qquad
    \includegraphics[width=0.33\linewidth]{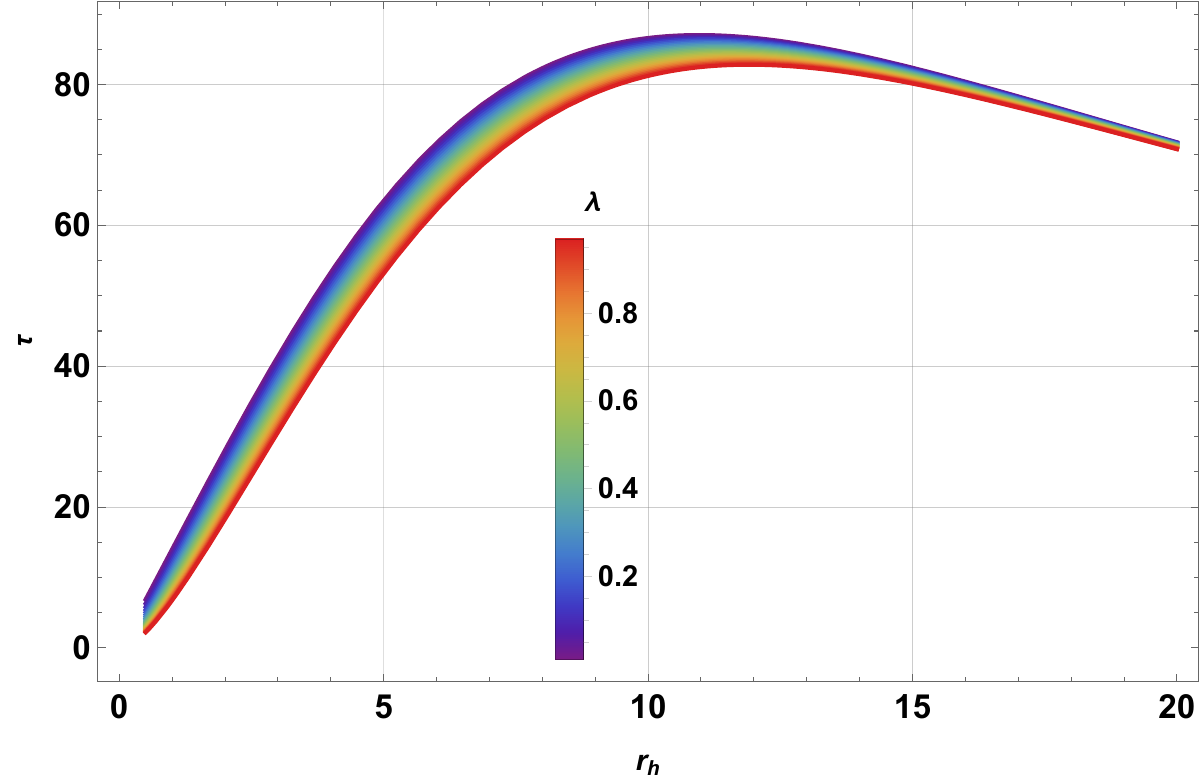}\\
    (i) $\lambda=0.1$ \hspace{6cm} (ii) $\alpha=0.1$
    \caption{\footnotesize Behavior of the inversion temperature $\tau$ as a function of the horizon radius. Each panel shows $G$ while sweeping a single parameter with a colorbar: (a) string-cloud strength $\alpha$, (b) perfect fluid dark matter parameter $\lambda$. Here $w=-2/3,\,N=0.01,\,\ell_p=\sqrt{\tfrac{3}{8\pi P}}=20$.}
    \label{fig:inversion-temperature}
\end{figure*}
From Eqs.~(\ref{ff4a})--(\ref{ff7}), it is evident that the ADM mass \(M(r_h)\), the Hawking temperature \(T_H\), the Gibbs free energy \(G\), and the specific heat capacity \(C_P\) all depend on the key parameters \((\alpha, N, w, \lambda)\), as well as on the thermodynamic pressure \(P\). Consequently, the thermodynamic quantities are modified by these geometric parameters, leading to deviations from the results known for the standard Schwarzschild black hole.

Figure \ref{fig:specific-heat} shows the behavior of the specific heat capacity as a function of horizon by varying string parameter $\alpha$, perfect fluid dark matter parameter $\lambda$, and state parameter $w$ of the quintessence-like field. Different parameters affect specific heat capacity differently, as shown in this figure.

\section*{Modified first law of thermodynamics}

Considering $\alpha,\,\lambda$ and $N$ as intensive thermodynamic variables, the first law of thermodynamics can be expressed as
\begin{equation}
dM=T_H\,dS+V\,dP+\Phi_{\alpha}\,d\alpha+\Phi_{\lambda}\,d\lambda+\Phi_{N}\,dN,\label{ff8}
\end{equation}
where $V$ is the thermodynamic volume and $\Phi_i$ is the thermodynamic potential associated with variables $\alpha, \lambda, N$ and is given by
\begin{align}
    V&=\left(\frac{\partial M}{\partial P}\right)\Big{|}_{S, \alpha, \lambda, N}=\tfrac{4\pi}{3}\,r^3_h=\tfrac{4\pi}{3}\,\left(\tfrac{S}{\pi}\right)^{3/2},\label{ff9a}\\
    \Phi_{\alpha}&=\left(\frac{\partial M}{\partial \alpha}\right)\Big{|}_{P, S, \lambda, N}=-\tfrac{r_h}{2}=-\tfrac{1}{2}\,\sqrt{\tfrac{S}{\pi}},\label{ff9}\\
    \Phi_{\lambda}&=\left(\frac{\partial M}{\partial \lambda}\right)\Big{|}_{P, S, \alpha, N}=\tfrac{1}{2}\,\ln\frac{r_h}{|\lambda|} - \tfrac{1}{2}=\tfrac{1}{2}\,\ln\frac{\sqrt{S/\pi}}{|\lambda|} -\tfrac{1}{2},\label{ff10}\\
    \Phi_{N}&=\left(\frac{\partial M}{\partial N}\right)\Big{|}_{P, \alpha, \lambda, N}=-\frac{1}{2r_h^{3w}}=-\tfrac{1}{2}\,\left(\tfrac{S}{\pi}\right)^{-3 w/2}.\label{ff11}
\end{align}

\section{Thermodynamic Topology}\label{sec:6}

Thermodynamic topology introduces topology into black hole thermodynamics by assigning topological numbers to the zero points in the phase diagram. These numbers, defined as the residues of the generalized Helmholtz free energy at critical points, reveal features and classifications that conventional analyses may overlook. This approach distinguishes between conventional and novel critical points, offering deeper insight into first-order phase transitions. The off-shell Helmholtz free energy, a generalization of the standard Helmholtz free energy, accounts for non-equilibrium states, while the Helmholtz free energy itself represents the system’s internal energy minus the product of temperature and entropy, quantifying the useful work obtainable at constant temperature and volume.

To study the thermodynamic topology of the black hole, we define a potential dependent on the Hawking temperature of the black hole in the form of \cite{Liu2022,Alipour2023}
\begin{eqnarray}
\Phi =\frac{1}{\sin \theta}\, T_{\rm H},\label{ee1}
\end{eqnarray}
which is represented in vector space using the vectors \cite{D2023,Rizwan2023,D2025}
\begin{equation}
\phi_{\Phi}^r=\partial_{r_h}\Phi,\qquad
\phi_{\Phi}^\theta=\partial_\theta\Phi.
\end{equation}
The zero points of this vector space are located at $\theta=\pi/2$ and $\partial_{r_h}T_{\rm H}\big|_{r_h=r_c}=0$.

Researchers have developed several approaches to compute the generalized Helmholtz free energy from different perspectives. To examine the topological structure of thermodynamics, we consider the following form of the generalized Helmholtz free energy \cite{Wei2022,Di2024,Liu2025} given by
\begin{equation}
\mathcal{F}=M(r_h)-\tfrac{S}{\tau},\label{ee2}
\end{equation}
where $\tau$ is the inverse temperature outside the horizon and $S=\pi r_h^2$ represents the entropy of the black hole.

The vector space $\phi^a_{\mathcal{F}}$ of this energy can be represented using the vectors
\begin{align}
\phi^{r}_{\mathcal{F}}&=\partial_{r_h}\mathcal{F}=\frac{1}{2}\,\left[1 - \alpha+\frac{\lambda}{r_h} + \frac{3 w N}{r_h^{3w+1}}+\frac{3}{\ell^2_p}\,r^2_h\right]-\frac{2\,\pi\,r_h}{\tau},\\
\phi^{\theta}_{\mathcal{F}}&=-\cot \theta \csc \theta,\label{ee4}
\end{align}
where $a=1,2$ and $\phi^{1}_{\mathcal{F}}=\phi^{r}_{\mathcal{F}}$,\, $\phi^{2}_{\mathcal{F}}=\phi^{\theta}_{\mathcal{F}}$. 

The zero points of this vector space $\phi_{\mathcal{F}}^a$ are located at $\theta=\pi/2$ and $\partial_{r_h}\mathcal{F}\big|_{r=r_c}=0$. Moreover, the vector space has infinite magnitude and points away from the origin when $\theta=0$ and  $\theta=\pi$. The ranges for $r_h$ and $\theta$ are $0 \leq r_h \leq \infty$ and $0 \leq \theta \leq \pi$, respectively. We can rewrite the vector space as $\phi^a_{\mathcal{F}}=||\phi^a_{\mathcal{F}}||\,e^{i\,\Theta}$, where $||\phi^a_{\mathcal{F}}||$ is the magnitude. Based on this, the normalized vector is defined as
\begin{equation}
n^{r}_{\mathcal{F}}=\frac{\phi^{r}_{\mathcal{F}}}{||\phi^a_{\mathcal{F}}||},\; n^{\theta}_{\mathcal{F}}=\frac{\phi^{\theta}_{\mathcal{F}}}{||\phi^a_{\mathcal{F}}||},\; ||\phi^a_{\mathcal{F}}||=\sqrt{(\phi^{r}_{\mathcal{F}})^2+(\phi^{\theta}_{\mathcal{F}})^2}.\label{ee6}
\end{equation}

Additionally, we determine $\tau$, using the condition $\partial_{r_h}\mathcal{F}=0$. This yields a relation for $\tau$ in terms of the event horizon $r_h$ as
\begin{eqnarray}
\tau=\frac{2\pi r_h}{M'(r_h)}=\tfrac{4\pi r_h}{1-\alpha+\frac{\lambda}{r_h}+\frac{ 3 w N}{r_h^{3w+1}}+\frac{3}{\ell^2_p}\,r^2_h}=T^{-1}_H.\label{ee5}
\end{eqnarray}

For the static black hole solution, the normalized vector field $\boldsymbol{n}_{\mathcal{F}}$ of free energy $\mathcal{F}$ is plotted on a portion of the $r$–$\theta$ plane, as shown in Figure~\ref{fig:unit-vector-3}, with parameters $M = 1$, $N = 0.1$, $w = -2/3$, $\lambda = 0.1$, $\ell_p = 50$, and $\tau=20\pi$. In this figure, the event horizon is located at the point indicated by the black dot at $(r, \theta) = (r_h, \pi/2)$, where (i) $r_h = 2.5$ for $\alpha = 0.1$, and (ii) $r_h = 2.0$ for $\alpha = 0.2$. 

Figure~\ref{fig:inversion-temperature} illustrates the behavior of the inversion temperature $\tau$ as a function of the horizon radius for different values of the parameters $\alpha$ and $\lambda$, while keeping the quintessence-like field parameters fixed. Figure~\ref{fig:inversion-temperature} (i) shows that the inversion temperature increases with increasing values of $\alpha$. In contrast, Fig.~\ref{fig:inversion-temperature} (ii) indicates that the inversion temperature decreases as $\lambda$ increases.

\section{Conclusions} \label{sec:7}

We analyzed a static, spherically symmetric Schwarzschild-AdS black hole dressed by a cloud of strings, surrounded by perfect-fluid dark matter and a quintessence-like field. We showed how these external sectors reshape both the geometry and the observable dynamics. The string cloud acts as an angular deficit, systematically enlarging the horizon and pushing the photon region outward. The dark-matter contribution introduces a mild, radius-dependent deformation that produces non-monotonic shifts in turning points and barrier heights, while the quintessence term lowers the lapse at intermediate and large radii and competes with the AdS confinement. Together, these effects displace the photon-sphere radius and the critical impact parameter, modifying capture versus scattering and the expected shadow scale. A topological diagnosis based on the $H$-potential confirms a single standard, unstable light ring with unit charge in the physical branch.

For timelike motion, we derived the specific energy and angular momentum for circular orbits and numerically identified the ISCO. The trends mirror the null case: the string cloud and quintessence move the ISCO outward, whereas the dark-matter parameter produces subtler, non-monotonic changes. Despite these geometric shifts, the radiative efficiency at the ISCO remains close to the Schwarzschild benchmark across wide parameter ranges, varying at the sub-percent level with the string-cloud and dark-matter sectors and only modestly with the quintessence normalization. This indicates that the external fields primarily relocate characteristic radii without significantly impacting thin-disk energetics.

Thermodynamically, we computed the ADM mass and Hawking temperature and surveyed their behavior along the physical branch defined by positive mass and non-negative temperature. Increasing the string-cloud strength lowers the temperature, the dark-matter parameter induces a non-monotonic response, and larger quintessence normalization reduces the temperature and can narrow the admissible horizon domain; extremality is reached as the temperature vanishes. A complementary thermodynamic-topology construction yields the expected index at the outer horizon, consistent with a single physical black-hole branch. Known limits are recovered by dialing off individual sectors, providing nontrivial checks of the framework.

These results establish a compact, tractable model that links external matter fields to optical, dynamical, and thermodynamic observables in AdS black holes. On the observational side, shadow and photon-ring sizes are more sensitive to the string-cloud and quintessence sectors. In contrast, accretion efficiency is comparatively robust, suggesting a pathway to break parameter degeneracies when imaging data are combined with independent disk constraints. Natural next steps include extending the analysis to rotating backgrounds, computing quasinormal modes and ringdown systematics tied to the photon region, incorporating more realistic dark-sector profiles and time variability, and performing radiative-transfer calculations to translate geometric trends into multiwavelength signatures. We expect the present setup to serve as a practical baseline for confronting external-field black-hole phenomenology with current and forthcoming observations.

\section*{Acknowledgments}

F.A. acknowledges the Inter University Centre for Astronomy and Astrophysics (IUCAA), Pune, India for granting visiting associateship. E. O. Silva acknowledges the support from grants CNPq/306308/2022-3, FAPEMA/UNIVERSAL-06395/22, and (CAPES) - Brazil (Code 001).

\section*{Data Availability Statement}

This manuscript has no associated data.

\section*{Conflict of Interests}

Authors declare no conflict of interest.

\end{document}